\documentclass[letterpaper,twocolumn,english,aps,rmp,superscriptaddress,nofootinbib,floatfix]{revtex4}
\usepackage{graphics}
\usepackage{amstext}
\usepackage{graphicx}
\usepackage{epsfig}
\usepackage{hyperref}

\def\Ang{\buildrel _{\circ} \over {\mathrm{A}}}
\def\beq{\begin{equation}}
\def\eeq{\end{equation}}
\def\beqa{\begin{eqnarray}}
\def\eeqa{\end{eqnarray}}

\def\e{\epsilon}

\def\half{{\ss 1\over 2}}

\def\D{\Delta}
\def\del{\delta}
\def\e{\epsilon}
\def\cH{{\mathcal H}}

\def\T{{\mathcal T}}
\def\ss{\scriptstyle}

\def\Tr{\mathrm{Tr}}
\def\L{\mathrm{L}}
\def\R{\mathrm{R}}

\def\d{\mathrm{d}}

\def\ss{\scriptstyle}
\def\al{\alpha}

\def\si{\sigma}
\def\etal{{\sl et al.}}
\def\o{\omega}

\def\nonum{\nonumber \\}

\def\del{\delta}

\def\dag{\dagger}

\def\nonum{ \nonumber \\}
\def\TL{ T_\mathrm{L}}
\def\TR{ T_\mathrm{R}}
\def\Teff{T_{\text{eff}}}
\def\rhot{\stackrel {\leftrightarrow}{\rho}}

\newcommand{\ket}[1]{| #1 \rangle}
\newcommand{\bra}[1]{\langle #1 |}

\begin{document}
\title{Heat flow and thermoelectricity in atomic and molecular junctions}

\author{Yonatan Dubi \footnote{current address: School of Physics and Astronomy, Tel-Aviv University, Tel Aviv, Israel}}
\email{jdubi@lanl.gov}
\author{Massimiliano Di Ventra}
\email{diventra@physics.ucsd.edu}
\affiliation{Department of Physics, University of California - San
Diego, La Jolla, CA 92093}

\begin{abstract}
Advances in the fabrication and characterization of nanoscale
systems now allow for a deeper understanding of one of the most
basic issues in science and technology: the flow of heat at the
microscopic level. In this Colloquium we survey recent advances
and present understanding of physical mechanisms of energy transport
in nanostructures, focusing mainly on molecular junctions and atomic wires. We examine basic issues such as thermal conductivity,
thermoelectricity, local temperature and heating, and the relation between heat current density and temperature gradient - known as Fourier's law. We critically report on both theoretical and experimental progress in each
of these issues, and discuss future research opportunities in the
field.
\end{abstract}

\maketitle
\tableofcontents
\section{Introduction} \label{sec:introduction}

Understanding how heat is carried, distributed, stored and
converted in various systems has occupied the minds of many scholars
for centuries. Recently, the problem has garnered even more
attention and has grown considerably in importance. This is not due
only to purely academic reasons: its practical impact in society has been recognized as one of the most critical
programs for the development of the necessary
resources to sustain the future welfare of mankind
\cite{DOE}.

In conjunction with these motivations, research seems to suggest that nanoscale systems (such as
carbon-based nanostructures, organic molecules, etc.) may be good candidates for such
technological advances. For instance, the flow of heat in nanoscale systems may be harnessed via
thermoelectric effects \cite{Majumdar:02062004,Bell:09122008,Rodgers:1748} to generate
heat-voltage converters, which (if their efficiency can be improved) may have real impact on
global energy consumption. Other interesting applications, such as nanoscale local refrigerators
\cite{Shakouri:1705146}, thermal transistors
\cite{saira:027203,Franceschi:538,giazotto:217,Lo:054402,Li:143501}, thermal rectifiers
\cite{Casati:184301,Terraneo:094302,segal:034301,Li:184301,Yang:033107,Wu:085424} and nanoscale radiation detectors \cite{giazotto:217} and even thermal memory and logic gates \cite{Wang:267203,Wang:177208}   add to
the importance and interest of this research field.

In spite of the recent advances, this research program still
presents quite a few challenges related to the intrinsic
non-equilibrium nature of the problem. In the
presence of a heat current, quite generally, both electrons and
ions may be very far from their equilibrium state. In addition, they
are in interaction with each other and, at the same time, in {\it
dynamical} interaction with one or more environments.

To complicate matters, heat flow is in many ways (as we will
discuss in detail in the following sections) fundamentally
different from charge flow. Therefore, many of the theoretical tools
which are used to describe charge transport cannot be
straightforwardly and uncritically extended to the study of heat
transport. From an experimental perspective, studying energy flow at
the nanoscale is in several ways more challenging than studying
charge transport, one reason being that no simple device analogous
to an ``ammeter'' is at hand to measure energy currents. Furthermore, the scale of
achievable thermal conductivities is generally much smaller than that of electrical conductivities~\cite{Majumdar:02062004}. Consequently, one has to necessarily introduce models by which the thermal conductance can be \emph{deduced} from measurable quantities such as charge current, voltage and temperature. In addition, measurement schemes with macroscopic probes
are necessarily used so that the channeling of heat only across the junction is difficult to achieve.

\begin{figure}[ht]
\vskip 0.5truecm
\includegraphics[width=8truecm]{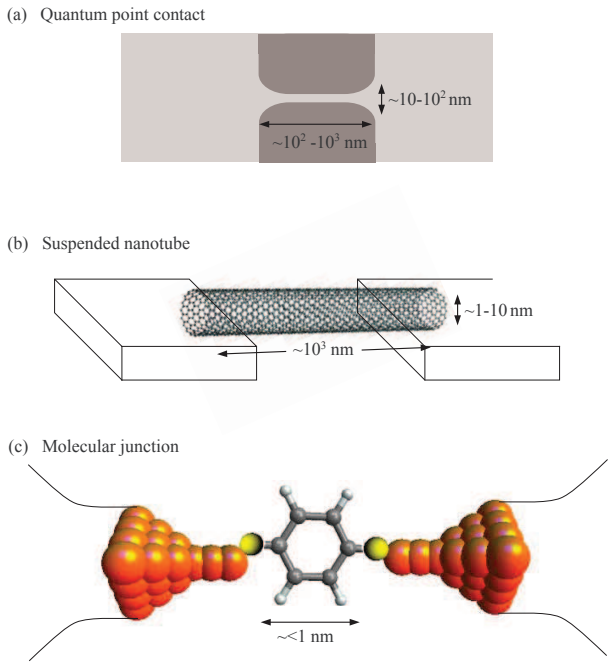}
\caption{(Color online) Schematic representation of the different
systems we consider in this Colloquium, ranging from metallic point contacts to molecular junctions.}\label{systemsscheme}
\end{figure}
In this Colloquium we will discuss all these issues
at the microscopic level. The basic systems we
will consider consist of a nanoscale junction, namely two leads
connected by a nanoscale element, with possibly a third lead
controlling some state variable of the system, e.g., its local
temperature. Typical examples are point contacts or quantum
dots placed between a two-dimensional electron gas
\cite{vanHouten:215,Staring:57,Molenkamp:903,Godijn:2927,scheibner:041301},
a molecule trapped between a substrate and a scanning tunneling
microscope (STM) tip \cite{Reddy:1568,Baheti:715}, metallic wires
\cite{Ludoph:12290}, carbon nanotubes \cite{Kim:215502,Yu:1842} or
silicon nanowires \cite{Hochbaum:163,Boukai:168} between two metal
contacts, etc. Fig.~\ref{systemsscheme} shows a schematic representation of the
different systems we consider. The leads are held at
different temperatures, which allow for the flow of energy (and
possibly charge) through the junction. Here, we point out that, due to space limitations, we will not
be able to discuss
the entire class of systems collectively known as ``nanomaterials'' - composite layers of various materials fabricated
on nanometer scales, which show unique electronic properties, often engineered by adding scattering mechanisms (for instance boundary scattering), that may be beneficial for
energy applications \cite{Majumdar:02062004, Volz:2009}. The interested reader may refer to Chen \cite{Chen:2005} for systems other than those presented here.

To make the review easier to follow for the reader, we have
 divided it into three main (yet closely related)
sub-topics. The first one is the transport of heat through the
system by phonons (lattice vibrations) and electrons, which (in
linear response) is mainly characterized by the \emph{thermal
conductivity} $\kappa$. This issue has already been reviewed
elsewhere~\cite{Galperin:155312,Wang:381}, emphasizing the effects of vibrations and focusing primarily on the method
of non-equilibrium Green's functions. To make the
present review complete, and in order to highlight the various theoretical methods and the open
questions that still pertain to this subject, we give it some space
here too. In particular, we will discuss the different processes
that contribute to $\kappa$ and their importance in nanoscale
junctions.

The second subject is that of the {\em local temperature} and {\em
heating} inside the nanoscale system. This issue is particularly
subtle, precisely because we are dealing with a non-equilibrium
process where a temperature difference is set at the two sides of
the nanojunction. We will address several experimental and
theoretical issues and fundamental open questions, such as: How does one define a local temperature at the nanoscale in a non-equilibrium situation?
What determines the local temperature and the temperature profile along the system?

As a corollary of the above studies we are finally led to analyze a nearly two-century old and important
physical law, which so far has eluded a satisfactory theoretical understanding, namely {\it
Fourier's law} (FL). This law, as originally formulated, states that in the presence of a
temperature difference between the two leads, (i) a temperature gradient develops, (ii)
the energy current density is proportional to it, and (iii) the constant of proportionality is independent of
system size. While FL was empirically postulated for bulk systems almost two
centuries ago \cite{Fourier:1882} and has been derived phenomenologically for phonons
more than eighty years ago~\cite{Peierls:1055}, no simple proof of its validity (or invalidity) has ever been
derived from first principles, nor do we have a well-defined set of conditions to determine its
validity for a given system \cite{Bonetto:2000}. As we will emphasize later, the issue has
everything to do with the difficulty in defining the basic quantities that enter its formulation
-- namely the local temperature and heat current -- from a microscopic, quantum mechanical point
of view.

The final issue is that of the inter-relation between the heat
flow and the electron transport through the junction, which can be
collected under the general name of ``thermoelectricity''. The
central quantity here is the \emph{thermopower} (or Seebeck
coefficient) $S$, which describes the voltage drop generated by a
temperature difference. A sample of important open questions for
this topic are: What are the different mechanisms contributing to thermoelectricity? Are they properly taken into account in the present theories?
What are the state-of-the-art experiments, and are their results
interpreted satisfactorily?

All these issues and open questions will accompany us for the full
length of this Colloquium. We will stress their importance for both
their fundamental character as well as their impact in possible
technological applications. We will finally point out possible
future research directions that could explore them in more depth.

The Colloquium is organized as follows. In Sec.~\ref{Sec1} we
discuss heat flow in nanoscale systems due to phonons,
electrons and their mutual interaction, and describe the different processes
which contribute to it. We review both theoretical tools and
state-of-the art experiments for measuring heat flow in nanostructures. We devote Sec.~\ref{Sec3} to local
temperature effects, and proceed to discussing
Fourier's law. In Sec.~\ref{Sec2} we discuss thermoelectric
effects in
nanoscale junctions. We give a detailed account of present
theoretical tools, and discuss recent experiments, with emphasis on
open issues in the field.  Finally, we conclude in Sec.~\ref{summary} with some prospects on the future
of the field.
\section{Heat current and thermal conductivity} \label{Sec1}
Let us start by reviewing the topic of heat current and thermal
conductivity of nanoscale junctions. We will not present full
derivations of the methods and results. Rather, we will outline only
the main theoretical tools. The interested reader may find extensive
accounts in recent reviews \cite{Galperin:103201,Wang:381,Dhar:457}
or books~\cite{Akkermans:2007,DiVentra:2008} where these methods are
discussed in detail. In addition, we will review recent experimental
advances in measurements of the thermal conductivity in nanoscale
systems, with emphasis on the measurement process itself and open
questions.

\subsection{Definitions}\label{def1}
When a nanoscale junction is placed in contact with leads held at
different temperatures, energy flows through it. The original
qualitative description for this phenomenon in bulk materials is
attributed to Fourier \cite{Fourier:1882}, and amounts to
\emph{Fourier's law} which states that a temperature gradient
$\nabla T$ induces a thermal current density linearly proportional
to it, namely \beq j_{th}=-\kappa \nabla T~~, \label{FourierLaw}\eeq
where $j_{th}$ is the heat current density (which may contain both
phonon and electron contributions, see below) and $\kappa$ is the
thermal conductivity (such an equation is usually valid only in the linear regime).

In Secs.~\ref{Ionheat} and~\ref{Fourier} we will expand more on the significance of the term ``temperature'' for a system out of equilibrium, and its different definitions.
Here, we anticipate that whenever we
do not discuss its meaning explicitly we
call {\it temperature that which is measured by a local thermal probe weakly coupled to the system and whose temperature has been adjusted so that the system dynamics is minimally perturbed}~\cite{DiVentra:2008}. This defines what we will later call a {\em temperature floating probe}~\cite{Dubi:97,Dubi:115415}. Note that we do {\em not} define it in terms of a probe adjusted so that the thermal
current between the system and probe is zero, precisely because we do not have means to
measure directly the thermal current (although these two definitions may give the same quantitative results).
In addition, the reader needs to keep in mind that while this is an operational definition of temperature out of equilibrium, its actual experimental determination is far from
trivial at present.

The validity of Eq.~(\ref{FourierLaw}) in nanoscale
junctions is discussed in detail in Sec.~\ref{Fourier}. Here, we are
mainly interested in the theoretical understanding and measurement
of $j_{th}$ and $\kappa$, assuming that Fourier's law is indeed
valid. A relation between the formalism described below (Landauer's formula~(\ref{LandauerFormula:1})) and Fourier's law can be determined, which requires calculation of thermal conductances at larger and larger length scales. Such derivation, discussed in other reviews \cite{ Dhar:457} implies going beyond the realm of nanoscale junctions and will thus not be discussed in detail here.

It is also convenient to introduce the \emph{thermal conductance}, which is
the ratio between the
total heat current $J_{th}$ and temperature difference $\D T=T_R-T_L$,  \beq \si_{th}=-\lim_{\D T \to 0}
\frac{J_{th}}{\D
T}~~. \label{thermal conductance} \eeq  If the sample is uniform
with a constant cross section $A$ and length $L$, the thermal
conductance is related to thermal conductivity $\kappa$ via
$\si_{th}= \frac{A}{L} \kappa$. If the sample is
not uniform, then the relation between thermal conductance and
conductivity depends on the microscopic details of the system.
In addition, in analogy with electric circuit theory, it is convenient to define the {\em thermal resistance}, being the
reciprocal of the thermal conductance: $\rho_{th}=\si_{th}^{-1}$.

Energy can be carried through a nanoscale junction (or through a
solid) either by lattice vibrations (phonons) or by electrons, or
both \footnote{At low temperatures energy can also be carried by the electromagnetic environment (photons), an effect which was studied in mesoscopic systems \cite{Schmidt:045901} but was not systematically addressed in nanoscale junctions.}. In insulating bulk materials the electronic contribution is
negligible, while it is sizeable in bulk metals. This simple
distinction is less obvious in nanoscale junctions, where, due to
the large current densities they can carry\footnote{For instance, in an atomic quantum point contact of a
nominal cross section of 0.1 nm$^2$, to a typical current of 1 $\mu$A corresponds a current density of about 10$^9$ A/cm$^2$. This is several
orders of magnitude larger than in mesoscopic or bulk systems.\label{largecurrd}}, the two contributions
may be equally important and need to be discussed on equal footing.
For bulk insulating materials, the theory of phonon thermal
conductivity based on the Boltzmann equation was derived by Peierls
\cite{Peierls:1955} (see also the detailed
review~\cite{Carruthers:1961}). The main idea is that $\kappa$ is
governed by phonon scattering, especially the so-called Umklapp
scattering (processes that do not conserve crystal momentum),
whereby phonons scatter between states which are separated (in
reciprocal space) by a reciprocal lattice vector.~\footnote{For a
homogeneous bulk system in which the Umklapp processes are suppressed and only
``normal'' processes occur (namely, processes that conserve crystal
momentum) energy can flow undisturbed, giving rise to a diverging
$\kappa$, and such a system cannot reach local or global equilibrium.}
Considering a phonon mean-free path $l$ (mainly due to scattering by impurities), simple arguments lead to
the following relation at high temperatures (in three dimensions)~\cite{Ashcroft-Mermin} \beq \kappa \approx
\frac{1}{3} l v c_v~~, \label{kappa:1}\eeq where $v$ is the velocity
of sound and $c_v$ is the phonon heat capacity at constant volume (in the above equation optical and acoustic phonons are considered on equal footing, although only the latter ones participate in heat transport). In a
bulk metal a similar relation can be derived \cite{Ashcroft-Mermin},
where now $l$ stands for the electronic mean-free path, $c_v$ is the
electronic heat capacity at constant volume, and $v$ is the electron
drift velocity. Here, a comment is in order. In the case of electrons the heat (or thermal) current
contains also a contribution from the variation of number of particles. In fact, let us consider the thermodynamic relation
(at constant volume)
$\delta Q=d{\cal E} -\mu dn$, where $Q$ and ${\cal E}$ are the heat and energy per unit volume, respectively, $n$ is the particle number
density, and $\mu$ the chemical potential. From this relation, dividing by the infinitesimal time interval $dt$, we obtain ($e$ is the electron charge)
\begin{equation}
J_{th}=J_E-\frac{\mu}{e}J_e,\label{heate}
\end{equation}
namely, for electrons the heat current has both a contribution from the energy current, $J_E$, and from the charge
current $J_e$ (there is no such term for phonons,
since their number is not conserved). In this review, we will use the terms ``energy current'' and
``heat (thermal) current'' interchangeably, but with the understanding that, in the case of
electrons, one must generally include a contribution from the variation of the number of particles (see also discussion after
Eq.~(\ref{LandauerFormula:1})).

It is now natural to ask whether these arguments can be extended to
the regime in which strong material inhomogeneities are the norm,
like in nanoscale systems. Before we embark in this quest, however,
it is worth asking why $\kappa$ is such an important quantity in the
first place, especially since measuring the thermal
conductivity at the nanoscale is all but trivial.
The answer is that $\kappa$ contains information regarding two main
processes relevant to the future applicability of nanoscale systems.
The first is the rate at which energy is dissipated in and removed
from the junction. This has an effect on the heating of the system,
which may affect its structural stability. The second is that
$\kappa$ is an important (and limiting) factor in the efficiency of
nanoscale systems as heat-voltage converters (as it will be
discussed more at length in Sec.~\ref{Sec2}). Therefore, according
to the desired use, an ideal nanosystem should have opposite thermal
properties: for current-carrying wires one wishes a high thermal
conductance that would allow heat to pass through the wire and
prevent over-heating, and for thermoelectric conversion one requires
a thermal conductance as small as possible. These requirements make
the understanding, predictability, and control of $\kappa$ highly
desirable.

\subsection{Experiment}
In this section we focus on the experimental measurements of the
thermal conductivity in nanoscale systems. As already pointed out, a
major difficulty in measuring $\kappa$ (other than the usual ones
related to any measurements at the nanoscale) stems from the simple
fact that there is {\em no direct way} to measure a heat current.
Indeed, the only directly measurable quantities are electrical
currents, voltages and temperatures (the latter also typically
measured via resistance measurements), and from these one deduces
$\kappa$. The main limitation is that the value of $\kappa$ as extracted from the experiment may then depend on
the model used to describe the whole experimental setup or device,
which may generate some ambiguity. Here, we will describe some
recent experiments, discuss the methods employed in deducing
$\kappa$, and review some of the main results.

\begin{figure}[ht]
\vskip 0.5truecm
\includegraphics[width=8truecm]{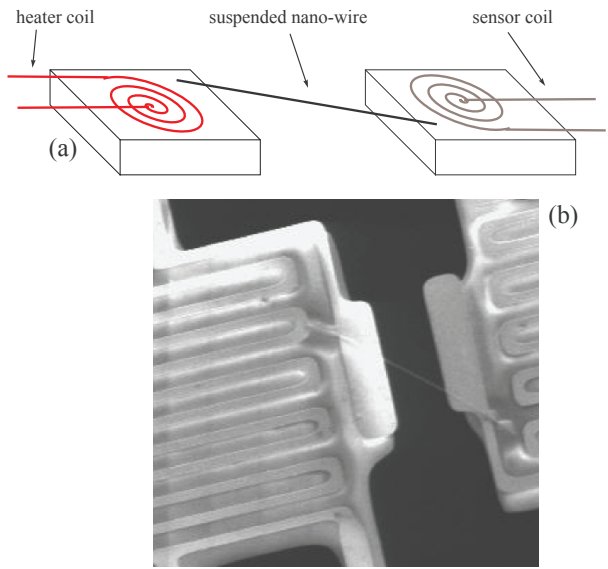}
\caption{(Color online) (a) A schematic representation for a simple
setup to measure the thermal conductance. (b) An actual device to
measure the thermal conductance of boron nitrade nanotubes (from \cite{Chang:085901}).}\label{fig1}
\end{figure}

A conceptually simple way to measure the {\em thermal conductance}
of a suspended nanojunction is the following. Consider the schematic
system of Fig.~\ref{fig1}. The ``heater'' coil is heated by passing
a current through it. By measuring the current and the voltage
through the heater coil, the power transferred through it is given
by the well-known relation, $P=I V$. This power increases the
temperature of the coil to $T_h$. At the same time, the temperature
of the ``sensor'' coil, $T_s$, is evaluated (by measuring its
resistance, which is pre-calibrated to correspond to a given
temperature). If the wire is suspended, then the entire heat
current should be equal to the power supplied by the heater coil,
$\dot{Q}=P$, which is related, in linear response, to the
temperature difference by \beq \dot{Q}=-\sigma_{th} (T_h-T_s)\;,\label{Qdot}\eeq
from which the thermal conductance $\sigma_{th}$ can be evaluated, under the assumption that all the power supplied by the electric circuit flows through the junction without loss, and the thermal conductivity $\kappa$ can then be extracted from a microscopic model that relates thermal conductivity to thermal conductance
(see Sec.~\ref{Sec1theory}). If, as indeed is the case in many experiment, some of the power is lost due to heat diffusion away from the contacts (e.g. into the substrate) then the Joule heating is the sum of the heat flowing away through the contacts and that flowing through the wire.

This method seems very simple, and was indeed employed to measure
the quantum of thermal conductance \cite{Schwab:974}. However, it
needs to be acknowledged that it has obvious limitations. For one, dissipative effects at surfaces or local thermal
gradients in the heating and cooling parts of the
coils~\footnote{Recall that one can destroy and create phonons at
the surfaces of a material.} may reduce the heat flow in the
suspended wire. In addition, recent theoretical studies indicate that the contact thermal resistance between nanowires
and
substrate plays an important role in determining the overall thermal resistance \cite{Zhong:125403,Chalopin:233309}.

More difficult is the determination of the thermal conductivity
$\kappa$ from a model that includes all the effects of device
geometry and dissipation through the contacts and substrates. Such
models vary for different devices and geometries
\cite{Shi:881,Chang:085901,Chang:075903}, but share the common
feature that thermal conductances are treated on the same footing as
classical (charge) conductances, with the same Kirchhoff-like laws
for the addition of resistances in series ($\rho_{th}=\sum_i \rho_{th,i}$, with $\rho_{th,i}$ the thermal
resistance of a single element of the circuit) and parallel ($\rho_{th}^{-1}=\sum_i \rho_{th,i}^{-1} $). Thus, the
measured thermal conductivity may depend on the circuit model
used, which makes it hard to compare between different experiments.
This means that when performing a measurement, one is in fact
measuring the thermal conductance of the system of interest {\em
embedded in that specific device}. Nevertheless, this method was
used to study the thermal conductivity of many nanoscale structures, mainly carbon nanotubes
\cite{Brown:023107,Chang:085901,Chang:075903,Shi:881,Chiu:226101,Fujii:065502,Kim:215502,Yu:1842}
but also nanotubes of other materials
\cite{chen:105501,Chang:075903,Li:2934}. Some experimental features
are universal, like ballistic thermal conductance
\cite{Brown:023107,Chiu:226101}, a value of $\kappa$ which is orders
of magnitude larger than the bulk value for carbon nanotubes ($\sim
3000 W/K$ at room temperature), an increase of thermal conductance with nanowire
diameter, or a peak of the thermal conductance at $\sim 320K$
\cite{Fujii:065502,Kim:215502}, attributed to the onset of Umklapp phonon scattering processes. However, other
features, such as the
detailed power-law dependence of $\kappa$ on temperature vary
between experiments, indicating that this is not a universal
feature, and depends on the details of the experimental setup.

Other experimental approaches to measure $\kappa$ have been
introduced in the literature. For instance, Pop \etal~\cite{Pop:96}
have used high currents to induce heating in a single-walled carbon
nanotube, with a model to relate the current-voltage (I-V)
characteristics to the high-temperature thermal conductance. In another
example, the so-called $3\o$ method \cite{Cahill:802,Lu:2996}, was
used to study nanotubes
\cite{Bourgeois:016104,Choi:013108,Choi:1589}. In this method, an
a.c. current is applied to the sample which also acts
as a heater. From a simple derivation one finds that the third
harmonic of the voltage drop across the sample is related to the
thermal conductivity of the sample (at small frequencies of the current).
Using this method, the authors found a deviation of the thermal
conductance from a cubic dependence on temperature for Si nanowires,
indicating a dimensional crossover at low temperatures. Both these
methods rely on current-induced self-heating of the sample (rather
than direct heating by an external source). In a third example,
laser-induced heating and Raman spectroscopy (already used in various nanoscale systems such as graphene ribbons \cite{Balandin:902,Calizo:2645}) have been used to
determine the local temperatures \cite{Hsu:590,Deshpande:105501} and extract the
thermal conductance of carbon nanotube bundles. The main
disadvantage of this method is that to obtain the thermal conductance
one needs to assume a value for the optical absorbtion of the
sample, which is usually unknown.

\subsection{Theoretical Methods}\label{Sec1theory}
We now provide a brief description of the theoretical methods most
commonly employed to describe energy flow, with an eye on their
strengths and limitations.

\subsubsection{Single-particle scattering
approach}\label{Sec1theorysp}
 Many theoretical calculations of
thermal conductance are based on an approach pioneered by
Landauer~\cite{LANDAUER:863,LANDAUER:223} in the context of charge
transport in mesoscopic and nanoscopic systems
\cite{Imry:1997,DiVentra:2008, Datta:1997}. The same ideas have been
generalized to phonon transport through a nanoscale junction
\cite{Dhar:805,Angelescu:673,Rego:232,Rego:239,Blencowe:4992,Blencowe:159,Segal:6840}.

The basic tenet of this approach is that one assumes the leads
non-interacting (otherwise no closed form for the current can be
obtained~\cite{DiVentra:2008}), so that a convenient basis, such as
plane-waves, can be chosen to develop state vectors for both types
of particles, either phonons or electrons. As a further conceptual
simplification, the leads are thought to be adiabatically
``connected'' to reservoirs whose only role is to define the
occupation of the scattering states according to a local equilibrium
Bose-Einstein (BE) distribution for phonons or a Fermi-Dirac (FD)
distribution for electrons. Once this occupation is set, the
particles are free to propagate in the leads before scattering at
the lead-system interface. Charge and/or energy current is then
determined by an electrochemical potential difference and/or a
temperature difference between the reservoirs.

Most of the calculations also assume that the particles in the
sample are either truly non-interacting or interacting at a
mean-field level (which is the same from a formal point of view). In
this case the current is simply proportional to the probability for the particles to cross the sample from one electrode
to the other. For instance, in the case of phonon transport, phonon
states at a given energy $\hbar \o$, scatter off the junction
and may be either transmitted through it or reflected back. The
probability to be transmitted through the junction is characterized
by the transmission coefficient ${\cal T}(\o)$. The expression for the
heat current is then simply \beq J_{th}=\int^\infty_0
\frac{d\o}{2\pi} \hbar \o {\cal T}(\o)(g_L-g_R)
~~,\label{LandauerFormula:1} \eeq where $g_{L(R)}=1/\left(
\exp\left( \frac{\hbar \o}{k_B T_{L(R)}}\right)-1\right)$ are the
distribution functions of phonons in the left (right) lead. From
$J_{th}$ one can then evaluate the thermal conductance according to Eq.~(\ref{thermal conductance}).

Within this approach the electronic contribution to the heat current
is calculated similarly, where in Eq.~(\ref{LandauerFormula:1}) one
makes two changes, namely (i) the BE distribution functions are
replaced by FD distributions, and (ii) the energy in each reservoir
is measured from the respective electrochemical potential, $\mu_L$
and $\mu_R$ for the left and right reservoir, respectively, i.e.,
$\hbar \o \rightarrow \hbar \o-\mu_{L,R}$ (see Eq.~(\ref{heate})). In linear response this leads to
the substitution $\hbar \o \rightarrow \hbar \o-(\mu_{L}+\mu_R)/2$ in the
energy term that multiplies ${\cal T}(\o)$ in
Eq.~(\ref{LandauerFormula:1}).

To actually evaluate $\si_{th}$, one has to compute the
transmission coefficient ${\cal T}(\o)$. To this aim several
methods have been employed, such as the use of continuum models
\cite{Angelescu:673}, boundary condition method \cite{wang:054303},
mode-matching method \cite{Khomyakov:195402,Ando:8017,Ting:3583} and
scattering or transfer matrices \cite{Tong:8639,DiVentra2002}. All
these methods are fundamentally equivalent, and in fact have their
origin in the single-particle elastic scattering theory of
conduction (see, e.g., \cite{DiVentra:2008}), whereby one can write
the transmission coefficient as a sum of all the partial
probabilities of transmission $T_{if}(\o)$ from one of the momentum
states of the incoming ($i$) particle (whether electron or phonon) at
energy $\hbar\o$ to one of the momentum states of the outgoing ($f$)
particle at the same energy, namely~\cite{Butt1} \beq {\cal T}(\o)= \sum_i\sum_f
T_{if}(\o) =\mathrm{Tr} \{\tau \tau^{\dag}\},\label{Scattering
Formula}\eeq where $\tau$ is a sub-matrix of the scattering matrix
with dimensions $N_R\times N_L$, with $N_R$ and $N_L$ the number of
channels in the right and left leads, respectively, at energy $\hbar
\o$.
This result can be cast in another equivalent form in terms of
single-particle Green's functions via \cite{Meir:2512} \beq {\cal T}(\o)=\Tr \left\{
G^r \Gamma_L G^a \Gamma_R \right\},\label{Caroli Formula} \eeq where
$G^{r(a)}$ is the retarded (advanced) single-particle Green's
function corresponding to the interaction of a ``central'' part of the junction
with the electrodes and
$\Gamma_{L(R)}$ describe the ``rate'' at which particles scatter between the leads and the
central part of the junction. It has been re-derived for thermal
transport by many authors
\cite{Ozpineci:125415,Segal:6840,Wang:033408,Mingo:245406,Mingo:125402,Yamamoto:255503,Galperin:155312,Wang:061128,Dhar:457}.

Arguably the most universal result obtained from the Landauer
formula~(\ref{LandauerFormula:1}) is that of thermal conductance
quantization. Similarly to the quantization of electrical conductance in
ideally one-dimensional (1D) electronic systems~\cite{Wees:848}, at
low temperatures the thermal conductance (per phonon mode) was
predicted to acquire a quantized value \beq \si_{0}=\frac{\pi^2
k^2_B T}{3 h}~~,\label{Quantized Conductance}\eeq where $h$ is
Planck's constant \cite{Pendry:2161,Maynard:5440,Greiner:1114,Rego:232}. This
result is readily derived from Eq.~\ref{LandauerFormula:1} in linear response by
setting the number of modes to unity, and letting the transmission
coefficient to be one, i.e., ${\cal T}(\o)=1$.

The fact that this conductance is material-independent relies on the
fact that, like in the electronic case, in 1D the phonon density of
states is exactly proportional to the inverse of the group velocity.
Remarkably, thermal conductance quantization does not depend on the
statistics of the carriers \cite{Rego:239}. Indeed, $\si_0$ was
experimentally measured for phonons \cite{Schwab:974}, electrons
\cite{Nicholls:164210,Chiatti:056601} and even photons
\cite{Meschke:187}.

Another application of the Landauer
formula~(\ref{LandauerFormula:1}) has been in the study of
geometrical and temperature effects on thermal transport. To give a
few examples, this approach has been used to understand the role of
defects on the thermal conductance of a nanowire \cite{Chen:045422},
the effects of different geometries such as stubs, T-junctions and
concavities \cite{Peng:193502,Tang:163505,Xie:084501}, periodic
modulations \cite{Tang:1497}, and surface roughness
\cite{Kambili:15593,Santamore:184306}. As a general rule, disorder and temperature
are found to have competing roles: disorder tends to reduce the
thermal conductance (by decreasing the transmission coefficients of
the different transport modes), and a temperature increase usually
results in a larger thermal conductance, due to an increased number of
modes which participate in the thermal transport.

\begin{figure}[ht]
\vskip 0.5truecm
\includegraphics[width=8truecm]{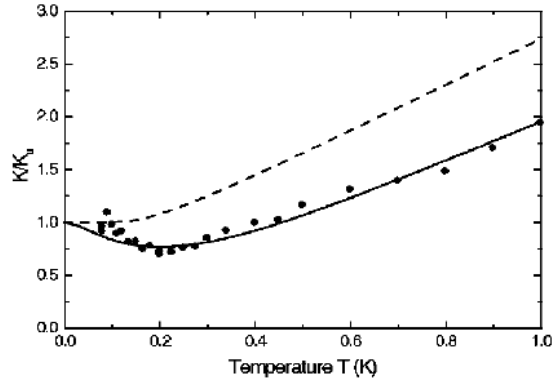}
\caption{Thermal conductance of a quasi-1D wire with surface roughness, exhibiting inhomogeneous thermal conductance.
Points correspond to experimental data of Schwab \etal \cite{Schwab:974}, and the solid line is the theoretical curve.
From~\cite{Santamore:184306}.}\label{fig-Santamore}
\end{figure}

The interplay
between the two processes can result in interesting phenomena. For
instance, Santamore \etal~\cite{Santamore:184306} showed that
disorder in the form of surface roughness may generate a
non-monotonicity in $\sigma_{th}$ with increasing temperature, with a
slight decrease (below the quantum of thermal conductance) followed
by a rise of  $\sigma_{th}$ with increasing temperature, in similarity to
the experimental results of Schwab \etal~\cite{Schwab:974}. Their
results (shown in Fig.~\ref{fig-Santamore}) are explained as
follows: at very low temperatures, there is only one mode which
contributes to the thermal conductance. As temperature increases,
scattering of that mode off the surface roughness increases,
generating a decrease in the thermal conductance. As the temperature
is raised even higher, higher modes start to participate in the
thermal transport, giving rise to an increase in the thermal
conductance.

This ties with the use of the scattering approach to thermal
conduction in real materials, which comes about from using realistic
phonon spectra (e.g., as obtained from experiment or
first-principles approaches) in combination with ground-state
density-functional theory (DFT) calculations to obtain the
scattering coefficient ${\cal T}(\o)$. To give several examples,
Tanaka \etal~have combined geometrical structure (i.e., realistic shape of the wire) with real material
parameters to study the onset of the thermal conductance
quantization in GaAs and silicon nitride wires \cite{Tanaka:205308}.
The thermal conductance of nanowires made of, e.g., Si, Ge and GaAs
was studied by several authors \cite{Tanaka:205308,
Mingo:245406,Mingo:1713,Wang:241908}. Much attention has been given
to carbon-based structures, such as carbon nanotubes, graphene and
graphite
\cite{Zhang:114714,Mingo:096105,Mingo:1221,Yamamoto:075502,Yamamoto:255503,Zimmermann:045410,Lan:115401,Lu:235436}.
Another example is the recent study of isotope and disorder
effects \cite{Murphy:155313}, specifically in carbon and
boron-nitride nanotubes \cite{Stewart:81,Savic:165502,Savic:235434}.

Some universal conclusions arise from these calculations. For
instance, a dimensional crossover from three- to one-dimensional
transport (manifested by, e.g. a change in the temperature dependence of the thermal conductance) occurs in many systems as the diameter of the nanotube
decreases, the length scales determined by the wavelengths
of the typical phonon modes \cite{Wang:241908}. Also, disorder in various forms (local
defects, surface roughness, etc.) has a dramatic effect on the
thermal conductance, as it influences the scattering of the
different modes~\cite{Dhar:2008}. Due to the translational invariance
of the lattice, long wavelength (or zero-frequency) modes are always conducting, while short wavelength modes are
scattered by disorder. Since the
short wavelength modes participate in the thermal transport only at
high temperatures, it is found that the low-temperature thermal
conductance is less affected by disorder and defects. Finally, the
thermal conductance of molecular junctions has also been widely
studied \cite{Segal:6840,Galperin:155312,Mingo:125402}. It is found
to be strongly dependent on a multitude of factors, among which the
phonon spectrum of the molecules, the degree of localization of the
molecular modes, the molecule-lead coupling, non-harmonicity (i.e.
phonon interactions), etc.

It is important to stress once more that
Eqs.~(\ref{LandauerFormula:1}),~(\ref{Scattering Formula}),
~(\ref{Caroli Formula}), and indeed the whole Landauer approach, are
based on some strong assumptions, which may breakdown in nanoscale
junctions and under certain experimental conditions. The first
assumption is that the system is ``closed'', in the sense that it
does not dynamically interact with its environment. The latter only
provides the boundary conditions and the relevant parameters (like
the temperatures, chemical potentials, etc.). The second assumption
is that the leads are ideal, i.e., are unaffected by the proximity
to the junction (either in their structure or in the distribution of
particles) and support well-defined single-particle states. In
addition, it is assumed that ``dissipation'' takes place at the
(infinitely far) edges of the leads and that the temperature (and
chemical potential for electrons) is uniform in them. Most
critically, the approach does not provide information on the {\em
dynamics} of the system. Therefore, transient, memory and non-linear
dynamical phenomena are beyond its reach. A further issue arises
when one uses ground-state DFT in combination with the Landauer
approach: one is effectively using a ground-state theory for a
non-equilibrium problem. This issue cannot be
solved by knowledge of the exact ground-state exchange-correlation functional,
and as such, the use of ground-state DFT in this context can only be
viewed as a mean-field approximation. This has been explicitly demonstrated in~\cite{Vignale2009}, where for
the case of electrical conductance it was shown that the exact resistivity tensor $\rhot$ can be written as
\beq\label{rhoDyson} \rhot = \rhot_s +
\rhot_{xc} \eeq where $\rhot_s$ is is the resistivity tensor of a noninteracting
system in the presence of a static potential $V_s$ that reproduces the {\em exact} ground-state
density, and $\rhot_{xc}$  is a {\em dynamical} contribution related to dynamical exchange-correlation effects,
and which does not vanish even
in the zero-frequency (d.c.) limit. A possible way out would be to
use a fully dynamical approach (e.g., the microcanonical picture of
transport as suggested in~\cite{Ventra:8025}) combined with
time-dependent DFT~\cite{Runge:997}. This approach (recently implemented to study charge transport \cite{Cheng:155112})
would provide, in
principle, the exact thermal total current, if the exact {\em
dynamical} exchange-correlation potential is known. However, we are
not aware of any calculation of thermal current along these lines.

\subsubsection{The role of interactions}

Up to this point the system Hamiltonian has been assumed to describe
single particles with interactions included at most at the
mean-field level. As briefly mentioned above, many-body correlations
can be accounted for within a time-dependent DFT approach, namely
within an effective single-particle picture. Alternatively, the
effect of interactions beyond mean-field, could be explicitly
included via the so-called non-equilibrium Green's functions
formalism (NEGF) (see, e.g., N. Mingo, chapter $3$ in \cite{Volz:2009}). In this approach one solves equations of motion
for appropriate single-particle Green's functions that can be
conveniently defined on the Keldysh contour \cite{KadanoffBaym,
Keldysh:1515}. In its exact formulation, the NEGF has however limited
practical utility, since if one assumes particles interacting -
beyond mean-field - in the {\em whole} system (leads plus
nanostructure) no closed equation of motion for the single-particle
Green's functions can be obtained~\cite{DiVentra:2008}. Instead, it
is common to assume (as in the Landauer approach) that the leads
contain non-interacting particles and interactions are confined
within a ``central'' region containing the nanostructure. This is a
strong assumption and may not always correspond to the physical
problem at hand and/or its experimental realization.

If one makes the assumption of non-interacting particles in the
leads, and assumes that a steady-state has been reached in the
long-time limit (not an obvious statement either), the equation of
motion for the different single-particle Green's functions can be
closed and the NEGF provides a compact expression for the total
current similar to that derived for electron transport \cite{Meir:2512}, given by  \beqa J&=&\frac{1}{4 \pi} \int^\infty_0 \hbar \o d
\o \left[ (G^r-G^a)(\Sigma^<_R-\Sigma^<_L)+ \right. \nonum & &
~~~~~~~~\left.+i G^<(\Gamma_R-\Gamma_L) \right]
~~,\label{NEGF1}\eeqa where $G^{r,a,<}$ are the retarded, advanced
and ``lesser'' single-particle Green's functions, respectively;
$\Sigma^<_\al$ are the ``lesser'' self energies of the $\al=L,R$
leads and $\Gamma_\al=i(\Sigma^r_\al-\Sigma^a_\al)$ namely the difference
between ``retarded'' and ``advanced'' self energies (the explicit
$\o$-dependence of all these quantities has been omitted). The first term on the right hand side of Eq.~(\ref{NEGF1}) may be interpreted as describing the current from the bias-induced difference in the coupling to the leads, while the second is related to the non-equilibrium distribution function in the interacting region. The
single-particle Green's functions can represent either phonons or
electrons, and should be calculated in the presence of interactions.
In the mean-field approximation, Eq.~(\ref{NEGF1}) reduces to
Eq.~(\ref{LandauerFormula:1}) (or its equivalent form for fermions).
Many-body perturbation expansions to compute these Green's functions
have been performed for simple model
Hamiltonians~\cite{Galperin:155312,Lu:165418} but it is no easy task
to introduce interactions (beyond mean field) in realistic systems.

The NEGF could also be used to study the effects of electron-phonon
interactions. In that case as well, however, quite strong
approximations need to be made in order to have an analytically
tractable theory. For instance, if one assumes electrons interacting
with each other at a mean-field level, but interacting in a
``central'' region with non-interacting phonons, the heat current
can be approximated as a sum of contributions from both electrons
and phonons, $J=J_{el}+J_{ph}$ , each component calculated with the
help of Eq. ~(\ref{NEGF1}). The key ingredient here is that, due to
the electron-phonon interaction, the self-energy of phonons includes
an electronic contribution and vice versa. These contributions can
be calculated in a perturbative way. However, this is clearly an
idealization, since it neglects correlated electron-ion motion,
which, in principle, does not even allow the total thermal current
$J$ to be separated into two distinct contributions from the two
particle species. Along the same lines of reasoning, the effects of
phonon-phonon interaction have been
studied~\cite{Xu:224303,Liu:3194,Mingo:125402}. According to these
results both electron-phonon and phonon-phonon
interactions decrease the thermal conductivity. However, we
need to stress once more that due to the large current densities
nanoscale systems carry - and hence the large number of scattering
events per unit time and unit volume - it is not a simple task to
include all the relevant physical scattering mechanisms in the present non-equilibrium case.
An example of this is the possibility of phonon modes in the junction which are weakly coupled to
the bulk modes of the electrodes. In this case, these ``localized'' modes may be energetically ``pumped'' by scattering
with electrons or other phonons before
energy could efficiently be dissipated away. This physical situation is beyond second-order perturbation theory and more work
in this direction is thus highly desirable.

\subsubsection{Molecular dynamics}\label{MolecDynam}
Another method to evaluate the thermal conductivity which is gaining
increasing popularity is that of molecular dynamics (MD). Basically,
molecular dynamics comes down to solving the classical equations of
motion of the system numerically. The origin of the method in the
present context can be traced back to the seminal work of
Fermi, Pasta and Ulam~\cite{Fermi:1940}, where the energy transfer
in non-harmonic lattices has been studied numerically. Since then it
has been widely used to study heat transport in classical 1D systems
\cite{Dhar:457,Lepri:1}. It has also been generalized to study
quantum effects, by providing appropriate boundary conditions
\cite{Wang:381}. These approximations, however, should be thought of
as quasi-classical, since the microscopic dynamics of the system is
described by classical Newtonian equations of motion,
and the quantum nature is only introduced via indirect
conditions (such as the noise in a Langevin term). A big advantage
of molecular dynamics is the ability to model realistic systems and
geometries in a rather straightforward way. The forces between atoms
are evaluated from realistic parameters, so that different
geometries, impurities, structures, etc. are easily taken into
account.

In order to calculate the heat transport directly from MD, one needs to account for a finite temperature in the
system. This is usually done in linear response by adding to the
Newtonian equations of motion a Langevin fluctuating term which
satisfies the fluctuation-dissipation relation, i.e., the two-time
correlation function of the current is proportional to the
temperature (see, e.g.,~\cite{VanKampen:2001}). Alternatively, a Nos\'{e}-Hoover thermostat is introduced, in
which a fictitious coordinate is added to the real coordinate to
maintain a finite temperature~\cite{Nose:511,Hoover:1695}.

Once a finite temperature is set, there are two main methods to
calculate the thermal conductivity. The first (sometimes called equilibrium MD) is via the
linear-response Green-Kubo formula \cite{Luttinger:1505,Dhar:457,Lepri:1}\beq
\kappa=\frac{1}{3V k_B T^2} \int^\infty_0 \langle J_{th}(t)
J_{th}(0) \rangle dt ~~,\label{Green-Kubo}\eeq where $V$ is the
volume, $k_B$ the Boltzmann constant, $T$ is the system temperature,
$J_{th}(t)=\int d{\bf r} j_{th}({\bf r},t)$ is the integral of the
heat current density, $j_{th}({\bf r},t)$, over the entire system,
and the brackets denote equilibrium ensemble averaging in the
absence of a thermal gradient. However, the Green-Kubo equation has two main
weaknesses. The first is that it is derived in the thermodynamic
limit and therefore its use in finite systems is not well
justified~\cite{Kundu2009}. Secondly, one needs to assume that a small temperature gradient (the external
perturbation)
ensues in the system, which may not be the case in every experiment.
However, its relative simplicity makes it a good starting point in many
cases.

An alternative method (also known as nonequilibrium MD), still based on molecular dynamics, is the one
in which the system is held in contact between two heat baths of
different temperatures. Once the dynamics reaches a steady state,
the temperature profile and the local heat currents can be calculated, from which the thermal
conductivity is extracted. Here lies one of the disadvantages of the
model, since the definition of the local heat current requires
defining a local energy operator, which is not always a unique
quantity \cite{Lepri:1,Wu:025302}. Likewise, a local
temperature needs to be defined and evaluated; a somewhat tricky
issue to which we will come back in Sec.~\ref{Sec3}. At high temperatures (where the distribution function is practically classical and quantum effects are negligible; say at
temperatures higher than the typical vibrational mode temperature) one may define the
local temperature as the kinetic energy of the atoms (via the
equipartition function), but this assumption breaks down at low
temperatures, and one needs to use a definition of temperature which
rests on the equilibrium distribution of phonons~\cite{Wang:381}.
This yields a quasi-classical treatment (which is somewhat better
than a fully classical treatment at low temperatures), but leans on
the assumption that the phonon distribution resembles its
equilibrium form, which may not be the case in this non-equilibrium
problem. On the other hand, the obvious advantage of this method is
that it does not rely on any thermodynamic-limit assumptions and is thus
applicable for any system size, which is important for the study of
realistic nanoscale systems. For instance, Yang \etal~\cite{Yang:201085} recently used the method to
study Fourier's law and thermal conductance of realistic Si nanowires, and showed that Fourier's law breaks down in these systems (see Sec.~\ref{Fourier}).
Studies along similar lines have been recently performed to
investigate the thermal conductance of carbon nanotubes
 \cite{Berber:4613,Hu:083503,Padgett:1051}, Si wires \cite{Ponomareva:1155,Yang:276,Henry:141}, diamond nano-rods
 \cite{Padgett:1827} and polyethylene chains \cite{Henry:235502}, to name only a
 few recent studies.

An additional method, related to MD, is that of {\it lattice dynamics} models. In this method the phonon dispersion relations are obtained by calculating the direct change in energy due to atom displacements, using force fields obtained from DFT calculations \cite{Ren:103505,Feldman:9209,Turney:064301}.

The abundance of literature makes it hard to describe universal
features of the thermal conductance, which seems to strongly depend
on the details of the model and/or material. Specifically, $\sigma_{th}$
is very sensitive to the phonon spectra and to phonon localization \cite{Dhar:134301,Zhernov:308},
which are in turn sensitive to material, geometry and disorder,
surface roughness, and more. The rationale behind these studies is that by uncovering the
detailed influence of these parameters on $\sigma_{th}$, theory may
provide guidance to experiments and even suggest new materials with
optimized thermal properties.
\section{Local temperature and heating} \label{Sec3}
\subsection{General remarks}
When a current passes through a classical resistor, the latter heats up. This phenomenon is known as ``Joule heating''.
It is a consequence of the inelastic relaxation of electrons in the resistor which transfer energy to the surrounding lattice~\cite{Ashcroft-Mermin}.
In a nanoscale system, such as a molecular junction or an atomic wire, electrons can analogously
scatter inelastically off the phonons (i.e., the vibrational modes of the structure).
However, since electrons typically spend very little time in the junction region, one might naively think
that their inelastic scattering rate is negligible with consequent little heating of the junction itself. This conclusion, which is, for instance, at the heart of the Landauer scattering approach where all dissipation is assumed to occur only in the ``reservoirs'', does not take into account the fact that due to the small cross-section of nanoscale systems, the \emph{current density} at the junction can be very large (typically much larger than in mesoscopic and macroscopic systems, see footnote~\ref{largecurrd}).
This implies that the power \emph{per atom} in the junction can be very large, possibly leading to large local heating~\cite{Todorov:965,Chen:1691,Chen:621}. The rate at which this power is then dissipated back to the electrodes determines the effective local (and out of equilibrium) temperature of the junction.

In addition, current-carrying electrons can transfer energy, via inelastic electron-electron interactions, to other electrons in the Fermi gas~\cite{D'Agosta:2935}.
This effect is generally small
in macroscopic systems. However, similarly to the increased rate of electron-phonon scattering in nanoscale junctions due to the large current densities, the inelastic scattering
rate of electron-electron interactions may increase in nanoscale systems leading to a local heating of the electron liquid~\cite{D'Agosta:2935}. This effective higher temperature of the electrons may influence the local ionic heating due to electron-phonon interactions and thus can be indirectly measured, by measuring the local
temperature of the ions or the broadening of inelastic conductance features~\cite{D'Agosta:374102}.

An obvious reason why local temperatures and heating are such important phenomena lies in the fact that substantial heating of a nanoscale system leads to the system instability and eventually to the breaking of atomic bonds \cite{Teramae:083121,Tsutsui:3293,Ward:213108}. A different and even more fundamental interest in these phenomena arises in the context
of Fourier's law, Eq.~(\ref{FourierLaw}), that we will discuss in Sec.~\ref{Fourier}.
Of course, at the nanoscale, it seems inappropriate to
discuss the scaling of the thermal conductance with length, since this
is an asymptotic (in terms of system size) property \cite{Lepri:1}. Thus, one is left with the simple
question: under which physical conditions does a uniform temperature
gradient develops in a nanoscale system held in contact between two
heat baths of different temperatures?

In this section we discuss all these issues. We review the various mechanisms which give rise to heating in current-carrying junctions,
using simple arguments and models, followed by some basic results obtained from more elaborate models. We then turn to discuss the onset of a temperature gradient, analyzing a molecular wire junction in terms of the theory of open quantum systems, discussed in some detail in Sec.~\ref{Theoryopen}.

\subsection{Heating in current-carrying nanostructures: theory}\label{Ionheat}

\subsubsection{Various definitions of out-of-equilibrium temperature}
 In order to discuss local heating, the first question one should ask is: how is a \emph{local} temperature defined and calculated? Since temperature is a thermodynamic quantity, some caution is needed~\cite{Hartmann:080402,Hartmann:579,Hartmann:066148}. Apart from the definition of temperature that
we have given in Sec.~\ref{def1}, and which we will use also in Sec.~\ref{Fourier}, we here report several other notions of local temperature (not necessarily leading to the same quantitative results) and their microscopic origin, which were used to study local ionic heating in atomic junctions, each with its own pros and cons.

{\em Kinetic definition -} An intuitive definition of local temperature is to relate it to the local kinetic energy of the ions, i.e. $\langle \half m v^2 \rangle \sim 3k_B T/2$. However, this definition, mainly used in molecular dynamics simulations (see Sec.~\ref{MolecDynam}), has several drawbacks: (i) it relies on the equipartition theorem which is strictly proven in the thermodynamic limit
only for systems whose energy is quadratic in the particle momenta (as for non-interacting systems), and does not encompass any quantum effects. (ii) One needs to define an average kinetic energy over some length scale, while the quantum nature of particles may preclude such definition.

{\em Local phonon mode -} Consider a phonon mode somehow coupled to the system and vary its temperature in such a way that no heat flows between that mode and the system. This idea is somewhat similar to the idea of connecting an external bath to a system and imposing that no heat current flows between the system and bath, which was suggested to study the onset of Fourier's law in one-dimensional systems, both classical and quantum~\cite{Bonetto:783,Dhar:457,Dhar:805,Roy:062102}. This idea was recently used to study the local temperature of a model molecular junction  using the NEGF formalism \cite{Galperin:155312,Galperin:103201}. The main result is the existence of two voltage thresholds. The first is at the voltage which corresponds to the vibrational energy of the phonon, $eV\sim \hbar \omega_0$, at which local heating starts to occur and the temperature increases abruptly. The local temperature then remains roughly constant, until it rises again when the bias is so large as to encompass the molecular conduction window (i.e., both the HOMO and LUMO states). The disadvantage of this method is that the temperature of the mode depends on the microscopic details, i.e., the phonon excitation energy $\hbar \omega_0$ and/or the electron-phonon coupling.

{\em Distribution function definition -} A slightly different model of local temperature is to connect a phonon mode to the nanoscale system, but instead of determining its temperature self-consistently, its distribution function
is compared to an equilibrium distribution function with a given temperature, which is tuned to give the best comparison. Clearly, the disadvantage of this method is that the non-equilibrium distribution
function may be very different from the equilibrium one~\cite{Pekola:056804,Koch:155306}. The last two methods were compared, and were found to give similar local temperatures at large bias (compared to the typical vibrational modes, implying strong non-equilibrium and population of higher modes), but deviated from each other substantially at low biases. In fact, the second method turned out to give erroneous results in the zero-bias limit, when one expects the temperature to be the same as that of the leads. This is precisely because an equilibrium form for the phonon mode was assumed, although even with no current the distribution function of the mode may have contributions arising from
the coupling to the electronic (and other phononic) degrees of freedom in the junction \cite{Galperin:155312}.

\begin{figure}[ht]
\vskip 0.5truecm
\includegraphics[width=8truecm]{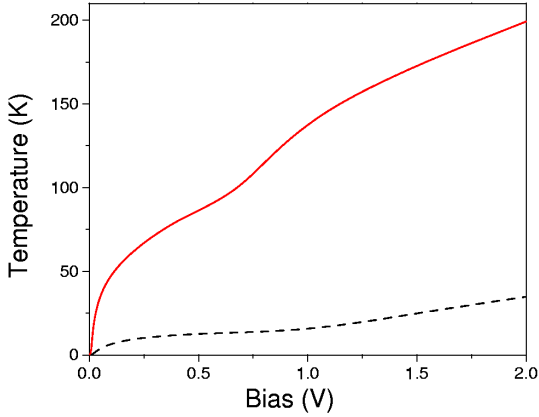}
\caption{(Color online) Local temperature as a function of bias, calculated from a scattering theory approach, for a benzene-dithiol molecular junction (dashed line) and a gold-atom point contact (solid line). From~\cite{Chen:1691}.}\label{Local_T_Benzene_Gold}
\end{figure}

{\em Definition from dissipated power -} A microscopic theory which relies on first-principles was suggested by Chen and coauthors \cite{Chen:1691,Chen:621}. The method is as follows. As a starting point, the electronic scattering states are calculated using ground-state DFT. The electron-phonon coupling for the different modes is also calculated using first-principles approaches. Using perturbation theory then one can calculate the power dissipated into the junction from current-carrying states. This power is then compared to the rate at which heat escapes the junction, typically assumed as $\sigma_{th} T^4$ with $\sigma_{th}$ the thermal coefficient that can be estimated from a microscopic model, and $T$ the effective temperature of the junction~\cite{Chen:1691,Chen:621}. A result of these calculations is presented in Fig.~\ref{Local_T_Benzene_Gold}, where the local temperature as a function of bias was calculated for a benzene-dithiol (BDT) junction and a gold-atom point contact. The results indicate that, for a
given bias, the BDT junction heats up less than the gold-atom junction, due to better thermal coupling with the electrodes and larger resistance to electrical
currents (see Eq.~\ref{local ionic temperature 1}). This result is also confirmed by experiments on similar
systems~\cite{Tsutsui:133121,Teramae:083121,Huang:1240}. While not visible from Fig.~\ref{Local_T_Benzene_Gold}, theoretical results of the threshold voltage for heating - see Eq.~\ref{local ionic temperature 1} - are also in good agreement with experiments~\cite{Chen:1691,Chen:621}. The same method was used to study local heating in alkane chains of different lengths~\cite{Chen:621}. It was predicted that, at fixed voltage, heating decreases with increasing chain length, which is due to increased resistance to electron flow; a result also confirmed experimentally~\cite{Huang:698}. More recently, the same approach was used to study the effect of different isotope substitutions on the heating in hydrogen molecular junctions \cite{Chen:233310}. It was found that local heating is very sensitive to isotope effects since the
electron-phonon coupling constant is inversely proportional to the ionic mass.

The method described above has the advantage that it can treat realistic systems. However, its main drawback is that it relies on the assumptions of the Landauer
approach - see Sec.~\ref{Thermolinear} - and its
practical implementation employs ground-state DFT, which, as we have discussed at length in this review, does not take into account properly all dynamical effects.

\subsubsection{Ionic heating}
After discussing various definitions of local temperature, we are now in a position to discuss local heating. As described previously, we consider here a junction, composed of leads (which are assumed to be held at local equilibrium), and a nanoscale system which has both electronic and vibrational degrees of freedom. Even in the presence of current, we can assume that in the leads, far away from the nanojunction, electrons and phonons reach the same temperature
$T_0$~\footnote{The extent to which this statement is correct depends on the current density in the leads. If this current density can be assumed to be zero, then the leads are
at an ideal global thermal equilibrium, with electrons and ions sharing the same temperature. Otherwise some difference (albeit extremely small) may arise between the lead
temperature of the ions and electrons.}. In the junction, however, the electrons and phonons may have different temperatures, $T_e$ and $T_{ph}$, respectively. These temperatures depend on bias, strength of electron-phonon and electron-electron interactions, the coupling of phonons with the bulk phonons in the leads, as well
as the transmission properties of the electrons.

\begin{figure}[ht]
\vskip 0.5truecm
\includegraphics[width=8truecm]{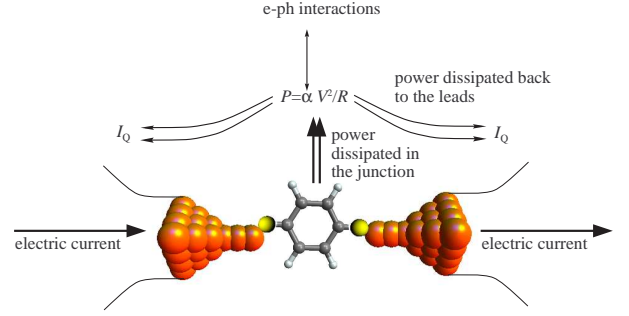}
\caption{(Color online) A schematic representation of the mechanism of ionic heating in nanoscale junctions. The electric current dissipates a fraction $\al V^2/R$ of its power in the junction, depending on the strength of the electron-phonon interaction. This power is then dissipated to the phonons in the electrodes in the form of a heat current. The balance between the power flowing into the junction and the heat current $I_Q$ flowing out of the junction determines the ionic temperature of the junction. }\label{Junction_Heating}
\end{figure}

Let us start by discussing the temperature of the ions in the junction, or the phenomenon of \emph{local ionic heating} (see schematic in Fig.~\ref{Junction_Heating}). We start from some simple considerations assuming first no inelastic electron-electron scattering occurs~\cite{Todorov:965,DiVentra:2008}. The power of the entire circuit (nanojunction plus power source) is given by $V^2/R$, where $V$ is the source bias and
$R$ is the junction resistance (assuming zero impedance of the external circuit). Only a small fraction $\alpha$ of this power, i.e., $\alpha V^2/R$, is dissipated into the ionic degrees of freedom in the junction due to the electron-phonon coupling. The value of $\alpha$ needs to be determined from a microscopic theory \cite{Todorov:965}. Since the spectrum of modes of the junction is typically
discrete, one expects a minimal bias (we call $V_c$) necessary to excite the lowest-energy phonon mode of the structure, and hence $\alpha(V<V_c)=0$~\footnote{In molecules,
this bias may be very close to zero, due to the longitudinal ``acoustic'' mode of the whole molecule vibrating against the bulk electrodes.}. Therefore we write $P=\Theta (V-V_c) \alpha V^2/R$, where $\Theta$ is the step-function. Now, if the power $P$ were not dissipated away from the junction, the latter would heat up substantially and eventually break down. Therefore, there must be a heat current $I_Q$ which dissipates this power into the electrodes. Since the leads are much bigger than the junction and are three-dimensional in nature, one can assume that this energy is carried away at a bulk rate $I_Q=\sigma_{th} T_{\text{eff}}^4$ \cite{Ashcroft-Mermin}, with $T_{\text{eff}}$ an average effective temperature of the junction ions and $\sigma_{th}$ the lattice thermal conductance. At steady state the condition $P=I_Q$ then yields for the effective temperature \beq T_{\text{eff}}=\Theta (V-V_c) \left( \frac{\alpha}{\sigma_{th} R} \right)^{1/4} \sqrt{V} ~~.\label{local ionic temperature 1} \eeq Here, we have considered the bulk electrode temperature $T_0=0$. If both electrodes are at finite temperature, then there is also a heat current $\sim \sigma_{th} T_0^4$ flowing into the junction, and hence the balance equation $P=I^{\text{out}}_Q- I^{\text{in}}_Q$ gives $\Teff=(T_0^4+T^4_V)^{1/4}$, where $T_V= \Theta (V-V_c) \left( \frac{\alpha}{\sigma_{th} R} \right)^{1/4} \sqrt{V}$ is the contribution to the temperature from the finite voltage bias.

In the above considerations we have assumed that heat can be dissipated away from the junction rather easily. The results may change depending on the heat-transport properties of the leads and the coupling between the leads and the junction. For instance, if the leads are strongly disordered heat is carried away with a different exponent of the temperature difference \cite{Yudson:155310}. If the nanojunction has poor thermal coupling to the leads, or in the
presence of localized phonon modes~\cite{Lepri:1}, namely modes that have a very weak coupling with the bulk modes, then the local ionic temperature can reach very large values, even at relatively
small biases. The reason is simple: in the above cases, due to the bias $V$, the current-carrying electrons are away from their ground state, and they are thus ``seen''
by the local modes of the nanostructure at an effective finite temperature. Thus, this situation provides the possibility for inelastic electron-ion scattering in
an energy window $\sim eV$, with consequent ion temperatures of the same order of magnitude~\cite{Todorov:965,DiVentra:2008,Yang:041402}. We note that similar results were recently obtained from microscopic considerations \cite{Mozyrsky:035104}. That being the case, a voltage bias of $0.1 V$ would generate an effective temperature of $\sim 1000 K$. This seems to have been observed in atomic quantum point contacts at the breaking point \cite{Ward:213108}. Thus, good thermal coupling to the electrodes is essential for maintaining junction stability.

\subsubsection{Electron heating}
Up to now we have discussed the heating of the phonons in the junction due to their interaction with the current-carrying electrons. But what about the temperature of the electrons themselves? To be precise, we refer here to the temperature $T_e$ of the Fermi sea of electrons of the nanojunction and those in its proximity. This temperature is affected by both inelastic electron-electron interactions and electron-phonon coupling~\cite{D'Agosta:2935}. Clearly, the local electron temperature influences the local ionic temperature of the junction. However, accounting for both electron-electron and electron-phonon interactions is a challenging task. While attempts have been made to account for both in calculating charge transport \cite{Galperin:103201} and recently even heat currents \cite{Liu:161309}, we are unaware of any calculation of the local temperature where these interactions are considered on equal footing.

In a recent work, D'Agosta \etal~\cite{D'Agosta:2935} have predicted the bias dependence of the local electron temperature in quasi-ballistic nanoscale junctions and its
effect on ionic heating, treating the electron liquid as a viscous fluid. The general argument, which was accompanied by a microscopic theory based on the quantum hydrodynamic equations for the interacting electron liquid \cite{D'Agosta:11059}, is as follows. Assuming no electron-phonon interaction is present, to first approximation, the thermal electronic conductance of the electron liquid can be taken to be proportional to the temperature, $\sigma_{th}=\gamma T_e$. Therefore, the heat current, given by $I_{Q}=\sigma_{th} T$ is quadratic in temperature, $I_{Q}=\gamma  T^2_e$. As in the case of local ionic heating, at steady state this thermal current has to balance the power dissipated in the junction, which is a small fraction of the total power of the circuit, $P=\alpha V^2/R$. One thus obtains
\beq T_e=\gamma_{e-e} V \label{T_e_Local} ~~,\eeq where $\gamma_{e-e}$ is to be determined from a microscopic calculation. Assuming the coefficient $\gamma_{e-e}$ weakly dependent
on bias, this simple argument shows that the local electron temperature grows linearly with bias. This result clearly hinges on the assumption that electronic heat is dissipated away from the junction at a bulk rate, which may not hold for all systems and under all experimental conditions.

From a microscopic hydrodynamic theory D'Agosta \etal~\cite{D'Agosta:2935} have also calculated the local temperature profile, $T_e(x)$, along the junction. From the maximal value of $T_e$, an estimate of $\gamma_{e-e}$ was supplied for various junctions. For instance, for a 3D gold quantum point contact (QPC) with effective cross section of 7 $\Ang$$^2$, these authors evaluated  $\gamma_{e-e}(QPC) = 65 $ K/V. For a 2DEG, assuming a cross section of $20$ nm they found $\gamma_{e-e}(2DEG) = 1.2 \times 10^2$ K/V, suggesting that heating from inelastic
electron-electron interactions is generally smaller than the corresponding heating due to electron-ion interaction.


\subsubsection{Ionic cooling}
A direct measurement of local electron temperatures, however, seems a very difficult task, and in fact we are not aware of such a direct method. On the other hand, local ionic temperatures are relatively easier to obtain (see Sec.~\ref{heatexp}). It is then relevant to ask what is the effect of the local electronic temperature on the ionic heating. Since part of the total power dissipated in the junction goes into heating electrons via electron-electron interactions, that power is no longer available to induce ionic heating. Since the initial energy is always
that of the current-carrying electrons, the ionic temperature must be smaller if electron heating takes place. The power of this electron-phonon scattering process can be assumed to have a form
$P_{e-ph}=\Sigma (T^n_{ph} - T^n_e)$ with $\Sigma$ a system-specific constant, and $n>0$ \cite{Schmidt:045901}. This ionic energy is then dissipated away from the junction. If we assume again a bulk dissipation
law, $I_Q=\sigma_{th} T_{\text{eff}}^4$, for electronic temperatures much smaller than the ionic ones, the steady state condition $P_{e-ph}=I_Q$ is satisfied by $\Sigma\sim \sigma_{th}$ and $n=4$.
By taking into account a background temperature $T_0$ we then get the relation~\cite{D'Agosta:2935}
\beq T_{\text{eff}} =\left( T^4_0+\gamma^4_{e-ph} V^2 -\gamma^4_{e-e} V^4 \right)^{1/4} \label{phonon_T_with_ee},\eeq which is valid for $V<(\gamma_{e-ph}/\gamma_{e-e})^2$. The meaning of Eq.~\ref{phonon_T_with_ee} is that at sufficiently large biases, the effective phonon temperature is reduced, i.e., the phonons ``cool down''. As we will discuss in the
following Sec.~\ref{heatexp}, this result has been recently confirmed experimentally (see Figs.~\ref{effective_T_experiment} and~\ref{Local_T_Raman}). It is important to point out, however, that the exact power-law dependence in
Eq.~(\ref{phonon_T_with_ee}) and the value of the various coefficients may depend strongly on the details of the nanostructure and its contact with the leads.

Another interesting idea to obtain reduced ionic heating is to use a nanostructure with an appreciable Peltier coefficient. In this situation passing current through the junction would result in the cooling of one side of the junction, which may induce cooling of the molecule.
The idea of local cooling of a junction has received renewed attention in recent years in the context of mesoscopic systems~\cite{saira:027203,giazotto:217} and molecular junctions \cite{Zippilli:096804,Pistolesi:199,Galperin:0905.2748,McEniry:195304}. While the details vary, the main concept is the same: the system is tuned in such a way that hot electrons (i.e., those with large kinetic energy) find it easier to tunnel through the junction, thus depleting the lead up-stream in voltage from hot electrons, thus cooling it. The cooling of the molecule is achieved either by its proximity to a cooler lead, or in more subtle cases, by the fact that electrons "borrow" energy from the localized phonon modes to assist transport, thus cooling them in the process \cite{Galperin:0905.2748}.

\subsection{Heating in current-carrying nanostructures: experiment}\label{heatexp}
\begin{figure}[t]
\vskip 0.5truecm
\includegraphics[width=8truecm]{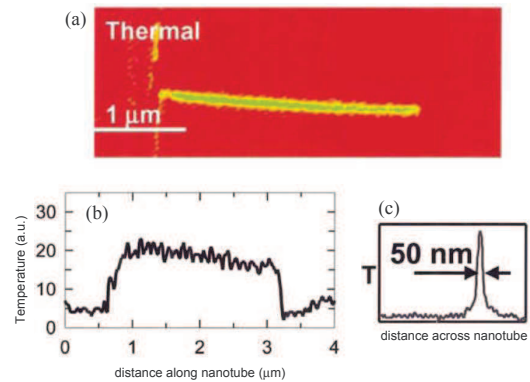}
\caption{(Color online) Scanning thermal microscopy (SThM) of a 10nm diameter multiwall carbon nanotube. (a) Full thermal image. (b) A cut along the nanotube. (c) A cut across the nanotube. Taken from \cite{Cahill:793}.}\label{Local_T_ee}
\end{figure}

Despite the difficulty in measuring directly local temperatures of nanoscale systems, we have witnessed much progress in this
direction over the last years.
The first concepts of local temperature measurements are reviewed by Cahill \etal~\cite{Cahill:793}. Especially noteworthy are experiments where a thermocouple (serving as a thermostat) is mounted on top of an STM tip, thus creating a ``scanning thermal microscope'' (SThM). This device was then used to study the local temperature of a carbon nanotube placed on a
substrate (see Fig.~\ref{Local_T_ee}). The authors of this work discuss several possible shortcomings and limitations of SThM studies: the dependence of the measured temperature on topography of the sample and surface chemistry, the fact that the tip itself might perturb the sample (e.g., via near-field radiation, or by effectively cooling it), only surfaces can be measured,  some of the heat is delivered through the air between the sample and tip, etc. These issues render this method hard for quantitative analysis, although some progress has been achieved~\cite{Grover:233501,Kim:203115}. We are unaware of any theoretical work (other than the one presented in this review) which has been directly related to SThM measurements.

In mesoscopic systems (e.g., quantum dots etched in 2D electron systems) which are of typical sizes of microns, tremendous advance in local thermometry has been achieved, as summarized in the thorough review by Giazotto \etal~\cite{giazotto:217}. In these systems, thermometry is achieved by analyzing some temperature-dependent function (current, conductance, etc.) from which, by using known properties of the electronic surroundings, the temperature can be extracted.
To give a specific recent example, by analyzing the derivatives of the current as a function of temperature and voltage, Hoffmann \etal~\cite{Hoffmann:779} were able to measure the temperature gradient across a current-carrying quantum dot of $15$nm length, with the conclusion that the heat flow is mediated by phonons in the quantum dot.

\begin{figure}[ht]
\vskip 0.5truecm
\includegraphics[width=8truecm]{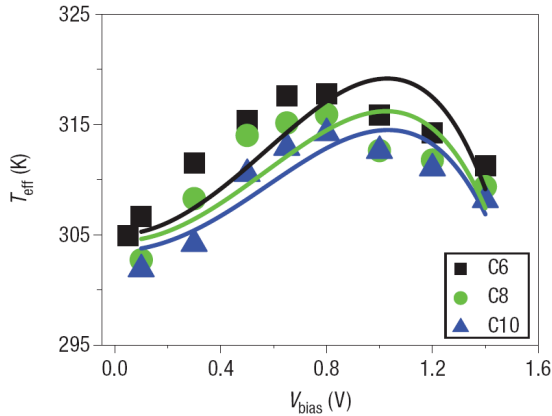}
\caption{(Color online) Effective temperature of a molecular junction, for three different types of molecules $n$-alkanedithiol, with $n=6$ (squares), $n=8$ (circles) and $n=10$ (triangles). The solid lines are theoretical
estimates from  Eq.~\ref{phonon_T_with_ee}. From Huang \etal~\cite{Huang:698}.}\label{effective_T_experiment}
\end{figure}

Other options for measuring the local temperature are available. One method is to study the force at which a molecular junction breaks as a function of current. The idea behind this method is that the higher the temperature of the structure, the less external force is needed to break it~\cite{Huang:698,Tsutsui:223110,Tsutsui:3293}. From
this force one can then extract an effective temperature. For example, Schulze \etal~\cite{Schulze:136801} have studied the breakdown of a molecular junction composed of a C$_{60}$ molecule, and showed directly that better cooling of the junction is achieved when the coupling between the molecule and the leads is improved.

Using the above method, in a recent series of experiments, Huang \etal~\cite{Huang:698,Huang:1240} have studied the local temperature of single-molecule (alkanethiols) junctions as a function of voltage bias. Their results, shown in Fig.~\ref{effective_T_experiment} (points) indicate that with increasing voltage, the local temperature first increases, saturates, and then
slightly decreases. This is in agreement with the prediction of Eq.~\ref{phonon_T_with_ee} (solid lines), and suggests that electron-electron interactions indeed occur in these junctions. The same experiment also confirms that longer alkanethiol molecules heat up less due to increased
electronic resistance, at fixed voltage~\cite{Chen:621}.

 \begin{figure}[ht]
\vskip 0.5truecm
\includegraphics[width=8truecm]{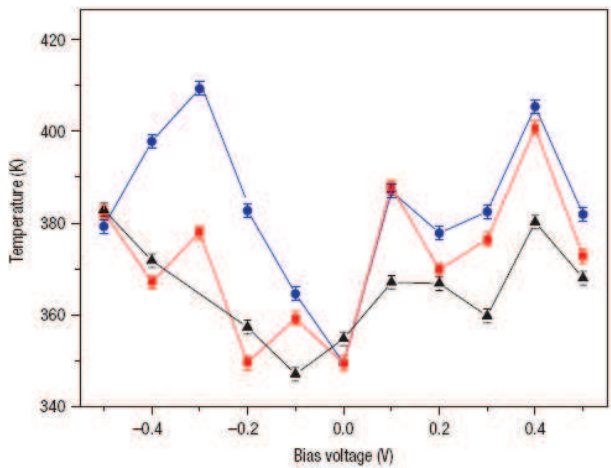}
\caption{(Color online) Effective temperature of a molecular junction measured using Raman scattering; from Ioffe \etal~\cite{Ioffe:727}. The different points correspond to different modes, and while the temperature is slightly different, the overall voltage dependence shows roughly similar features. Note also the apparent decrease in the local temperature with bias, which is in line with the results of Huang \etal~\cite{Huang:698}.}\label{Local_T_Raman}
\end{figure}
An alternative method to study the local temperature has been suggested recently. It makes use of Raman spectroscopy and it was first applied to the study of a suspended nanotube \cite{Bushmaker:3618,Deshpande:105501}. In these measurements, the local temperature was deduced from the shifts in the local Raman  $G_+$ and $G_-$ bands of the nanotubes. The authors compared two nanotubes of lengths  $2\mu$m and $5 \mu$m, and found that the longer nanotube was less heated, an effect which was attributed to the thermalization of hot phonons at the center of the nanotube.

Along the same lines, the local temperature of a molecular junction has been investigated via Raman spectroscopy \cite{Ioffe:727}. The idea is that the ratio between the Stokes and anti-Stokes intensities is directly related to their non-equilibrium populations in the presence of electronic current. The method has been discussed theoretically in detail~\cite{Ioffe:727,Galperin:758}. In Fig.~\ref{Local_T_Raman} the effective temperature is plotted as a function of voltage bias, for different Raman modes. From the figure one can see that although the temperature is slightly different between different modes (due to the different electron-phonon coupling strengths
and symmetries), the overall voltage dependence shows roughly similar features. Note also the apparent decrease in the local temperature, which is in line with the results of Huang \etal~\cite{Huang:698}. This study, along with the one described before, indicate that local Raman spectroscopy may serve as a valuable tool for the study of local temperatures at the nanoscale.

\subsection{Fourier's law at the nanoscale}
\label{Fourier} Let us now end this section with a somewhat different issue, that of
the onset of Fourier's law, Eq.~(\ref{FourierLaw}), in nanostructures. As previously noted, in the context of nanoscale junctions, there
is not much point in discussing the scaling of the thermal conductivity $\kappa$, which pertains to an
asymptotic relation, strictly valid in the limit of large system
lengths. Therefore, here we limit our discussion to the simple
question of what is the temperature profile along the junction.

In the context of Fourier's law, this question has been widely studied for both classical \cite{Lepri:1} and quantum systems \cite{Michel:4855}. The main focus has been on either spin-chains (i.e., Ising-like models) \cite{Michel:4855,Wu:060101} or harmonic oscillator chains. The local temperatures are usually evaluated by calculating the averages of some local energy operators \cite{Saito:34,Mejia:113,Michel:4855,Michel:325}, or by using self-consistent thermal baths \cite{Dhar:805,Dhar:457,Roy:062102,Jacquet:709}. In the first case, one assumes that the local energy is related to the temperature via a local Boltzmann relation \cite{Dubi:042101}, or directly proportional to the temperature via a local equipartition law \cite{Michel:325}. The disadvantage of this method is two-fold: (i) there is some arbitrariness in choosing the local energy operator, since one can represent the same Hamiltonian in different ways \cite{Wu:025302}, and (ii) this method \emph{assumes} from the outset that the system is in a local thermodynamic equilibrium, which may not always be the case.

In the second approach, the system is attached to local heat baths. The heat current between the junction (or quantum wire) and the local baths is calculated, and the temperatures of the heat baths are determined in such a way that the heat current between the wire and the baths vanishes. This method was recently described in detail and applied to a quantum chain of non-interacting harmonic oscillators \cite{Roy:062102} and a chain of quantum dots~\cite{Jacquet:709}. For instance, in~\cite{Roy:062102}, using quantum Langevin equations, the local temperature as a function of position was calculated for different chain lengths and for different coupling between the wire and the local baths.
 The conclusion of this work is that the coupling between the wire and the baths determines a length scale (mean-free path), and the heat transport crosses over from a diffusive regime (uniform temperature gradient) to a ballistic regime (uniform temperature, vanishing gradient) depending on the system length being longer or shorter than the mean free path, respectively.
Since the dynamics of the system is calculated in the presence of the local baths, this shows that the properties calculated (i.e., local temperature) pertain to the combined system of quantum chain \emph{and} thermal baths, and thus naturally depends on, e.g., the coupling strength between them.
%
%
%
%

Recently, a method has been suggested to calculate the local temperature of electrons in a nanoscale junction \cite{Dubi:97,Dubi:115415}. Its starting point is the stochastic Schr\"odinger equation (see Eq.~(\ref{SSE})), which for non-interacting electrons reduces to a quantum master equation \cite{Pershin:054302}. In this approach the
{\em finite} electronic
system is coupled to two local heat baths at the edges of the system, in similarity to the study presented above for a chain of harmonic oscillators. In order to evaluate the local temperature, the definition we introduced in Sec.~\ref{def1} has been used. Namely, a third environment is coupled locally to the system at the position where the temperature needs to be evaluated. The properties of the system are then evaluated twice: once with the additional environment (so-called ``tip'', as it mimics, e.g., the operation of a thermostat mounted on an STM tip) and once without the probe.
The temperature of the probe is then varied ({\em floated}) such that a minimal change in some local (or global) properties of the system, such as its local electron density, occurs. A scan of the local temperature of the whole system can then be obtained with this method. The advantage of this approach is that it can, in principle, be implemented
experimentally, and it provides the local temperature of the electrons without further scattering from other sources. In addition, it can be shown analytically that the above definition reduces to the standard thermodynamic temperature in limiting cases, for instance, in local equilibrium (see also~\cite{DiVentra:40004}) or for two-level systems.

For the case of a wire coupled to two electrodes at different temperatures, it was found that the local temperature
of the wire may exhibit quantum oscillations for intermediate lead-wire couplings \cite{Dubi:97}. Similar oscillations were later observed for a driven quantum wire \cite{Caso:041301} and reflect the quantum coherent nature of the system. When the lead-wire coupling is large enough a uniform temperature ensues. In this limiting (ballistic) case, one also finds that the non-equilibrium distribution function of the system is an average of the distribution functions of the left and right baths. The fact that the temperature is uniform in the wire demonstrates the known result that for a clean system, Fourier's law is invalid.

In order to reconstruct Fourier's law (with an associated temperature gradient), diagonal disorder was introduced in the wire (which localizes the electronic wave-functions), and the local temperature was averaged over disorder realizations \cite{Dubi:115415}. It was found that for large enough disorder, a local uniform temperature gradient ensues, giving rise to Fourier's law. This result was interpreted in terms of an effective thermal length which controls the scale of the temperature gradient \cite{Dubi:042101}. By adding the effect of dephasing the model was also able to explain the results by Roy \cite{Roy:062102} described above.
We finally conclude that for the above model the onset of Fourier's law coincides with the onset of chaos~\cite{Dubi:115415}. This has also been found in other model systems \cite{Michel:4855,Gaul:011111}, but not in all cases \cite{Lepri:1896,Li:223901}. Thus, this result does not appear to be universal.
\section{Thermopower} \label{Sec2}
\subsection{Introduction and basic definitions}
In this section we discuss the concept of thermopower in nanoscale
junctions. As prototypical examples that show all main features of the problem we will be focusing mainly on experiments in molecular junctions \cite{Reddy:1568} and briefly mention experiments in mesoscopic systems and nanowires. The thermopower phenomenon corresponds to the case in which a
temperature difference at two sides of a given junction induces a
voltage drop across it. From a technological point of view, this
effect is of great importance, since it may be used to recover part
of the heat wasted in physical processes and generate electrical
power with no moving mechanical parts. It is also of basic
scientific interest, since, by combining energy and charge flow, it
may encode information about the system dynamics which is
unavailable in charge transport experiments \cite{Segal:165426}.

The configuration we have in mind is again a junction composed of
two leads separated by a nanoscale element - a quantum dot, a
molecule, nanotube, etc. Consider such a junction, where the two
leads are held at different temperatures, $\TL$ and $\TR$. The corresponding temperature difference $\D T=\TR-\TL$
gives rise to both a heat current (discussed
in Sec.~\ref{Sec1}) and a charge current. If the circuit is closed,
after a transient time charges accumulate on one side of the
junction and deplete on the other, so that a zero charge current is
achieved and a voltage drop across the junction is formed. If the
circuit is open (namely it is connected to an electron source), and
a voltage difference $\D V$ is applied between the two leads with
appropriate sign, a bias-induced electric current can cancel out the
thermally-induced current. Note, however, that given a temperature difference the two procedures
may not yield the same voltage difference. In fact, the voltage difference may also depend on the
location along the system where it is probed.

The thermopower $S$ is defined as (minus)
the amount of voltage $\D V$ at the state of vanishing current, \beq S=-\left.\frac{\D V}{\D
T}\right|_{I=0} ~~, \label{Thermopower_definition}\eeq in the limit of $\D T \to
0$.

This definition can also be understood from the current expressed in linear response. This is defined as \beq
I=G \D V+L_T \D T, \label{linear-response-for-thermopowr}\eeq where $G$ is the electrical conductance and $L_T$ is a response coefficient related to the energy flow. From this
expression one readily sees that
$S=L_T/G$. Therefore, in order to determine $S$, one has to determine the conductance
$G$ and the thermal response $L_T$.

Before we proceed to discuss different theoretical and experimental aspects of the thermopower, it is important to
ask the following question: is knowledge of the thermopower $S$ sufficient to design devices that operate as efficient heat-voltage converters?
In fact, in a real device, it is not at all clear that the system is, under the given experimental conditions, in the linear
response regime. Nor it is obvious that the best conversion should be achieved in
that regime \cite{Esposito:60010,Dubi:97}.

To this end, it is useful to define the unit-less ``figure
of merit'', \beq ZT=\frac{G S^2}{\sigma_{th}/T} \label{ZT}~~,\eeq
where $T$ is the temperature of the system \cite{Mahan:7436}. The quantity $ZT$
describes the efficiency of a real device or material as a
thermo-electric converter. While an exact relation between $ZT$ and
thermodynamic efficiency is available \cite{Muller:044708}, this
choice can be intuitively understood: $S$ measures how large a
voltage drop can develop for a given temperature gradient, $G$
measures how easy charges can cross the junction to generate that
voltage drop, and $\sigma_{th}$ measures how hard it is to maintain
a temperature gradient.

It is commonly stated that for applications, one must achieve $ZT>>1$ (in
fact $ZT>4$ would already be a great advance \cite{Majumdar:02062004}). However, such a situation is hard to obtain: in
most cases the electrical conductivity $\sigma$ and thermal conductivity $\kappa$ are related via the Wiedemann-Franz law \cite{Ashcroft-Mermin}, which states that \beq
\frac{\kappa}{\si}=\left (\frac{\pi^2 k_B^2}{3e^2} \right )T,\eeq with the quantity in parenthesis commonly referred
to as the ``Lorenz number''. This means that it is difficult to increase $\si$ and $S$ without
also increasing $\kappa$, and vice versa. However, deviations from the Wiedemann-Franz law have been observed
\cite{Appleyard:16275} and discussed theoretically in various systems, including Luttinger liquids \cite{Rejec:235301,
Kubala:066801,Murphy:161406,garg:096402,Kane:3192,Li:432}. These deviations are attributed to interactions, where the simple single-particle theory
fails (see Sec.~\ref{Thermotheory} on theoretical methods), and are exactly what is required in order to increase
the efficiency of thermoelectric devices.

\subsection{Experiments on thermopower at the
nanoscale}\label{Thermoexp} Measurements of thermopower are
conceptually easier than those of thermal conductance: one applies a temperature gradient
across the junction, and measures the ensuing voltage in a closed
circuit when the transient current vanishes. Or, in an open
circuit, one supplies a voltage to compensate for the
thermally-induced current. The slope of the
resulting voltage-temperature gradient curve gives the thermopower.
However, in an actual experiment, particular care needs to be taken
to extract this quantity. The reason is because the voltage probe
that is connected to the system in order to measure the thermopower
is necessarily invasive, since the applied thermal gradient would
induce, locally at the voltage probe contact, an extra voltage
difference. This extra effect needs to be subtracted to get the
actual thermopower of the nanojunction. In addition, the ensuing voltage (including its sign) is very sensitive to the junction
geometry and thus may fluctuate considerably in
an actual experiment, providing non-trivial distributions of the
voltage as a function of thermal gradient, from which a single
voltage value may not always be easy to extract.

Before reviewing some recent experiments on nanoscale junctions, it is
of interest to briefly survey some of the older experiments on mesoscopic systems as well. We point out that while most
of these experimental results may be understood in terms of a linear
response scattering theory (see next Sec.~\ref{Thermotheory}), some recent results,
such as the appearance of additional peaks in the distribution of
induced voltages versus temperature gradient (Scheibner {\em et al.}, 2005), are yet to be completely accounted for. The
discovery of pronounced mesoscopic effects such as Coulomb blockade
and conductance quantization prompted the study of
thermopower in quantum point contacts
\cite{Molenkamp:1052,Houten:B215,Molenkamp:903} and quantum dots
\cite{Staring:57,Godijn:2927,Scheibner:176602}. These devices are
defined by depositing gates on top of a two-dimensional electron gas
formed in a semiconductor interface (typically GaAs/AlGaAs). Heating
of one side of the device is achieved by passing current through it
with consequent Joule heating and temperature rise. Most of the
results of these experiments are well understood within the simple,
single-particle picture of thermopower \cite{Houten:B215}, which
will be described below.

Another, more recent batch of thermopower experiments are those
conducted on nanowires, namely wires with nanoscale diameter, but
extending in the longitudinal direction as long as a few microns.
Various experiments were performed on wires of different materials
\cite{Boukai:864,Boukai:168,Hochbaum:163,Duarte:617,Seol:023706,Lee:022106},
as well as carbon nanotubes
\cite{Kong:1923,Small:256801,Sumanasekera:166801}. These experiments
suggest that in these systems it is possible to either
increase $S$ (by designing the system to have an increased
electronic density of states) or reduce the thermal conductance independently by, e.g., designing a system
which is smaller than the phonon mean-free-path but still larger
than the corresponding mean-free-path of the electrons or holes, thus increasing the figure of merit. Specifically, in Si nanowires these are obtained by the combined effect of the change in phonon spectra and enhanced scattering off the boundary, both having little effect on the electronic part (see recent review by Rurali, \cite{Rurali:427}).
Along similar lines, boundary effects seem to highly reduce the
thermal conductance but leave the charge conductance roughly
unchanged \cite{Majumdar:02062004}.

\begin{figure}[ht]
\vskip 0.5truecm
\includegraphics[width=8truecm]{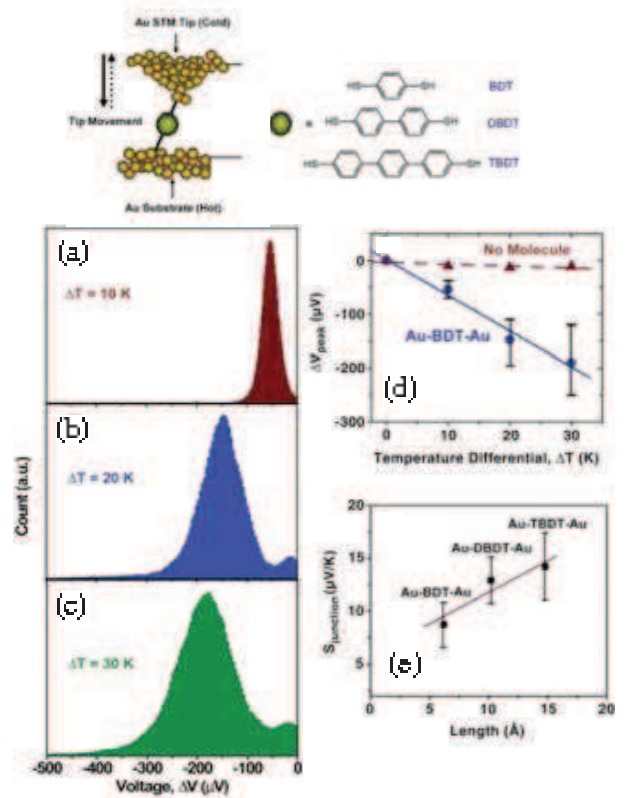}
\caption{(Color online) Upper panel: schematic representation of
thermopower experiments on molecular junctions. (a-c) Distribution
of thermo-voltages obtained at different temperature gradients. Note
the widening of the distributions and their non-trivial structure.
(d) The most-probable thermo-voltage obtained from (a-c) as a
function of the temperature gradient. The derivative of the linear
fit of this curve yields the thermopower $S$. (e) $S$ of various
molecules, in terms of the molecule's length (adapted from
 Reddy \etal \cite{Reddy:1568}).}\label{fig2}
\end{figure}

A set of experiments which is of great interest both from an
academic and technological point of view, are those performed on
junctions of nanometer length, such as metallic contacts
\cite{Ludoph:12290} or molecular junctions
\cite{Reddy:1568,Baheti:715,Malen:1164,Malen:3406,tan:013110}. The latter ones are of
interest since, as we will discuss in Sec.~\ref{Thermotheory},
theoretical arguments suggest that molecular junctions may exhibit
large thermopower. In these latter experiments, a gold STM tip is placed on
top of a gold substrate which is covered with various molecules. As
the STM tip touches (and is attached to) a molecule, a thermal
gradient is applied and the thermopower is measured by applying a
voltage so that no current passes through the junction (see upper
panel of Fig.~\ref{fig2}). This procedure is repeated many times and
a histogram of the voltage required to achieve vanishing current is
obtained (for different temperature gradients, $\D T=10K,20K,30K$,
Fig.~\ref{fig2}(a-c)). The authors of this work then plot the
voltage with maximum probability (i.e., the position of the peak in
the voltage histograms) as a function of $\D T$, and by fitting the
resulting curve with a linear fit the thermopower is obtained
(Fig.~\ref{fig2}(d)). These experiments were performed with various
kinds of molecules, and interesting phenomena such as a length
dependence of the thermopower (Fig.~\ref{fig2}(e)) or strong
dependence of the molecular end-groups were observed. The
experiments indicate that indeed molecular junctions have favorable thermoelectric properties, suggesting that devices incorporating molecular junctions or arrays (for instance metallic plates separated by a molecular layer) may be good
candidates for thermoelectric applications.

Let us, however, point out some features of the experiments which at present do not have a satisfactory
theoretical explanation. For instance, as seen in Fig.~\ref{fig2}(c), the
voltage histograms have a well-defined structure, with a not so negligible secondary
peak near $\D V=0$. Note also that the
distributions cross the $\D V$=0 line into negative values, not
shown in the figure. An additional feature of the histograms is their apparent widening
with increasing temperature gradient. These fluctuations effects have been recently studied experimentally \cite{Malen:3406}
and are attributed to variations in contact geometry and orbital hybridization, as well as intermolecular interactions,
in accord with theoretical studies \cite{Dubi:97}.

The analysis of the above results has been done based on the
single-particle (non-interacting) Landauer approach to thermopower.
As it will be discussed in Sec.~\ref{Thermotheory}, in the linear
response single-particle theory of thermopower, $S$ can be simply
related to the electronic properties of the junction, and
specifically in molecular junctions, to the position of the
electrochemical potentials of the leads with respect to the gap
between the highest occupied molecular orbital (HOMO) and the lowest
unoccupied molecular orbital (LUMO). Since the position of the
HOMO-LUMO gap affects charge transport in molecular junctions
\cite{Nitzan:1384}, measurements of $S$ were suggested as a way to
probe the energy position of these levels
\cite{Paulsson:241403,Baheti:715}. In the
experiments~\cite{Reddy:1568,Baheti:715}, a comparison of the
thermopower and conductance with numerical simulations using
ground-state DFT within the Landauer approach has been performed.
From this comparison it was then concluded that the position of the
HOMO-LUMO gap can be thus determined. This analysis, however, raises
several questions. The applicability of a
linear-response single-particle theory was questioned by the authors
themselves \cite{Reddy:1568}, following the fact that the
temperature differences can be tens of degrees Kelvin. In fact, one could
ask if this is the smallest energy scale in the experiment.
Specifically, is this energy smaller than, say, the coupling energy
between the molecule and the substrate? The answer to this question
is unclear, especially in light of the large error bars shown in
Fig.~\ref{fig2}(d). In addition, the non-trivial structure of the
fluctuations in the voltage histogram implies that non-equilibrium
effects may come into play, which are not taken into account in the
linear response theory. Finally, electron-electron and
electron-phonon interactions may play a crucial role in this problem.

Despite these open questions, the experiments described above are very impressive and
important for the field and there are many interesting future directions to which they could be taken. For example, it would be interesting to
study the change in the width of the distributions and their
structures as the overall temperature is reduced. This would
determine, e.g., if these distributions are due to static or dynamic
effects. Another interesting direction would be to study, for
several molecular structures, not just the most-probable voltage,
but the real (statistical) average of the distributions, and infer
from this whether the resulting thermopower displays the same
features as reported above (e.g., length dependence, etc.), and
whether this quantity matches calculations based on single-particle
theories.

\subsection{Theoretical methods}\label{Thermotheory}
In this subsection we describe the present theoretical methods
available to describe the phenomenon of thermopower. The most common
one is based on the Landauer approach with its most common implementation
within ground-state DFT. As we will discuss, this approach has
several advantages, being rather computationally straightforward,
and having a rather simple physical interpretation. However, we will
argue that in many cases of actual experimental interest, it may be
inadequate, since it is based on the usual assumptions of scattering
theory of non-interacting electrons. In addition, as we have
emphasized also in Sec.~\ref{Sec1theorysp}, the use of ground-state
DFT is questionable in an intrinsically non-equilibrium problem as
that discussed here. We will then introduce an approach based on the
theory of open quantum systems, which is ideally suited for the
present problem and can, in principle, account for
interactions (beyond mean-field). The latter point has its most
practical implementation in an extension of time-dependent DFT to
open quantum systems~\cite{Ventra:226403,D'Agosta:165105}.

\subsubsection{Single-particle theory of
thermopower}\label{Thermolinear} The starting point for calculating
the thermopower within a single-particle picture is the Landauer
expression for the electrical current~\cite{Butcher1990}, \beq
I=\frac{e}{\pi \hbar} \int^\infty_{-\infty} \d \e \T(\e)\left[
f_\L(\e)-f_\R(\e)\right] \label{Landauer_charge_current}~~, \eeq
where $\T(\e)$ is the transmission coefficient at energy $\e$ and
$f_{\L,\R}$ are the Fermi distributions of the left and right leads.
In the limit of small bias and temperature gradient (i.e., $|\D
T|<<T$ and $|e \D V|<<\mu$, where $T$ is the background temperature
and $\mu$ is the equilibrium chemical potential) the distribution
functions are approximately given by~\cite{Butcher1990} ($i=L,R$) \beq
f(\e,\mu_i,T_i) \simeq f(\e,\mu,T)\pm \frac{\d f}{\d \e}(\mu-\mu_i)\mp
\frac{\d f}{\d \e}(\e-\mu) \frac{(T_i-T)}{T}~,\label{Linear_response_f} \eeq where now $f(\e)$
is the equilibrium distribution, and the $+$ and $-$ sign correspond
to which electrochemical potential is higher or lower in energy with
respect to the equilibrium chemical potential. Inserting this into
Eq.~(\ref{Landauer_charge_current}) and equating the current to
zero, one obtains \beq S(T)=\frac{1}{e T}
\frac{\int^\infty_{-\infty} \d\e \T(\e) (\e-\mu)
(-f'(\e))}{\int^\infty_{-\infty} \d\e \T(\e)(-f'(\e))} ~~.
\label{S1}\eeq Already from this result several features may be
seen. First, since at $T=0$ we have $-f'(\e)=\del (\e-\mu)$, the
numerator of $S$ vanishes and $S(T=0)=0$. Second, even at finite
temperatures $f'(\e)$ is symmetric around $\mu$, and therefore $S=0$
unless $\T(\e)$ is \emph{not} symmetric around $\mu$. This is
similar to the condition in bulk materials that requires the
particle-hole symmetry be broken to have a finite
thermopower~\cite{Ashcroft-Mermin}.

One can further simplify $S(T)$ by taking the low-temperature limit
and by assuming that there are no electronic resonances close to the
equilibrium chemical potential. Using the
Sommerfeld expansion to first order around $\mu(T=0)=\e_F$~\cite{Ashcroft-Mermin} one has
\beqa
\int^{\infty}_{-\infty}\T(\e)(\e-\mu)f'(\e)&\approx& \left. \frac{\pi^2}{6}k_B^2T^2\frac{\d^2[ \T(\e)(\e-\mu)]}{\d \e^2} \right|_{\e_F} \nonum &=&\frac{\pi^2}{3} k_B^2T^2 \T '(\e) \eeqa
(where the second derivative comes from integration by parts) and
one arrives at the
expression for the thermopower, \beq
S=\frac{\pi^2}{3}\frac{k_B}{e}k_B T \left. \frac{\d \ln(\T(\e))}{\d
\e} \right|_{\e_F} \label{S_T0}~~,\eeq which is similar to Mott's
semiclassical formula for bulk metals
\cite{Ashcroft-Mermin,Lunde:3879}. We stress once more that this
approximation is only valid at low temperatures and away from
transmission resonances, so that the variation in $\T(\e)$ is small.

The advantages of using the Landauer formalism are evident: it
provides both a simple interpretation of the thermopower in terms of
single-particle properties such as the transmission coefficient
$\T(\e)$, and a rather straightforward computational procedure. In
fact, one only needs to determine the transmission coefficient
$\T(\e)$, which can be done as discussed in Sec.~\ref{Sec1theorysp}.
These reasons have made this approach extremely popular and widely
used. An early use of Eq.~(\ref{S_T0}) is in the study of
thermopower in quantum point contacts
\cite{Molenkamp:1052,Houten:B215,Molenkamp:903} and quantum dots
\cite{Staring:57}. In these mesoscopic systems, a gate voltage is
used to tune either the width of the quantum point contacts or the
energy levels in the quantum dots, giving rise to quantized
conductance and Coulomb blockade. It turns out that in the cases above,
the Landauer approach yields reasonably good agreement between
theory and experiment \cite{Molenkamp:1052}, and knowledge of $\T(\e)$ reasonably
describes both the conductance and the thermopower. This would
naively suggest that for these types of systems, the above
single-particle picture accounts for most of the thermopower.
However, more recent investigations which include effects of
interactions, show that in both types of systems interactions may
induce deviations from the Wiedmann-Franz law, thus reducing the
agreement with
experiments~\cite{Kubala:066801,Lunde:256802,Turek:220503,Turek:115332,Zhang:086214},
suggesting that the agreement in the single-particle case may be the
result of cancelation of errors.

In fact, despite its simplicity, the above approach suffers several
shortcomings of particular relevance in nanoscale systems. The most
prominent is the fact that it is formulated for non-interacting
electrons. This means that any inclusion of interaction effects
directly into $\T(\e)$ can only be done at the mean-field
level~\cite{Vignale2009}. To correct this, one should abandon the
Landauer formula for the current, and, alternatively, use
expressions for the currents obtained by using, e.g., the NEGF
\cite{Meir:2512} or rate equations \cite{Koch:195107}.
To our knowledge, in its fully interacting form NEGF was never
employed to study the effects of electron interactions on the
thermopower of molecular junctions.

Another limitation of the Landauer approach is the erroneous result
it supplies in the zero-coupling limit. To demonstrate this,
consider the simplest model for a nanoscale junction: a single
resonant level symmetrically coupled to leads with spinless
electrons (adding spin simply introduces a factor of two). The
transmission is given by a Breit-Wigner formula,
$\T(\e)=\Gamma^2/(\Gamma^2+(\e-\e_F)^2)$, where $\Gamma$ is the
lead-induced level broadening. Substituting into the expression for
$S$ (Eq.~(\ref{S_T0})) and taking the limit of $\Gamma \to 0$ gives
a finite value, $S=-\frac{2\pi^2}{3}\frac{k_B^2}{e}
\frac{T}{\e-\e_F}$. However, if one would consider a real device, it is
clear that by detaching the leads would result in no
temperature-induced voltage drop. The reason for this discrepancy is
simple: in the linear response calculation one assumes that the
temperature difference is the smallest energy scale, yet in the limit
$\Gamma \to 0$, $\Gamma$ becomes comparable to such a scale, and the
approximation breaks down. One should thus be careful both in using
perturbation theory in the coupling between the leads and, say, a
molecule in the junction and in comparing such calculations to
experiments (see Sec.~\ref{Theoryopen}).

\begin{figure}[ht]
\vskip 0.5truecm
\includegraphics[width=8.5truecm]{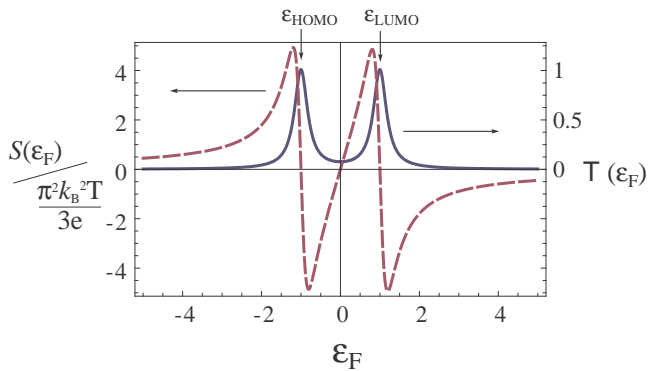}
\caption{(Color online) Transmission (solid line) and normalized
thermopower (dashed line) as a function of the position of the Fermi
energy with respect to the HOMO-LUMO gap, based on the Landauer
formula, Eq.~(\ref{S_T0}), with approximation Eq.~(\ref{HOMO LUMO
transimssion}).}\label{fig3}
\end{figure}

Much of the recent theoretical work on thermopower has been devoted
to molecular junctions. Before we review some recent results, it is
important to understand the origin of the specific interest in such
systems, which may be understood from analyzing the Landauer
formula~(\ref{S_T0}). In a molecular junction, the Fermi energy of
the leads is placed somewhere between the HOMO and the LUMO (i.e.,
in the HOMO-LUMO gap) \cite{Nitzan:1384}. The question is where
exactly? The answer to this question cannot be answered by studying
the conductance (or transmission) alone, which can be demonstrated
through a simple example \cite{Paulsson:241403}. Consider such a
molecular junction, with HOMO and LUMO energies $\e_{\text{HOMO}}$
and $\e_{\text{LUMO}}$, respectively. The transmission function can
be modeled by a double Lorenzian, corresponding to tunneling via
each of these levels independently, and assumes the following form,
\beq \T(\e)=\frac{\Gamma_\L \Gamma_\R
}{\tilde{\Gamma}^2+\left(\epsilon _F-\epsilon
_{\text{HOMO}}\right){}^2}+\frac{
  \Gamma_\L \Gamma_\R}{\tilde{\Gamma}^2+\left(\epsilon _F-\epsilon _{\text{LUMO}}\right){}^2}\label{HOMO LUMO transimssion}~~,
  \eeq
where $\Gamma_{\L,\R}$ is the level broadening due to the left
(right) lead and $\tilde{\Gamma}=(\Gamma_\L+\Gamma_\R)/2$. (For
simplicity, we assume it to be the same for the two levels.) The
resulting thermopower (in units of
$\frac{\pi^2}{3}\frac{k_B^2}{e}T$), along with the transmission
coefficient, is plotted in Fig.~\ref{fig3} (taking
$\tilde{\Gamma}/(\e_{\text{LUMO}}-\e_{\text{HOMO}})=0.1$). As seen,
according to this simple model, for a given value of transmission between $\e_{\text{LUMO}}$ and $\e_{\text{HOMO}}$,
there are two values of the Fermi energy which provide a solution to
Eq.~(\ref{HOMO LUMO transimssion}), and hence conductance alone does
not suffice to determine the position of the Fermi energy. From the
same model, however, one would infer that the sign of the
thermopower is determined by the position of the Fermi energy from
the center of the HOMO-LUMO gap, similarly to the fact that the sign
of the thermopower in bulk materials is determined by whether the
conductance is dominated by electrons or
holes~\cite{Ashcroft-Mermin}, and therefore the sign of thermopower
distinguishes between the two Fermi energies which solve
Eq.~(\ref{HOMO LUMO transimssion}).

This idea, along with the prospect of using molecular junctions as
efficient thermoelectric devices, has generated much theoretical
interest. To give a few examples,  Segal \cite{Segal:165426} showed
that by measuring the thermopower one can distinguish between
different electron transport mechanisms. Thermal and vibrational
effects were studied in detail \cite{Koch:195107} using rate
equations, and it was shown that at low temperatures the signature
of the vibrational modes on the thermopower can be measured. Murphy
\etal~\cite{Murphy:155313} used rate equations to study the
optimization of the figure of merit of a molecular junction, in
similarity to the optimization of the figure of merit in bulk
thermoelectrics \cite{Mahan:7436}.

Much recent attention has been devoted to studying thermopower of
molecular junctions using ground-state DFT to calculate $\T(\e)$
combined with the Landauer formula~(\ref{S1}) (or its even more
simplified version
Eq.~(\ref{S_T0}))~\cite{finch:033405,Paulsson:241403,Zheng:8537,Pauly:035315,Ke:1011,Muller:044708,
Wang:233406,Liu:193101,Liu:3497}.
In some cases, it has also been reported an impressive agreement
between the theoretical results and experiments~\cite{Ke:1011}.

However, caution has to be applied in making such claims. In fact, if the system is away from linear response - and many experiments
so far likely correspond to such case -
given a temperature difference, setting $I=0$ in Eq.~\ref{Landauer_charge_current} does
not necessarily provide a unique solution for the potential difference. In other words,
there may be more than one potential difference $\Delta V$ that gives rise to the same $\Delta T$ (and hence several values of thermopower
for the same temperature difference), when $I=0$, as it is permitted
by the non-linearity of Eq.~\ref{Landauer_charge_current}. In addition,
as already emphasized previously, even if the single-particle
equations~(\ref{S1}) and~(\ref{S_T0}) were good starting points to
describe the problem at hand, ground-state DFT is fundamentally
flawed in the present context (even if one knew the exact
ground-state functional) due to the fundamental non-equilibrium
nature of the problem~\cite{Vignale2009,Bushong:2569,Ventra:8025,DiVentra:2008}.
In this respect, even the interpretation of ground-state Kohn-Sham
orbitals is questionable, since the latter ones are auxiliary
quantities whose only role is to provide the correct density of the
corresponding many-body system in its ground-state.

To these limitations we must also add few more physical issues. When
a thermal gradient is applied to a junction, the transient dynamics
is fundamental in establishing the voltage difference that enters
the definition of thermopower. Since the {\em dynamical} formation
of local resistivity dipoles creates strong local fields at the
junction (especially at the nanoscale), these fields influence the
electron motion in a non-trivial way, and thus influence the
long-time behavior of the carrier dynamics, even in the dc limit. This is particularly
important {\em away} from linear response~\cite{DiVentra:2008}, which may be the experimental case.

It is the self-consistent formation of these fields that makes the
thermopower very sensitive (both in magnitude and sign) to atomic details, and thus to the
contact geometry between the nanostructure and bulk electrodes, as
demonstrated also experimentally~\cite{Ludoph:12290}. This precludes
an easy interpretation in terms of ``electron''
or ``hole'' excitations as in bulk metals, and thus an easy relation
with single-particle states (such as the HOMO and LUMO) as the
Landauer equation~(\ref{S1}) would imply. All this points to the fact
that, since the system is in {\em dynamical} interaction with two
different baths, one needs to go beyond the approximations
underlying Eq.~(\ref{S1}), and consider an open quantum system
approach.

\subsubsection{An open quantum system approach}\label{Theoryopen}
The present authors have precisely explored the problem of
thermopower within the theory of open quantum
systems~\cite{Dubi:97}. In analogy with the idea that electrical currents may be studied
using finite systems~\cite{Ventra:8025,Bushong:2569}, one can study a finite
system in contact with two heat baths held
at different temperatures (i.e., finite leads connected by a nanoscale constriction,
either a molecule, wire, etc.). If the system has a finite thermoelectric
response then charges would flow between the leads until the ensuing
electric potential ``balances'' the thermal gradient, and a charge
imbalance is created between the two leads (which is related to the
thermo-voltage via the Poisson equation). Note that in this approach
the system is allowed to find its own charge distribution via the
transient dynamics (unlike a static approach where a static
distribution is imposed {\em a priori} via boundary conditions), and
even when the charge current is zero an {\em energy current} is
still present, as in the actual experiments. Then, by calculating
the thermally-induced charge imbalance one obtains information on
the thermoelectric response of the junction via the usual
definitions. This approach is also not limited to linear response
thus providing information on the thermo-voltage even when the
temperature gradient is not the smallest energy scale.
\begin{figure}[ht]
\vskip 0.5truecm
\includegraphics[width=6.5truecm]{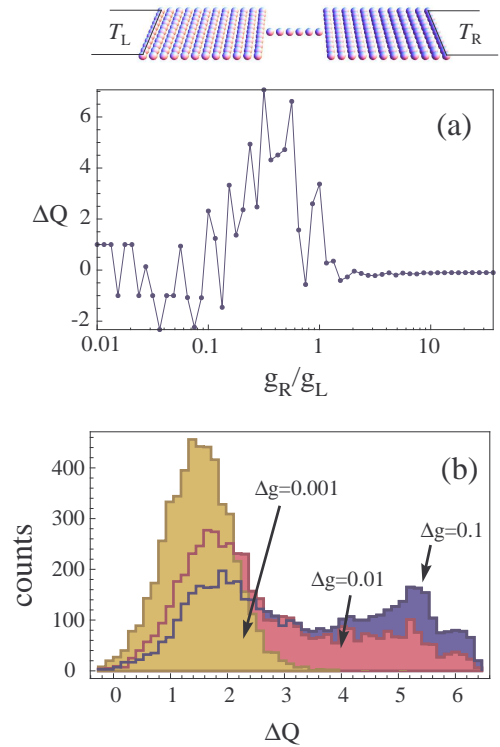}
\caption{(Color online) Upper panel: schematic representation of the
model molecular junction composed of two quasi-two-dimensional leads
connected with a molecular wire. The leads are coupled to external
heat baths, each at its own temperature. (a) Electron charge imbalance as a
function of the ratio between the couplings $g_{R(L)}$ between the wire and the
right (left) lead. A strong dependence can be observed and even a change of sign
(from \cite{Dubi:97}). (b) Distribution of charge imbalance, when
the couplings between the wire and the leads are drawn from a
Gaussian distribution, with an average $g=0.1$ and width $\D
g=0.001, 0.01$ and $0.1$ in units of the hopping parameter (see text).}\label{fig4}
\end{figure}

An implementation of such an open system approach can also be
formulated within time-dependent DFT, thus allowing to include, in
principle, all possible dynamical many-body effects in the
thermopower (recall in fact that given the baths that set the
temperature differences, the ensuing electrostatic voltage is a
well-defined functional of the density). Indeed, Di Ventra and
D'Agosta have recently proved that if the bath-electron interactions
are treated within a memory-less approximation (the thermal baths
being Ohmic) \cite{VanKampen:2001}, then there is a one-to-one
correspondence between the exact ensemble-averaged current density
and the external vector potential, therefore extending the theorem
(and Kohn-Sham scheme) of time-dependent current-DFT (TDCDFT) to
open quantum systems~\cite{Ventra:226403,D'Agosta:165105}. The
framework for this theory (named Stochastic TDCDFT) is the
stochastic Schr\"odinger equation, which describes a Hamiltonian
quantum system in the presence of a bath (the extension to several
baths is trivial) \cite{Breuer:2002} ($\hbar=1$), \beq \dot{\Psi}(t)=-i \cH \Psi
(t)-\half \hat{V}^\dagger \hat{V} \Psi (t)+l(t)\hat{V}
\Psi(t)\label{SSE} ~~.\eeq Here $\Psi(t)$ is the many-body state
vector, $\cH$ is the Hamiltonian of the system (describing both the
molecule, the leads and the coupling between them), $\hat{V}$ are
the so-called bath operators (which could in principle be position
and/or time-dependent) \cite{VanKampen:2001}, which describe
transitions between the different many-body states induced by the bath(s), and $l(t)$ is a
stochastic field which is taken to have zero mean and a
$\delta$-function autocorrelation, $\langle l(t)\rangle=0,~~\langle
l(t) l(t')\rangle=\del(t-t')$.

As a first demonstration, the method was recently used by the
present authors to study the thermopower of a simple model system of
spinless non-interacting electrons, for which the calculations can
be equivalently carried out with the density-matrix rather than the
state vector, by averaging over the stochastic
realizations~\cite{Pershin:054302}. In that model two planar leads,
each in contact with a thermal bath at a given temperature, are
connected via a nanoscale wire (see upper panel of Fig.~\ref{fig4}).
The bath-electron interactions are described by the operators \beq
\hat{V}^{L,R}_{kk'}=\sqrt{\gamma^{L,R}_{kk'} f^{L,R}(\e_k)}
\ket{k}\bra{k'}~~,\label{V-operators}\eeq where $\ket{k}$ are the
single-particle states of the Hamiltonian, $f^{L,R}$ is the Fermi
function containing information on the left (right) bath
temperature, and $\gamma^{L,R}_{kk'}$, which are the (inelastic)
transition rates between states $k$ and $k'$, depend on the bath
location (i.e., left or right) \cite{Dubi:97}. The corresponding
equations of motion are then solved to obtain the wave-function, and
hence the electron density, potential, and also the heat currents
(see also Sec.~\ref{Sec3}) at steady state.

Several interesting features are revealed by this model. For
instance, the obtained thermo-voltage shows non-linear
characteristics, which imply that the linear regime may not be the
best regime to operate a thermo-electric device \cite{Dubi:97}.
Another interesting feature is the strong dependence of the charge
imbalance in the leads (and hence the thermopower) on the coupling
between the wire (or molecule) and the leads. This confirms the
experimental findings in metallic quantum point
contacts~\cite{Ludoph:12290}. In Fig.~\ref{fig4}(a) \cite{Dubi:97}
the charge imbalance between the leads is plotted as a function of
the ratio between the coupling between the wire and the left and
right leads. From the figure it is obvious that the charge imbalance
strongly fluctuates, and can even change sign as a function of the
wire-lead coupling. To demonstrate the importance of these
fluctuations and to tie with the experimental results presented in
Sec.~\ref{Thermoexp}, we have performed a calculation for the same
system as discussed in~\cite{Dubi:97}, where the coupling
constants were drawn from a normal distribution around typical
values of $g=0.1t$, where $g$ is the lead-wire coupling and $t$ is
the tight-binding hopping parameter, which describes the bandwidth
of the leads. Other numerical parameters are the filling fraction
of electrons, $n=1/3$, and the temperatures of the left and right
heat baths, $T_L=0.1$ and $T_R=1$ (in units of  $t$). A histogram of
the resulting charge imbalance is plotted in Fig.~\ref{fig4}(b), for
three values of the width of the normal distribution, $\D g=0.001,
0.01$ and $0.1$ (in units of the hopping parameter). While more work
needs to be done to explain the experimental data presented in
Fig.~\ref{fig2}(c), these theoretical results clearly bear some
resemblance to experiments by showing a double structure in the
charge imbalance as a function of the coupling asymmetry.

Using Stochastic TDCDFT one can extend the above system
model to interacting systems, as well as to a multi-component
formulation~\cite{Appel:new}, whereby the Hamiltonian now
contains the correlated motion of electrons and (possibly quantum)
ions, with both components interacting with an external environment. Such
studies would shed new light on the role of interactions and ion dynamics on
thermopower, and enable a study of local heating effects in nanoscale systems (see
also Sec.~\ref{Sec3}). No results are, however, yet available for these cases.
\section{Summary and future prospects}\label{summary}
In this review we have discussed energy transport in nanoscale systems, such as molecular junctions, suspended nanotubes, quantum point contacts, etc. Our aim was to put under a unified theme the three major issues of thermal transport, namely thermal conductance, local temperature and heating, and thermo-electricity. We have critically examined both theoretical and experimental aspects of these topics. We have presented many theoretical methods based on the single-particle scattering approach, non-equilibrium Green's functions formalism, molecular dynamics, etc. From the experimental side we have reviewed state-of-the-art experiments, and stressed the difficulty and
open questions in analyzing such experiments.

\subsection{Future Prospects}
We wish to conclude this review by presenting three novel ideas related to energy transport in nanoscale systems. These ideas, which deviate somewhat from the usual path of thermoelectricity and heat transport, reflect to our opinion the richness and usefulness of studying energy flow in nanoscale systems, and we hope they will stimulate both the experimental and theoretical communities.

\emph{Thermo-spintronics} --~~
Thermo-spintronics (sometimes also called spin calorimetrics) refers to the manipulation of electron spins with thermal effects. Generating spin currents, that is the flow of electron spins, plays an eminent role in the field of spintronics, which is the spin analogue of electronics (see, e.g.,~\cite{Zutic}). However, manipulating spins in order to generate spin currents is quite difficult, and it is equally hard to generate a spin current without generating an accompanying charge current. In recent experiments \cite{Uchida:778,Uchida:07C908} a spin-analog to the Seebeck effect was used to generate a spin-voltage, induced by a temperature difference along a ferromagnetic slab. Although this effect is rather small (compared to its charge counterpart, but larger than expected in view of spin-flip scattering) and inherently induces an electric voltage as well, it has been suggested that these shortcomings may be overcome by applying a temperature gradient to a molecular junction placed between ferromagnetic leads \cite{Dubi:081302}, a setup which was further studied recently \cite{Lu:123111,Wang:057202,Ying:093104}. In another interesting work, a variety of thermoelectric effects in magnetic junctions have been studied \cite{Hatami:066603,Hatami:174426,Heikkila:100408}, with unusual features such as thermal spin-transfer torque, spin-polarized cooling and spin-heat coupling effects.

\emph{Enhanced thermopower in DNA} --~~
DNA, the basic building block of our genetic code, shows also large potential in nanotechnology applications \cite{Dekker:29,DiVentra:2004,Zwolak:141}. In a recent study it was shown \cite{Macia:035130,Macia:254}, using a model Hamiltonian for different DNA-like chains, that under certain conditions, the Seebeck coefficient and figure of merit of a lead-DNA-lead junction can be quite high, and can rise to hundreds of $\mu V/$K, to be compared with a few $\mu V/$K of other single-molecule junctions studied so far. These high values of the thermopower seem to stem from transport resonance effects, which can be tuned rather easily in DNA. This, with the fact that there is a lot of know-how regarding DNA manipulation and preparation, makes DNA-based systems interesting candidates for future thermoelectric and cooling devices at the nanoscale.

\emph{Thermoelectricity in superconducting wires} --~~
Raising the critical temperature, $T_c$, of superconducting materials is clearly a technologically important goal. However, most superconducting materials have a $T_c$ well below room temperature, even in the well-known high-$T_c$ materials (with $T_c \sim 80$K for wires and $T_c \sim 200$K for bulk). Recently, the present authors have suggested \cite{Dubi:new} to study a superconducting wire held at two different temperatures at its edges. Using the method introduced in Sec.~\ref{Theoryopen} combined with a self-consistent mean-field theory, it was shown that for an (ideally) clean superconducting wire, if one of the sides (the cold side) is held at low enough temperatures, the temperature of the hot side can be much larger than the equilibrium $T_c$, with the wire still in its superconducting
state. Although this study neglects some effects (such as phase fluctuations), the basic idea is simple: in order to have a superconducting wire, instead of cooling down the entire apparatus, one can \emph{locally} cool the wire by attaching to it a local refrigerator, for example one made from a Peltier cooling device~\cite{Shakouri:1705146}. This may pave the road for hybrid superconducting circuits which operate at relatively high temperatures.

\subsection{Final thoughts}
These last few examples - and what we have discussed in this Colloquium - clearly show that the quest to understand energy transport in nanostructures is far from over. In fact, it seems to us that we have barely scratched the
surface of this problem and more discoveries await us. Regarding thermal conductance, finding systems that show either very good (for nano-electronic applications) or very poor (for thermo-electric applications) thermal conductance is needed. As for thermoelectricity, there is a need to advance our theoretical tools quite substantially. For instance, theories that  account for the statistical nature of the experiments should be developed that include also electron-electron and electron-phonon interactions on equal footing. In addition, since the problem is intrinsically out of equilibrium (even at steady-state) these theories need to include dynamical effects. As for local heating and local temperatures, the handful of experiments that have appeared recently are certainly a great start, but more are needed in order to truly determine the processes leading to heating (and cooling) in nanoscale junctions. Similarly,
more experiments that could determine directly the validity (or invalidity) of Fourier's law are highly desirable.

Due to the rapid developments in science and technology it is difficult to predict where the field will go from here.
However, there is no doubt that novel and ingenious ideas will be put forward that will help us profit from energy flow, storage and conversion. Embarking in such a quest could not be more timely.

\section*{Acknowledgments}

We thank D. Roy for useful
discussions and critical reading of the manuscript. This work has been supported by DOE under grant DE-FG02-05ER46204
and University of California Laboratories.
\bibliographystyle{apsrmp}
\bibliography{Refs}

\begin{thebibliography}{265}
\expandafter\ifx\csname natexlab\endcsname\relax\def\natexlab#1{#1}\fi
\expandafter\ifx\csname bibnamefont\endcsname\relax
  \def\bibnamefont#1{#1}\fi
\expandafter\ifx\csname bibfnamefont\endcsname\relax
  \def\bibfnamefont#1{#1}\fi
\expandafter\ifx\csname citenamefont\endcsname\relax
  \def\citenamefont#1{#1}\fi
\expandafter\ifx\csname url\endcsname\relax
  \def\url#1{\texttt{#1}}\fi
\expandafter\ifx\csname urlprefix\endcsname\relax\def\urlprefix{URL }\fi
\providecommand{\bibinfo}[2]{#2}
\providecommand{\eprint}[2][]{\url{#2}}

\bibitem[{\citenamefont{Akkermans and Montambaux}(2007)}]{Akkermans:2007}
\bibinfo{author}{\bibnamefont{Akkermans}, \bibfnamefont{E.}}, and
  \bibinfo{author}{\bibfnamefont{G.}~\bibnamefont{Montambaux}},
  \bibinfo{year}{2007}, \emph{\bibinfo{title}{Mesoscopic Physics of Electrons
  and Photons}} (\bibinfo{publisher}{Cambridge University Press}).

\bibitem[{\citenamefont{Ando}(1991)}]{Ando:8017}
\bibinfo{author}{\bibnamefont{Ando}, \bibfnamefont{T.}}, \bibinfo{year}{1991},
  \bibinfo{journal}{Phys. Rev. B}
  \textbf{\bibinfo{volume}{44}}(\bibinfo{number}{15}), \bibinfo{pages}{8017}.

\bibitem[{\citenamefont{Angelescu} \emph{et~al.}(1998)\citenamefont{Angelescu,
  Cross, and Roukes}}]{Angelescu:673}
\bibinfo{author}{\bibnamefont{Angelescu}, \bibfnamefont{D.~E.}},
  \bibinfo{author}{\bibfnamefont{M.~C.} \bibnamefont{Cross}}, and
  \bibinfo{author}{\bibfnamefont{M.~L.} \bibnamefont{Roukes}},
  \bibinfo{year}{1998}, \bibinfo{journal}{Superlattices and Microstructures}
  \textbf{\bibinfo{volume}{23}}(\bibinfo{number}{3-4}), \bibinfo{pages}{673 }.

\bibitem[{\citenamefont{Appel and Di~Ventra}(2009)}]{Appel:new}
\bibinfo{author}{\bibnamefont{Appel}, \bibfnamefont{H.}}, and
  \bibinfo{author}{\bibfnamefont{M.}~\bibnamefont{Di~Ventra}},
  \bibinfo{year}{2009}, \bibinfo{journal}{Phys. Rev. B}
  \textbf{\bibinfo{volume}{80}}(\bibinfo{number}{21}), \bibinfo{pages}{212303}.

\bibitem[{\citenamefont{Appleyard} \emph{et~al.}(2000)\citenamefont{Appleyard,
  Nicholls, Pepper, Tribe, Simmons, and Ritchie}}]{Appleyard:16275}
\bibinfo{author}{\bibnamefont{Appleyard}, \bibfnamefont{N.~J.}},
  \bibinfo{author}{\bibfnamefont{J.~T.} \bibnamefont{Nicholls}},
  \bibinfo{author}{\bibfnamefont{M.}~\bibnamefont{Pepper}},
  \bibinfo{author}{\bibfnamefont{W.~R.} \bibnamefont{Tribe}},
  \bibinfo{author}{\bibfnamefont{M.~Y.} \bibnamefont{Simmons}}, and
  \bibinfo{author}{\bibfnamefont{D.~A.} \bibnamefont{Ritchie}},
  \bibinfo{year}{2000}, \bibinfo{journal}{Phys. Rev. B}
  \textbf{\bibinfo{volume}{62}}(\bibinfo{number}{24}), \bibinfo{pages}{R16275}.

\bibitem[{\citenamefont{Ashcroft and Mermin}(1976)}]{Ashcroft-Mermin}
\bibinfo{author}{\bibnamefont{Ashcroft}, \bibfnamefont{N.~W.}}, and
  \bibinfo{author}{\bibfnamefont{N.~D.} \bibnamefont{Mermin}},
  \bibinfo{year}{1976}, \emph{\bibinfo{title}{Solid State Physics}}
  (\bibinfo{publisher}{Brooks Cole}).

\bibitem[{\citenamefont{Baheti} \emph{et~al.}(2008)\citenamefont{Baheti, Malen,
  Doak, Reddy, Jang, Tilley, Majumdar, and Segalman}}]{Baheti:715}
\bibinfo{author}{\bibnamefont{Baheti}, \bibfnamefont{K.}},
  \bibinfo{author}{\bibfnamefont{J.~A.} \bibnamefont{Malen}},
  \bibinfo{author}{\bibfnamefont{P.}~\bibnamefont{Doak}},
  \bibinfo{author}{\bibfnamefont{P.}~\bibnamefont{Reddy}},
  \bibinfo{author}{\bibfnamefont{S.-Y.} \bibnamefont{Jang}},
  \bibinfo{author}{\bibfnamefont{T.~D.} \bibnamefont{Tilley}},
  \bibinfo{author}{\bibfnamefont{A.}~\bibnamefont{Majumdar}}, and
  \bibinfo{author}{\bibfnamefont{R.~A.} \bibnamefont{Segalman}},
  \bibinfo{year}{2008}, \bibinfo{journal}{Nano Lett.}
  \textbf{\bibinfo{volume}{8}}(\bibinfo{number}{2}), \bibinfo{pages}{715}.

\bibitem[{\citenamefont{Balandin} \emph{et~al.}(2008)\citenamefont{Balandin,
  Ghosh, Bao, Calizo, Teweldebrhan, Miao, and Lau}}]{Balandin:902}
\bibinfo{author}{\bibnamefont{Balandin}, \bibfnamefont{A.~A.}},
  \bibinfo{author}{\bibfnamefont{S.}~\bibnamefont{Ghosh}},
  \bibinfo{author}{\bibfnamefont{W.}~\bibnamefont{Bao}},
  \bibinfo{author}{\bibfnamefont{I.}~\bibnamefont{Calizo}},
  \bibinfo{author}{\bibfnamefont{D.}~\bibnamefont{Teweldebrhan}},
  \bibinfo{author}{\bibfnamefont{F.}~\bibnamefont{Miao}}, and
  \bibinfo{author}{\bibfnamefont{C.~N.} \bibnamefont{Lau}},
  \bibinfo{year}{2008}, \bibinfo{journal}{Nano Letters}
  \textbf{\bibinfo{volume}{8}}(\bibinfo{number}{3}), \bibinfo{pages}{902}.

\bibitem[{\citenamefont{Bell}(2008)}]{Bell:09122008}
\bibinfo{author}{\bibnamefont{Bell}, \bibfnamefont{L.~E.}},
  \bibinfo{year}{2008}, \bibinfo{journal}{Science}
  \textbf{\bibinfo{volume}{321}}(\bibinfo{number}{5895}),
  \bibinfo{pages}{1457}.

\bibitem[{\citenamefont{Berber} \emph{et~al.}(2000)\citenamefont{Berber, Kwon,
  and Tom\'anek}}]{Berber:4613}
\bibinfo{author}{\bibnamefont{Berber}, \bibfnamefont{S.}},
  \bibinfo{author}{\bibfnamefont{Y.-K.} \bibnamefont{Kwon}}, and
  \bibinfo{author}{\bibfnamefont{D.}~\bibnamefont{Tom\'anek}},
  \bibinfo{year}{2000}, \bibinfo{journal}{Phys. Rev. Lett.}
  \textbf{\bibinfo{volume}{84}}(\bibinfo{number}{20}), \bibinfo{pages}{4613}.

\bibitem[{\citenamefont{Blencowe}(1999)}]{Blencowe:4992}
\bibinfo{author}{\bibnamefont{Blencowe}, \bibfnamefont{M.}},
  \bibinfo{year}{1999}, \bibinfo{journal}{Phys. Rev. B}
  \textbf{\bibinfo{volume}{59}}(\bibinfo{number}{7}), \bibinfo{pages}{4992}.

\bibitem[{\citenamefont{Blencowe}(2004)}]{Blencowe:159}
\bibinfo{author}{\bibnamefont{Blencowe}, \bibfnamefont{M.}},
  \bibinfo{year}{2004}, \bibinfo{journal}{Physics Reports}
  \textbf{\bibinfo{volume}{395}}(\bibinfo{number}{3}), \bibinfo{pages}{159 }.

\bibitem[{\citenamefont{Bonetto} \emph{et~al.}(2004)\citenamefont{Bonetto,
  Lebowitz, and Lukkarinen}}]{Bonetto:783}
\bibinfo{author}{\bibnamefont{Bonetto}, \bibfnamefont{F.}},
  \bibinfo{author}{\bibfnamefont{J.}~\bibnamefont{Lebowitz}}, and
  \bibinfo{author}{\bibfnamefont{J.}~\bibnamefont{Lukkarinen}},
  \bibinfo{year}{2004}, \bibinfo{journal}{Journal of Statistical Physics}
  \textbf{\bibinfo{volume}{116}}(\bibinfo{number}{1-4}), \bibinfo{pages}{783}.

\bibitem[{\citenamefont{Bonetto} \emph{et~al.}(2000)\citenamefont{Bonetto,
  Lebowitz, and Rey-Bellet}}]{Bonetto:2000}
\bibinfo{author}{\bibnamefont{Bonetto}, \bibfnamefont{F.}},
  \bibinfo{author}{\bibfnamefont{J.~L.} \bibnamefont{Lebowitz}}, and
  \bibinfo{author}{\bibfnamefont{L.}~\bibnamefont{Rey-Bellet}},
  \bibinfo{year}{2000}, in \emph{\bibinfo{booktitle}{Mathematical Physics
  2000}}, edited by \bibinfo{editor}{\bibfnamefont{A.}~\bibnamefont{Fokas}},
  \bibinfo{editor}{\bibfnamefont{A.}~\bibnamefont{Grigoryan}},
  \bibinfo{editor}{\bibfnamefont{T.}~\bibnamefont{Kibble}}, and
  \bibinfo{editor}{\bibfnamefont{B.}~\bibnamefont{Zegarlinski}}
  (\bibinfo{publisher}{Imperial College Press}), p. \bibinfo{pages}{128}.

\bibitem[{\citenamefont{Boukai} \emph{et~al.}(2006)\citenamefont{Boukai, Xu,
  and Heath}}]{Boukai:864}
\bibinfo{author}{\bibnamefont{Boukai}, \bibfnamefont{A.}},
  \bibinfo{author}{\bibfnamefont{K.}~\bibnamefont{Xu}}, and
  \bibinfo{author}{\bibfnamefont{J.}~\bibnamefont{Heath}},
  \bibinfo{year}{2006}, \bibinfo{journal}{Advanced Materials}
  \textbf{\bibinfo{volume}{18}}(\bibinfo{number}{7}), \bibinfo{pages}{864}.

\bibitem[{\citenamefont{Boukai} \emph{et~al.}(2007)\citenamefont{Boukai,
  Bunimovich, Tahir-Kheli, Yu, III, and Heath}}]{Boukai:168}
\bibinfo{author}{\bibnamefont{Boukai}, \bibfnamefont{A.~I.}},
  \bibinfo{author}{\bibfnamefont{Y.}~\bibnamefont{Bunimovich}},
  \bibinfo{author}{\bibfnamefont{J.}~\bibnamefont{Tahir-Kheli}},
  \bibinfo{author}{\bibfnamefont{J.-K.} \bibnamefont{Yu}},
  \bibinfo{author}{\bibfnamefont{W.~A.~G.} \bibnamefont{III}}, and
  \bibinfo{author}{\bibfnamefont{J.~R.} \bibnamefont{Heath}},
  \bibinfo{year}{2007}, \bibinfo{journal}{Nature}
  \textbf{\bibinfo{volume}{451}}, \bibinfo{pages}{168}.

\bibitem[{\citenamefont{Bourgeois} \emph{et~al.}(2007)\citenamefont{Bourgeois,
  Fournier, and Chaussy}}]{Bourgeois:016104}
\bibinfo{author}{\bibnamefont{Bourgeois}, \bibfnamefont{O.}},
  \bibinfo{author}{\bibfnamefont{T.}~\bibnamefont{Fournier}}, and
  \bibinfo{author}{\bibfnamefont{J.}~\bibnamefont{Chaussy}},
  \bibinfo{year}{2007}, \bibinfo{journal}{Journal of Applied Physics}
  \textbf{\bibinfo{volume}{101}}(\bibinfo{number}{1}), \bibinfo{eid}{016104}.

\bibitem[{\citenamefont{Breuer and Petruccione}(2002)}]{Breuer:2002}
\bibinfo{author}{\bibnamefont{Breuer}, \bibfnamefont{H.~P.}}, and
  \bibinfo{author}{\bibfnamefont{F.}~\bibnamefont{Petruccione}},
  \bibinfo{year}{2002}, \emph{\bibinfo{title}{Theory of Open Quantum Systems}}
  (\bibinfo{publisher}{Oxford University Press}).

\bibitem[{\citenamefont{Brown} \emph{et~al.}(2005)\citenamefont{Brown, Hao,
  Gallop, and Macfarlane}}]{Brown:023107}
\bibinfo{author}{\bibnamefont{Brown}, \bibfnamefont{E.}},
  \bibinfo{author}{\bibfnamefont{L.}~\bibnamefont{Hao}},
  \bibinfo{author}{\bibfnamefont{J.~C.} \bibnamefont{Gallop}}, and
  \bibinfo{author}{\bibfnamefont{J.~C.} \bibnamefont{Macfarlane}},
  \bibinfo{year}{2005}, \bibinfo{journal}{Appl. Phys. Lett.}
  \textbf{\bibinfo{volume}{87}}(\bibinfo{number}{2}), \bibinfo{eid}{023107}.

\bibitem[{\citenamefont{Bushmaker} \emph{et~al.}(2007)\citenamefont{Bushmaker,
  Deshpande, Bockrath, and Cronin}}]{Bushmaker:3618}
\bibinfo{author}{\bibnamefont{Bushmaker}, \bibfnamefont{A.~W.}},
  \bibinfo{author}{\bibfnamefont{V.~V.} \bibnamefont{Deshpande}},
  \bibinfo{author}{\bibfnamefont{M.~W.} \bibnamefont{Bockrath}}, and
  \bibinfo{author}{\bibfnamefont{S.~B.} \bibnamefont{Cronin}},
  \bibinfo{year}{2007}, \bibinfo{journal}{Nano Lett.}
  \textbf{\bibinfo{volume}{7}}(\bibinfo{number}{12}), \bibinfo{pages}{3618}.

\bibitem[{\citenamefont{Bushong} \emph{et~al.}(2005)\citenamefont{Bushong, Sai,
  and {Di Ventra}}}]{Bushong:2569}
\bibinfo{author}{\bibnamefont{Bushong}, \bibfnamefont{N.}},
  \bibinfo{author}{\bibfnamefont{N.}~\bibnamefont{Sai}}, and
  \bibinfo{author}{\bibfnamefont{M.}~\bibnamefont{{Di Ventra}}},
  \bibinfo{year}{2005}, \bibinfo{journal}{Nano Lett.}
  \textbf{\bibinfo{volume}{5}}(\bibinfo{number}{12}), \bibinfo{pages}{2569}.

\bibitem[{\citenamefont{Butcher}(1990)}]{Butcher1990}
\bibinfo{author}{\bibnamefont{Butcher}, \bibfnamefont{P.}},
  \bibinfo{year}{1990}, \bibinfo{journal}{Journal of Physics: Condensed Matter}
  \textbf{\bibinfo{volume}{2}}(\bibinfo{number}{45}), \bibinfo{pages}{4869}.

\bibitem[{\citenamefont{B\"uttiker}
  \emph{et~al.}(1985)\citenamefont{B\"uttiker, Imry, Landauer, and
  Pinhas}}]{Butt1}
\bibinfo{author}{\bibnamefont{B\"uttiker}, \bibfnamefont{M.}},
  \bibinfo{author}{\bibfnamefont{Y.}~\bibnamefont{Imry}},
  \bibinfo{author}{\bibfnamefont{R.}~\bibnamefont{Landauer}}, and
  \bibinfo{author}{\bibfnamefont{S.}~\bibnamefont{Pinhas}},
  \bibinfo{year}{1985}, \bibinfo{journal}{Phys. Rev. B}
  \textbf{\bibinfo{volume}{31}}(\bibinfo{number}{10}), \bibinfo{pages}{6207}.

\bibitem[{\citenamefont{Cahill}(1990)}]{Cahill:802}
\bibinfo{author}{\bibnamefont{Cahill}, \bibfnamefont{D.~G.}},
  \bibinfo{year}{1990}, \bibinfo{journal}{Review of Scientific Instruments}
  \textbf{\bibinfo{volume}{61}}(\bibinfo{number}{2}), \bibinfo{pages}{802}.

\bibitem[{\citenamefont{Cahill} \emph{et~al.}(2003)\citenamefont{Cahill, Ford,
  Goodson, Mahan, Majumdar, Maris, Merlin, and Phillpot}}]{Cahill:793}
\bibinfo{author}{\bibnamefont{Cahill}, \bibfnamefont{D.~G.}},
  \bibinfo{author}{\bibfnamefont{W.~K.} \bibnamefont{Ford}},
  \bibinfo{author}{\bibfnamefont{K.~E.} \bibnamefont{Goodson}},
  \bibinfo{author}{\bibfnamefont{G.~D.} \bibnamefont{Mahan}},
  \bibinfo{author}{\bibfnamefont{A.}~\bibnamefont{Majumdar}},
  \bibinfo{author}{\bibfnamefont{H.~J.} \bibnamefont{Maris}},
  \bibinfo{author}{\bibfnamefont{R.}~\bibnamefont{Merlin}}, and
  \bibinfo{author}{\bibfnamefont{S.~R.} \bibnamefont{Phillpot}},
  \bibinfo{year}{2003}, \bibinfo{journal}{Journal of Applied Physics}
  \textbf{\bibinfo{volume}{93}}(\bibinfo{number}{2}), \bibinfo{pages}{793}.

\bibitem[{\citenamefont{Calizo} \emph{et~al.}(2007)\citenamefont{Calizo,
  Balandin, Bao, Miao, and Lau}}]{Calizo:2645}
\bibinfo{author}{\bibnamefont{Calizo}, \bibfnamefont{I.}},
  \bibinfo{author}{\bibfnamefont{A.~A.} \bibnamefont{Balandin}},
  \bibinfo{author}{\bibfnamefont{W.}~\bibnamefont{Bao}},
  \bibinfo{author}{\bibfnamefont{F.}~\bibnamefont{Miao}}, and
  \bibinfo{author}{\bibfnamefont{C.~N.} \bibnamefont{Lau}},
  \bibinfo{year}{2007}, \bibinfo{journal}{Nano Letters}
  \textbf{\bibinfo{volume}{7}}(\bibinfo{number}{9}), \bibinfo{pages}{2645}.

\bibitem[{\citenamefont{Carruthers}(1961)}]{Carruthers:1961}
\bibinfo{author}{\bibnamefont{Carruthers}, \bibfnamefont{P.}},
  \bibinfo{year}{1961}, \bibinfo{journal}{Rev. Mod. Phys.}
  \textbf{\bibinfo{volume}{33}}(\bibinfo{number}{1}), \bibinfo{pages}{92}.

\bibitem[{\citenamefont{Caso} \emph{et~al.}(2010)\citenamefont{Caso, Arrachea,
  and Lozano}}]{Caso:041301}
\bibinfo{author}{\bibnamefont{Caso}, \bibfnamefont{A.}},
  \bibinfo{author}{\bibfnamefont{L.}~\bibnamefont{Arrachea}}, and
  \bibinfo{author}{\bibfnamefont{G.~S.} \bibnamefont{Lozano}},
  \bibinfo{year}{2010}, \bibinfo{journal}{Phys. Rev. B}
  \textbf{\bibinfo{volume}{81}}(\bibinfo{number}{4}), \bibinfo{pages}{041301}.

\bibitem[{\citenamefont{Chalopin} \emph{et~al.}(2008)\citenamefont{Chalopin,
  Gillet, and Volz}}]{Chalopin:233309}
\bibinfo{author}{\bibnamefont{Chalopin}, \bibfnamefont{Y.}},
  \bibinfo{author}{\bibfnamefont{J.-N.} \bibnamefont{Gillet}}, and
  \bibinfo{author}{\bibfnamefont{S.}~\bibnamefont{Volz}}, \bibinfo{year}{2008},
  \bibinfo{journal}{Phys. Rev. B}
  \textbf{\bibinfo{volume}{77}}(\bibinfo{number}{23}), \bibinfo{pages}{233309}.

\bibitem[{\citenamefont{Chang} \emph{et~al.}(2006)\citenamefont{Chang,
  Fennimore, Afanasiev, Okawa, Ikuno, Garcia, Li, Majumdar, and
  Zettl}}]{Chang:085901}
\bibinfo{author}{\bibnamefont{Chang}, \bibfnamefont{C.~W.}},
  \bibinfo{author}{\bibfnamefont{A.~M.} \bibnamefont{Fennimore}},
  \bibinfo{author}{\bibfnamefont{A.}~\bibnamefont{Afanasiev}},
  \bibinfo{author}{\bibfnamefont{D.}~\bibnamefont{Okawa}},
  \bibinfo{author}{\bibfnamefont{T.}~\bibnamefont{Ikuno}},
  \bibinfo{author}{\bibfnamefont{H.}~\bibnamefont{Garcia}},
  \bibinfo{author}{\bibfnamefont{D.}~\bibnamefont{Li}},
  \bibinfo{author}{\bibfnamefont{A.}~\bibnamefont{Majumdar}}, and
  \bibinfo{author}{\bibfnamefont{A.}~\bibnamefont{Zettl}},
  \bibinfo{year}{2006}, \bibinfo{journal}{Phys. Rev. Lett.}
  \textbf{\bibinfo{volume}{97}}(\bibinfo{number}{8}), \bibinfo{eid}{085901}.

\bibitem[{\citenamefont{Chang} \emph{et~al.}(2008)\citenamefont{Chang, Okawa,
  Garcia, Majumdar, and Zettl}}]{Chang:075903}
\bibinfo{author}{\bibnamefont{Chang}, \bibfnamefont{C.~W.}},
  \bibinfo{author}{\bibfnamefont{D.}~\bibnamefont{Okawa}},
  \bibinfo{author}{\bibfnamefont{H.}~\bibnamefont{Garcia}},
  \bibinfo{author}{\bibfnamefont{A.}~\bibnamefont{Majumdar}}, and
  \bibinfo{author}{\bibfnamefont{A.}~\bibnamefont{Zettl}},
  \bibinfo{year}{2008}, \bibinfo{journal}{Phys. Rev. Lett.}
  \textbf{\bibinfo{volume}{101}}(\bibinfo{number}{7}), \bibinfo{eid}{075903}.

\bibitem[{\citenamefont{Chen}(2005)}]{Chen:2005}
\bibinfo{author}{\bibnamefont{Chen}, \bibfnamefont{G.}}, \bibinfo{year}{2005},
  \emph{\bibinfo{title}{Nanoscale Energy Transport and Conversion}}
  (\bibinfo{publisher}{Oxford University Press, USA}).

\bibitem[{\citenamefont{Chen}
  \emph{et~al.}(2005{\natexlab{a}})\citenamefont{Chen, Li, Duan, Shuai, and
  Gu}}]{Chen:045422}
\bibinfo{author}{\bibnamefont{Chen}, \bibfnamefont{K.-Q.}},
  \bibinfo{author}{\bibfnamefont{W.-X.} \bibnamefont{Li}},
  \bibinfo{author}{\bibfnamefont{W.}~\bibnamefont{Duan}},
  \bibinfo{author}{\bibfnamefont{Z.}~\bibnamefont{Shuai}}, and
  \bibinfo{author}{\bibfnamefont{B.-L.} \bibnamefont{Gu}},
  \bibinfo{year}{2005}{\natexlab{a}}, \bibinfo{journal}{Phys. Rev. B}
  \textbf{\bibinfo{volume}{72}}(\bibinfo{number}{4}), \bibinfo{pages}{045422}.

\bibitem[{\citenamefont{Chen} \emph{et~al.}(2008)\citenamefont{Chen, Hochbaum,
  Murphy, Moore, Yang, and Majumdar}}]{chen:105501}
\bibinfo{author}{\bibnamefont{Chen}, \bibfnamefont{R.}},
  \bibinfo{author}{\bibfnamefont{A.~I.} \bibnamefont{Hochbaum}},
  \bibinfo{author}{\bibfnamefont{P.}~\bibnamefont{Murphy}},
  \bibinfo{author}{\bibfnamefont{J.}~\bibnamefont{Moore}},
  \bibinfo{author}{\bibfnamefont{P.}~\bibnamefont{Yang}}, and
  \bibinfo{author}{\bibfnamefont{A.}~\bibnamefont{Majumdar}},
  \bibinfo{year}{2008}, \bibinfo{journal}{Phys. Rev. Lett.}
  \textbf{\bibinfo{volume}{101}}(\bibinfo{number}{10}), \bibinfo{eid}{105501}.

\bibitem[{\citenamefont{Chen}(2008)}]{Chen:233310}
\bibinfo{author}{\bibnamefont{Chen}, \bibfnamefont{Y.-C.}},
  \bibinfo{year}{2008}, \bibinfo{journal}{Phys. Rev. B}
  \textbf{\bibinfo{volume}{78}}(\bibinfo{number}{23}), \bibinfo{eid}{233310}.

\bibitem[{\citenamefont{Chen} \emph{et~al.}(2003)\citenamefont{Chen, Zwolak,
  and {Di Ventra}}}]{Chen:1691}
\bibinfo{author}{\bibnamefont{Chen}, \bibfnamefont{Y.-C.}},
  \bibinfo{author}{\bibfnamefont{M.}~\bibnamefont{Zwolak}}, and
  \bibinfo{author}{\bibfnamefont{M.}~\bibnamefont{{Di Ventra}}},
  \bibinfo{year}{2003}, \bibinfo{journal}{Nano Lett.}
  \textbf{\bibinfo{volume}{3}}(\bibinfo{number}{12}), \bibinfo{pages}{1691}.

\bibitem[{\citenamefont{Chen}
  \emph{et~al.}(2005{\natexlab{b}})\citenamefont{Chen, Zwolak, and {Di
  Ventra}}}]{Chen:621}
\bibinfo{author}{\bibnamefont{Chen}, \bibfnamefont{Y.-C.}},
  \bibinfo{author}{\bibfnamefont{M.}~\bibnamefont{Zwolak}}, and
  \bibinfo{author}{\bibfnamefont{M.}~\bibnamefont{{Di Ventra}}},
  \bibinfo{year}{2005}{\natexlab{b}}, \bibinfo{journal}{Nano Lett.}
  \textbf{\bibinfo{volume}{5}}(\bibinfo{number}{4}), \bibinfo{pages}{621}.

\bibitem[{\citenamefont{Cheng} \emph{et~al.}(2006)\citenamefont{Cheng, Evans,
  and Van~Voorhis}}]{Cheng:155112}
\bibinfo{author}{\bibnamefont{Cheng}, \bibfnamefont{C.-L.}},
  \bibinfo{author}{\bibfnamefont{J.~S.} \bibnamefont{Evans}}, and
  \bibinfo{author}{\bibfnamefont{T.}~\bibnamefont{Van~Voorhis}},
  \bibinfo{year}{2006}, \bibinfo{journal}{Phys. Rev. B}
  \textbf{\bibinfo{volume}{74}}(\bibinfo{number}{15}), \bibinfo{pages}{155112}.

\bibitem[{\citenamefont{Chiatti} \emph{et~al.}(2006)\citenamefont{Chiatti,
  Nicholls, Proskuryakov, Lumpkin, Farrer, and Ritchie}}]{Chiatti:056601}
\bibinfo{author}{\bibnamefont{Chiatti}, \bibfnamefont{O.}},
  \bibinfo{author}{\bibfnamefont{J.~T.} \bibnamefont{Nicholls}},
  \bibinfo{author}{\bibfnamefont{Y.~Y.} \bibnamefont{Proskuryakov}},
  \bibinfo{author}{\bibfnamefont{N.}~\bibnamefont{Lumpkin}},
  \bibinfo{author}{\bibfnamefont{I.}~\bibnamefont{Farrer}}, and
  \bibinfo{author}{\bibfnamefont{D.~A.} \bibnamefont{Ritchie}},
  \bibinfo{year}{2006}, \bibinfo{journal}{Phys. Rev. Lett.}
  \textbf{\bibinfo{volume}{97}}(\bibinfo{number}{5}), \bibinfo{eid}{056601}.

\bibitem[{\citenamefont{Chiu} \emph{et~al.}(2005)\citenamefont{Chiu, Deshpande,
  Postma, Lau, Mik\'o, Forr\'o, and Bockrath}}]{Chiu:226101}
\bibinfo{author}{\bibnamefont{Chiu}, \bibfnamefont{H.-Y.}},
  \bibinfo{author}{\bibfnamefont{V.~V.} \bibnamefont{Deshpande}},
  \bibinfo{author}{\bibfnamefont{H.~W.~C.} \bibnamefont{Postma}},
  \bibinfo{author}{\bibfnamefont{C.~N.} \bibnamefont{Lau}},
  \bibinfo{author}{\bibfnamefont{C.}~\bibnamefont{Mik\'o}},
  \bibinfo{author}{\bibfnamefont{L.}~\bibnamefont{Forr\'o}}, and
  \bibinfo{author}{\bibfnamefont{M.}~\bibnamefont{Bockrath}},
  \bibinfo{year}{2005}, \bibinfo{journal}{Phys. Rev. Lett.}
  \textbf{\bibinfo{volume}{95}}(\bibinfo{number}{22}), \bibinfo{pages}{226101}.

\bibitem[{\citenamefont{Choi} \emph{et~al.}(2005)\citenamefont{Choi,
  Poulikakos, Tharian, and Sennhauser}}]{Choi:013108}
\bibinfo{author}{\bibnamefont{Choi}, \bibfnamefont{T.~Y.}},
  \bibinfo{author}{\bibfnamefont{D.}~\bibnamefont{Poulikakos}},
  \bibinfo{author}{\bibfnamefont{J.}~\bibnamefont{Tharian}}, and
  \bibinfo{author}{\bibfnamefont{U.}~\bibnamefont{Sennhauser}},
  \bibinfo{year}{2005}, \bibinfo{journal}{Appl. Phys. Lett.}
  \textbf{\bibinfo{volume}{87}}(\bibinfo{number}{1}), \bibinfo{eid}{013108}.

\bibitem[{\citenamefont{Choi} \emph{et~al.}(2006)\citenamefont{Choi,
  Poulikakos, Tharian, and Sennhauser}}]{Choi:1589}
\bibinfo{author}{\bibnamefont{Choi}, \bibfnamefont{T.-Y.}},
  \bibinfo{author}{\bibfnamefont{D.}~\bibnamefont{Poulikakos}},
  \bibinfo{author}{\bibfnamefont{J.}~\bibnamefont{Tharian}}, and
  \bibinfo{author}{\bibfnamefont{U.}~\bibnamefont{Sennhauser}},
  \bibinfo{year}{2006}, \bibinfo{journal}{Nano Lett.}
  \textbf{\bibinfo{volume}{6}}(\bibinfo{number}{8}), \bibinfo{pages}{1589}.

\bibitem[{\citenamefont{D'Agosta and {Di Ventra}}(2006)}]{D'Agosta:11059}
\bibinfo{author}{\bibnamefont{D'Agosta}, \bibfnamefont{R.}}, and
  \bibinfo{author}{\bibfnamefont{M.}~\bibnamefont{{Di Ventra}}},
  \bibinfo{year}{2006}, \bibinfo{journal}{Journal of Physics: Condensed Matter}
  \textbf{\bibinfo{volume}{18}}(\bibinfo{number}{49}), \bibinfo{pages}{11059}.

\bibitem[{\citenamefont{D'Agosta and {Di
  Ventra}}(2008{\natexlab{a}})}]{D'Agosta:374102}
\bibinfo{author}{\bibnamefont{D'Agosta}, \bibfnamefont{R.}}, and
  \bibinfo{author}{\bibfnamefont{M.}~\bibnamefont{{Di Ventra}}},
  \bibinfo{year}{2008}{\natexlab{a}}, \bibinfo{journal}{Journal of Physics:
  Condensed Matter} \textbf{\bibinfo{volume}{20}}(\bibinfo{number}{49}),
  \bibinfo{pages}{374102}.

\bibitem[{\citenamefont{D'Agosta and {Di
  Ventra}}(2008{\natexlab{b}})}]{D'Agosta:165105}
\bibinfo{author}{\bibnamefont{D'Agosta}, \bibfnamefont{R.}}, and
  \bibinfo{author}{\bibfnamefont{M.}~\bibnamefont{{Di Ventra}}},
  \bibinfo{year}{2008}{\natexlab{b}}, \bibinfo{journal}{Phys. Rev. B}
  \textbf{\bibinfo{volume}{78}}(\bibinfo{number}{16}), \bibinfo{eid}{165105}.

\bibitem[{\citenamefont{D'Agosta} \emph{et~al.}(2006)\citenamefont{D'Agosta,
  Sai, and {Di Ventra}}}]{D'Agosta:2935}
\bibinfo{author}{\bibnamefont{D'Agosta}, \bibfnamefont{R.}},
  \bibinfo{author}{\bibfnamefont{N.}~\bibnamefont{Sai}}, and
  \bibinfo{author}{\bibfnamefont{M.}~\bibnamefont{{Di Ventra}}},
  \bibinfo{year}{2006}, \bibinfo{journal}{Nano Lett.}
  \textbf{\bibinfo{volume}{6}}(\bibinfo{number}{12}), \bibinfo{pages}{2935}.

\bibitem[{\citenamefont{Datta}(1997)}]{Datta:1997}
\bibinfo{author}{\bibnamefont{Datta}, \bibfnamefont{S.}}, \bibinfo{year}{1997},
  \emph{\bibinfo{title}{Electronic transport in mesoscopic systems}}
  (\bibinfo{publisher}{Cambridge University Press}).

\bibitem[{\citenamefont{Dekker and Ratner}(2001)}]{Dekker:29}
\bibinfo{author}{\bibnamefont{Dekker}, \bibfnamefont{C.}}, and
  \bibinfo{author}{\bibfnamefont{M.~A.} \bibnamefont{Ratner}},
  \bibinfo{year}{2001}, \bibinfo{journal}{Physics World}
  \textbf{\bibinfo{volume}{14}}, \bibinfo{pages}{29}.

\bibitem[{\citenamefont{Deshpande} \emph{et~al.}(2009)\citenamefont{Deshpande,
  Hsieh, Bushmaker, Bockrath, and Cronin}}]{Deshpande:105501}
\bibinfo{author}{\bibnamefont{Deshpande}, \bibfnamefont{V.~V.}},
  \bibinfo{author}{\bibfnamefont{S.}~\bibnamefont{Hsieh}},
  \bibinfo{author}{\bibfnamefont{A.~W.} \bibnamefont{Bushmaker}},
  \bibinfo{author}{\bibfnamefont{M.}~\bibnamefont{Bockrath}}, and
  \bibinfo{author}{\bibfnamefont{S.~B.} \bibnamefont{Cronin}},
  \bibinfo{year}{2009}, \bibinfo{journal}{Phys. Rev. Lett.}
  \textbf{\bibinfo{volume}{102}}(\bibinfo{number}{10}), \bibinfo{eid}{105501}.

\bibitem[{\citenamefont{Dhar}(2008)}]{Dhar:457}
\bibinfo{author}{\bibnamefont{Dhar}, \bibfnamefont{A.}}, \bibinfo{year}{2008},
  \bibinfo{journal}{Advances in Physics}
  \textbf{\bibinfo{volume}{57}}(\bibinfo{number}{5}), \bibinfo{pages}{457}.

\bibitem[{\citenamefont{Dhar and Lebowitz}(2008)}]{Dhar:134301}
\bibinfo{author}{\bibnamefont{Dhar}, \bibfnamefont{A.}}, and
  \bibinfo{author}{\bibfnamefont{J.~L.} \bibnamefont{Lebowitz}},
  \bibinfo{year}{2008}, \bibinfo{journal}{Phys. Rev. Lett.}
  \textbf{\bibinfo{volume}{100}}(\bibinfo{number}{13}),
  \bibinfo{pages}{134301}.

\bibitem[{\citenamefont{Dhar and Roy}(2006)}]{Dhar:805}
\bibinfo{author}{\bibnamefont{Dhar}, \bibfnamefont{A.}}, and
  \bibinfo{author}{\bibfnamefont{D.}~\bibnamefont{Roy}}, \bibinfo{year}{2006},
  \bibinfo{journal}{Journal of Statistical Physics}
  \textbf{\bibinfo{volume}{125}}(\bibinfo{number}{4}), \bibinfo{pages}{805}.

\bibitem[{\citenamefont{{Di Ventra}}(2008)}]{DiVentra:2008}
\bibinfo{author}{\bibnamefont{{Di Ventra}}, \bibfnamefont{M.}},
  \bibinfo{year}{2008}, \emph{\bibinfo{title}{Electrical Transport in Nanoscale
  Systems}} (\bibinfo{publisher}{Cambridge University Press}).

\bibitem[{\citenamefont{{Di Ventra} and D'Agosta}(2007)}]{Ventra:226403}
\bibinfo{author}{\bibnamefont{{Di Ventra}}, \bibfnamefont{M.}}, and
  \bibinfo{author}{\bibfnamefont{R.}~\bibnamefont{D'Agosta}},
  \bibinfo{year}{2007}, \bibinfo{journal}{Phys. Rev. Lett.}
  \textbf{\bibinfo{volume}{98}}(\bibinfo{number}{22}), \bibinfo{eid}{226403}.

\bibitem[{\citenamefont{{Di Ventra} and Dubi}(2009)}]{DiVentra:40004}
\bibinfo{author}{\bibnamefont{{Di Ventra}}, \bibfnamefont{M.}}, and
  \bibinfo{author}{\bibfnamefont{Y.}~\bibnamefont{Dubi}}, \bibinfo{year}{2009},
  \bibinfo{journal}{Europhys. Lett.}
  \textbf{\bibinfo{volume}{85}}(\bibinfo{number}{4}), \bibinfo{pages}{40004}.

\bibitem[{\citenamefont{{Di Ventra}} \emph{et~al.}(2004)\citenamefont{{Di
  Ventra}, Evoy, and James R.~Heflin}}]{DiVentra:2004}
\bibinfo{author}{\bibnamefont{{Di Ventra}}, \bibfnamefont{M.}},
  \bibinfo{author}{\bibfnamefont{S.}~\bibnamefont{Evoy}}, and
  \bibinfo{author}{\bibfnamefont{J.~E.} \bibnamefont{James R.~Heflin}},
  \bibinfo{year}{2004}, \emph{\bibinfo{title}{Introduction to Nanoscale Science
  and Technology}} (\bibinfo{publisher}{Kluwer Academic Publishers}).

\bibitem[{\citenamefont{{Di Ventra} and Lang}(2002)}]{DiVentra2002}
\bibinfo{author}{\bibnamefont{{Di Ventra}}, \bibfnamefont{M.}}, and
  \bibinfo{author}{\bibfnamefont{N.}~\bibnamefont{Lang}}, \bibinfo{year}{2002},
  \bibinfo{journal}{Phys. Rev. B}
  \textbf{\bibinfo{volume}{65}}(\bibinfo{number}{4}), \bibinfo{eid}{045402}.

\bibitem[{\citenamefont{{Di Ventra} and Todorov}(2004)}]{Ventra:8025}
\bibinfo{author}{\bibnamefont{{Di Ventra}}, \bibfnamefont{M.}}, and
  \bibinfo{author}{\bibfnamefont{T.~N.} \bibnamefont{Todorov}},
  \bibinfo{year}{2004}, \bibinfo{journal}{Journal of Physics: Condensed Matter}
  \textbf{\bibinfo{volume}{16}}(\bibinfo{number}{45}), \bibinfo{pages}{8025}.

\bibitem[{\citenamefont{Duarte} \emph{et~al.}(2009)\citenamefont{Duarte, Mahan,
  and Tadigadapa}}]{Duarte:617}
\bibinfo{author}{\bibnamefont{Duarte}, \bibfnamefont{N.~B.}},
  \bibinfo{author}{\bibfnamefont{G.~D.} \bibnamefont{Mahan}}, and
  \bibinfo{author}{\bibfnamefont{S.}~\bibnamefont{Tadigadapa}},
  \bibinfo{year}{2009}, \bibinfo{journal}{Nano Lett.}
  \textbf{\bibinfo{volume}{9}}(\bibinfo{number}{2}), \bibinfo{pages}{617}.

\bibitem[{\citenamefont{Dubi and {Di Ventra}}(2009{\natexlab{a}})}]{Dubi:new}
\bibinfo{author}{\bibnamefont{Dubi}, \bibfnamefont{Y.}}, and
  \bibinfo{author}{\bibfnamefont{M.}~\bibnamefont{{Di Ventra}}},
  \bibinfo{year}{2009}{\natexlab{a}}, \bibinfo{journal}{unpublished} .

\bibitem[{\citenamefont{Dubi and {Di
  Ventra}}(2009{\natexlab{b}})}]{Dubi:042101}
\bibinfo{author}{\bibnamefont{Dubi}, \bibfnamefont{Y.}}, and
  \bibinfo{author}{\bibfnamefont{M.}~\bibnamefont{{Di Ventra}}},
  \bibinfo{year}{2009}{\natexlab{b}}, \bibinfo{journal}{Phys. Rev. E}
  \textbf{\bibinfo{volume}{79}}(\bibinfo{number}{4}), \bibinfo{eid}{042101}.

\bibitem[{\citenamefont{Dubi and {Di
  Ventra}}(2009{\natexlab{c}})}]{Dubi:115415}
\bibinfo{author}{\bibnamefont{Dubi}, \bibfnamefont{Y.}}, and
  \bibinfo{author}{\bibfnamefont{M.}~\bibnamefont{{Di Ventra}}},
  \bibinfo{year}{2009}{\natexlab{c}}, \bibinfo{journal}{Phys. Rev. B}
  \textbf{\bibinfo{volume}{79}}(\bibinfo{number}{11}), \bibinfo{eid}{115415}.

\bibitem[{\citenamefont{Dubi and {Di Ventra}}(2009{\natexlab{d}})}]{Dubi:97}
\bibinfo{author}{\bibnamefont{Dubi}, \bibfnamefont{Y.}}, and
  \bibinfo{author}{\bibfnamefont{M.}~\bibnamefont{{Di Ventra}}},
  \bibinfo{year}{2009}{\natexlab{d}}, \bibinfo{journal}{Nano Lett.}
  \textbf{\bibinfo{volume}{9}}(\bibinfo{number}{1}), \bibinfo{pages}{97}.

\bibitem[{\citenamefont{Dubi and {Di
  Ventra}}(2009{\natexlab{e}})}]{Dubi:081302}
\bibinfo{author}{\bibnamefont{Dubi}, \bibfnamefont{Y.}}, and
  \bibinfo{author}{\bibfnamefont{M.}~\bibnamefont{{Di Ventra}}},
  \bibinfo{year}{2009}{\natexlab{e}}, \bibinfo{journal}{Phys. Rev. B}
  \textbf{\bibinfo{volume}{79}}(\bibinfo{number}{8}), \bibinfo{eid}{081302}.

\bibitem[{\citenamefont{Esposito} \emph{et~al.}(2009)\citenamefont{Esposito,
  Lindenberg, and den Broeck}}]{Esposito:60010}
\bibinfo{author}{\bibnamefont{Esposito}, \bibfnamefont{M.}},
  \bibinfo{author}{\bibfnamefont{K.}~\bibnamefont{Lindenberg}}, and
  \bibinfo{author}{\bibfnamefont{C.~V.} \bibnamefont{den Broeck}},
  \bibinfo{year}{2009}, \bibinfo{journal}{Europhys. Lett.}
  \textbf{\bibinfo{volume}{85}}(\bibinfo{number}{6}), \bibinfo{pages}{60010}.

\bibitem[{\citenamefont{Feldman} \emph{et~al.}(2000)\citenamefont{Feldman,
  Singh, Mazin, Mandrus, and Sales}}]{Feldman:9209}
\bibinfo{author}{\bibnamefont{Feldman}, \bibfnamefont{J.~L.}},
  \bibinfo{author}{\bibfnamefont{D.~J.} \bibnamefont{Singh}},
  \bibinfo{author}{\bibfnamefont{I.~I.} \bibnamefont{Mazin}},
  \bibinfo{author}{\bibfnamefont{D.}~\bibnamefont{Mandrus}}, and
  \bibinfo{author}{\bibfnamefont{B.~C.} \bibnamefont{Sales}},
  \bibinfo{year}{2000}, \bibinfo{journal}{Phys. Rev. B}
  \textbf{\bibinfo{volume}{61}}(\bibinfo{number}{14}), \bibinfo{pages}{R9209}.

\bibitem[{\citenamefont{Fermi} \emph{et~al.}(1955)\citenamefont{Fermi, Pasta,
  and Ulam}}]{Fermi:1940}
\bibinfo{author}{\bibnamefont{Fermi}, \bibfnamefont{E.}},
  \bibinfo{author}{\bibfnamefont{J.}~\bibnamefont{Pasta}}, and
  \bibinfo{author}{\bibfnamefont{S.}~\bibnamefont{Ulam}}, \bibinfo{year}{1955},
  \bibinfo{journal}{Los Alamos report LA-1940} .

\bibitem[{\citenamefont{Finch} \emph{et~al.}(2009)\citenamefont{Finch,
  Garc\'{\i}a-Su\'{a}rez, and Lambert}}]{finch:033405}
\bibinfo{author}{\bibnamefont{Finch}, \bibfnamefont{C.~M.}},
  \bibinfo{author}{\bibfnamefont{V.~M.} \bibnamefont{Garc\'{\i}a-Su\'{a}rez}},
  and \bibinfo{author}{\bibfnamefont{C.~J.} \bibnamefont{Lambert}},
  \bibinfo{year}{2009}, \bibinfo{journal}{Phys. Rev. B}
  \textbf{\bibinfo{volume}{79}}(\bibinfo{number}{3}), \bibinfo{eid}{033405}.

\bibitem[{\citenamefont{Fourier}(1822)}]{Fourier:1882}
\bibinfo{author}{\bibnamefont{Fourier}, \bibfnamefont{J.}},
  \bibinfo{year}{1822}, \emph{\bibinfo{title}{Theorie analytique de la
  chaleur}} (\bibinfo{publisher}{Didot, Paris}).

\bibitem[{\citenamefont{Franceschi and Mingo}(2007)}]{Franceschi:538}
\bibinfo{author}{\bibnamefont{Franceschi}, \bibfnamefont{S.~D.}}, and
  \bibinfo{author}{\bibfnamefont{N.}~\bibnamefont{Mingo}},
  \bibinfo{year}{2007}, \bibinfo{journal}{Nature Nanotechnology}
  \textbf{\bibinfo{volume}{2}}, \bibinfo{pages}{537}.

\bibitem[{\citenamefont{Fujii} \emph{et~al.}(2005)\citenamefont{Fujii, Zhang,
  Xie, Ago, Takahashi, Ikuta, Abe, and Shimizu}}]{Fujii:065502}
\bibinfo{author}{\bibnamefont{Fujii}, \bibfnamefont{M.}},
  \bibinfo{author}{\bibfnamefont{X.}~\bibnamefont{Zhang}},
  \bibinfo{author}{\bibfnamefont{H.}~\bibnamefont{Xie}},
  \bibinfo{author}{\bibfnamefont{H.}~\bibnamefont{Ago}},
  \bibinfo{author}{\bibfnamefont{K.}~\bibnamefont{Takahashi}},
  \bibinfo{author}{\bibfnamefont{T.}~\bibnamefont{Ikuta}},
  \bibinfo{author}{\bibfnamefont{H.}~\bibnamefont{Abe}}, and
  \bibinfo{author}{\bibfnamefont{T.}~\bibnamefont{Shimizu}},
  \bibinfo{year}{2005}, \bibinfo{journal}{Phys. Rev. Lett.}
  \textbf{\bibinfo{volume}{95}}(\bibinfo{number}{6}), \bibinfo{pages}{065502}.

\bibitem[{\citenamefont{Galperin}
  \emph{et~al.}(2007{\natexlab{a}})\citenamefont{Galperin, Nitzan, and
  Ratner}}]{Galperin:155312}
\bibinfo{author}{\bibnamefont{Galperin}, \bibfnamefont{M.}},
  \bibinfo{author}{\bibfnamefont{A.}~\bibnamefont{Nitzan}}, and
  \bibinfo{author}{\bibfnamefont{M.~A.} \bibnamefont{Ratner}},
  \bibinfo{year}{2007}{\natexlab{a}}, \bibinfo{journal}{Phys. Rev. B}
  \textbf{\bibinfo{volume}{75}}(\bibinfo{number}{15}), \bibinfo{eid}{155312}.

\bibitem[{\citenamefont{Galperin}
  \emph{et~al.}(2007{\natexlab{b}})\citenamefont{Galperin, Ratner, and
  Nitzan}}]{Galperin:103201}
\bibinfo{author}{\bibnamefont{Galperin}, \bibfnamefont{M.}},
  \bibinfo{author}{\bibfnamefont{M.~A.} \bibnamefont{Ratner}}, and
  \bibinfo{author}{\bibfnamefont{A.}~\bibnamefont{Nitzan}},
  \bibinfo{year}{2007}{\natexlab{b}}, \bibinfo{journal}{Journal of Physics:
  Condensed Matter} (\bibinfo{number}{10}), \bibinfo{pages}{103201}.

\bibitem[{\citenamefont{Galperin}
  \emph{et~al.}(2009{\natexlab{a}})\citenamefont{Galperin, Ratner, and
  Nitzan}}]{Galperin:758}
\bibinfo{author}{\bibnamefont{Galperin}, \bibfnamefont{M.}},
  \bibinfo{author}{\bibfnamefont{M.~A.} \bibnamefont{Ratner}}, and
  \bibinfo{author}{\bibfnamefont{A.}~\bibnamefont{Nitzan}},
  \bibinfo{year}{2009}{\natexlab{a}}, \bibinfo{journal}{Nano Lett.}
  \textbf{\bibinfo{volume}{9}}(\bibinfo{number}{2}), \bibinfo{pages}{758}.

\bibitem[{\citenamefont{Galperin}
  \emph{et~al.}(2009{\natexlab{b}})\citenamefont{Galperin, Saito, Balatsky, and
  Nitzan}}]{Galperin:0905.2748}
\bibinfo{author}{\bibnamefont{Galperin}, \bibfnamefont{M.}},
  \bibinfo{author}{\bibfnamefont{K.}~\bibnamefont{Saito}},
  \bibinfo{author}{\bibfnamefont{A.~V.} \bibnamefont{Balatsky}}, and
  \bibinfo{author}{\bibfnamefont{A.}~\bibnamefont{Nitzan}},
  \bibinfo{year}{2009}{\natexlab{b}}, \bibinfo{journal}{Phys. Rev. B}
  \textbf{\bibinfo{volume}{80}}(\bibinfo{number}{11}), \bibinfo{pages}{115427}.

\bibitem[{\citenamefont{Garg} \emph{et~al.}(2009)\citenamefont{Garg, Rasch,
  Shimshoni, and Rosch}}]{garg:096402}
\bibinfo{author}{\bibnamefont{Garg}, \bibfnamefont{A.}},
  \bibinfo{author}{\bibfnamefont{D.}~\bibnamefont{Rasch}},
  \bibinfo{author}{\bibfnamefont{E.}~\bibnamefont{Shimshoni}}, and
  \bibinfo{author}{\bibfnamefont{A.}~\bibnamefont{Rosch}},
  \bibinfo{year}{2009}, \bibinfo{journal}{Phys. Rev. Lett.}
  \textbf{\bibinfo{volume}{103}}(\bibinfo{number}{9}), \bibinfo{pages}{096402}.

\bibitem[{\citenamefont{Gaul and B\"{u}ttner}(2007)}]{Gaul:011111}
\bibinfo{author}{\bibnamefont{Gaul}, \bibfnamefont{C.}}, and
  \bibinfo{author}{\bibfnamefont{H.}~\bibnamefont{B\"{u}ttner}},
  \bibinfo{year}{2007}, \bibinfo{journal}{Phys. Rev. E}
  \textbf{\bibinfo{volume}{76}}(\bibinfo{number}{1}), \bibinfo{pages}{011111}.

\bibitem[{\citenamefont{Giazotto} \emph{et~al.}(2006)\citenamefont{Giazotto,
  Heikkil\"{a}, Luukanen, Savin, and Pekola}}]{giazotto:217}
\bibinfo{author}{\bibnamefont{Giazotto}, \bibfnamefont{F.}},
  \bibinfo{author}{\bibfnamefont{T.~T.} \bibnamefont{Heikkil\"{a}}},
  \bibinfo{author}{\bibfnamefont{A.}~\bibnamefont{Luukanen}},
  \bibinfo{author}{\bibfnamefont{A.~M.} \bibnamefont{Savin}}, and
  \bibinfo{author}{\bibfnamefont{J.~P.} \bibnamefont{Pekola}},
  \bibinfo{year}{2006}, \bibinfo{journal}{Reviews of Modern Physics}
  \textbf{\bibinfo{volume}{78}}(\bibinfo{number}{1}), \bibinfo{eid}{217}.

\bibitem[{\citenamefont{Godijn} \emph{et~al.}(1999)\citenamefont{Godijn,
  M\"oller, Buhmann, Molenkamp, and van Langen}}]{Godijn:2927}
\bibinfo{author}{\bibnamefont{Godijn}, \bibfnamefont{S.~F.}},
  \bibinfo{author}{\bibfnamefont{S.}~\bibnamefont{M\"oller}},
  \bibinfo{author}{\bibfnamefont{H.}~\bibnamefont{Buhmann}},
  \bibinfo{author}{\bibfnamefont{L.~W.} \bibnamefont{Molenkamp}}, and
  \bibinfo{author}{\bibfnamefont{S.~A.} \bibnamefont{van Langen}},
  \bibinfo{year}{1999}, \bibinfo{journal}{Phys. Rev. Lett.}
  \textbf{\bibinfo{volume}{82}}(\bibinfo{number}{14}), \bibinfo{pages}{2927}.

\bibitem[{\citenamefont{Greiner} \emph{et~al.}(1997)\citenamefont{Greiner,
  Reggiani, Kuhn, and Varani}}]{Greiner:1114}
\bibinfo{author}{\bibnamefont{Greiner}, \bibfnamefont{A.}},
  \bibinfo{author}{\bibfnamefont{L.}~\bibnamefont{Reggiani}},
  \bibinfo{author}{\bibfnamefont{T.}~\bibnamefont{Kuhn}}, and
  \bibinfo{author}{\bibfnamefont{L.}~\bibnamefont{Varani}},
  \bibinfo{year}{1997}, \bibinfo{journal}{Phys. Rev. Lett.}
  \textbf{\bibinfo{volume}{78}}(\bibinfo{number}{6}), \bibinfo{pages}{1114}.

\bibitem[{\citenamefont{Grover} \emph{et~al.}(2006)\citenamefont{Grover,
  McCarthy, Sarid, and Guven}}]{Grover:233501}
\bibinfo{author}{\bibnamefont{Grover}, \bibfnamefont{R.}},
  \bibinfo{author}{\bibfnamefont{B.}~\bibnamefont{McCarthy}},
  \bibinfo{author}{\bibfnamefont{D.}~\bibnamefont{Sarid}}, and
  \bibinfo{author}{\bibfnamefont{I.}~\bibnamefont{Guven}},
  \bibinfo{year}{2006}, \bibinfo{journal}{Appl. Phys. Lett.}
  \textbf{\bibinfo{volume}{88}}(\bibinfo{number}{23}), \bibinfo{eid}{233501}.

\bibitem[{\citenamefont{Hartmann and Mahler}(2005)}]{Hartmann:579}
\bibinfo{author}{\bibnamefont{Hartmann}, \bibfnamefont{M.}}, and
  \bibinfo{author}{\bibfnamefont{G.}~\bibnamefont{Mahler}},
  \bibinfo{year}{2005}, \bibinfo{journal}{Europhys. Lett.}
  \textbf{\bibinfo{volume}{70}}(\bibinfo{number}{5}), \bibinfo{pages}{579}.

\bibitem[{\citenamefont{Hartmann}
  \emph{et~al.}(2004{\natexlab{a}})\citenamefont{Hartmann, Mahler, and
  Hess}}]{Hartmann:080402}
\bibinfo{author}{\bibnamefont{Hartmann}, \bibfnamefont{M.}},
  \bibinfo{author}{\bibfnamefont{G.}~\bibnamefont{Mahler}}, and
  \bibinfo{author}{\bibfnamefont{O.}~\bibnamefont{Hess}},
  \bibinfo{year}{2004}{\natexlab{a}}, \bibinfo{journal}{Phys. Rev. Lett.}
  \textbf{\bibinfo{volume}{93}}(\bibinfo{number}{8}), \bibinfo{pages}{080402}.

\bibitem[{\citenamefont{Hartmann}
  \emph{et~al.}(2004{\natexlab{b}})\citenamefont{Hartmann, Mahler, and
  Hess}}]{Hartmann:066148}
\bibinfo{author}{\bibnamefont{Hartmann}, \bibfnamefont{M.}},
  \bibinfo{author}{\bibfnamefont{G.}~\bibnamefont{Mahler}}, and
  \bibinfo{author}{\bibfnamefont{O.}~\bibnamefont{Hess}},
  \bibinfo{year}{2004}{\natexlab{b}}, \bibinfo{journal}{Phys. Rev. E}
  \textbf{\bibinfo{volume}{70}}(\bibinfo{number}{6}), \bibinfo{pages}{066148}.

\bibitem[{\citenamefont{Hatami} \emph{et~al.}(2007)\citenamefont{Hatami, Bauer,
  Zhang, and Kelly}}]{Hatami:066603}
\bibinfo{author}{\bibnamefont{Hatami}, \bibfnamefont{M.}},
  \bibinfo{author}{\bibfnamefont{G.~E.~W.} \bibnamefont{Bauer}},
  \bibinfo{author}{\bibfnamefont{Q.}~\bibnamefont{Zhang}}, and
  \bibinfo{author}{\bibfnamefont{P.~J.} \bibnamefont{Kelly}},
  \bibinfo{year}{2007}, \bibinfo{journal}{Phys. Rev. Lett.}
  \textbf{\bibinfo{volume}{99}}(\bibinfo{number}{6}), \bibinfo{pages}{066603}.

\bibitem[{\citenamefont{Hatami} \emph{et~al.}(2009)\citenamefont{Hatami, Bauer,
  Zhang, and Kelly}}]{Hatami:174426}
\bibinfo{author}{\bibnamefont{Hatami}, \bibfnamefont{M.}},
  \bibinfo{author}{\bibfnamefont{G.~E.~W.} \bibnamefont{Bauer}},
  \bibinfo{author}{\bibfnamefont{Q.}~\bibnamefont{Zhang}}, and
  \bibinfo{author}{\bibfnamefont{P.~J.} \bibnamefont{Kelly}},
  \bibinfo{year}{2009}, \bibinfo{journal}{Phys. Rev. B}
  \textbf{\bibinfo{volume}{79}}(\bibinfo{number}{17}), \bibinfo{eid}{174426}.

\bibitem[{\citenamefont{Heikkil\"a}
  \emph{et~al.}(2010)\citenamefont{Heikkil\"a, Hatami, and
  Bauer}}]{Heikkila:100408}
\bibinfo{author}{\bibnamefont{Heikkil\"a}, \bibfnamefont{T.~T.}},
  \bibinfo{author}{\bibfnamefont{M.}~\bibnamefont{Hatami}}, and
  \bibinfo{author}{\bibfnamefont{G.~E.~W.} \bibnamefont{Bauer}},
  \bibinfo{year}{2010}, \bibinfo{journal}{Phys. Rev. B}
  \textbf{\bibinfo{volume}{81}}(\bibinfo{number}{10}), \bibinfo{pages}{100408}.

\bibitem[{\citenamefont{Henry and Chen}(2008{\natexlab{a}})}]{Henry:235502}
\bibinfo{author}{\bibnamefont{Henry}, \bibfnamefont{A.}}, and
  \bibinfo{author}{\bibfnamefont{G.}~\bibnamefont{Chen}},
  \bibinfo{year}{2008}{\natexlab{a}}, \bibinfo{journal}{Phys. Rev. Lett.}
  \textbf{\bibinfo{volume}{101}}(\bibinfo{number}{23}), \bibinfo{eid}{235502}.

\bibitem[{\citenamefont{Henry and Chen}(2008{\natexlab{b}})}]{Henry:141}
\bibinfo{author}{\bibnamefont{Henry}, \bibfnamefont{A.~S.}}, and
  \bibinfo{author}{\bibfnamefont{G.}~\bibnamefont{Chen}},
  \bibinfo{year}{2008}{\natexlab{b}}, \bibinfo{journal}{Journal of
  Computational and Theoretical Nanoscience}
  \textbf{\bibinfo{volume}{5}}(\bibinfo{number}{2}), \bibinfo{pages}{141}.

\bibitem[{\citenamefont{Hochbaum} \emph{et~al.}(2007)\citenamefont{Hochbaum,
  Chen, Delgado, Liang, Garnett, Najarian, Majumdar, and Yang}}]{Hochbaum:163}
\bibinfo{author}{\bibnamefont{Hochbaum}, \bibfnamefont{A.~I.}},
  \bibinfo{author}{\bibfnamefont{R.}~\bibnamefont{Chen}},
  \bibinfo{author}{\bibfnamefont{R.~D.} \bibnamefont{Delgado}},
  \bibinfo{author}{\bibfnamefont{W.}~\bibnamefont{Liang}},
  \bibinfo{author}{\bibfnamefont{E.~C.} \bibnamefont{Garnett}},
  \bibinfo{author}{\bibfnamefont{M.}~\bibnamefont{Najarian}},
  \bibinfo{author}{\bibfnamefont{A.}~\bibnamefont{Majumdar}}, and
  \bibinfo{author}{\bibfnamefont{P.}~\bibnamefont{Yang}}, \bibinfo{year}{2007},
  \bibinfo{journal}{Nature} \textbf{\bibinfo{volume}{451}},
  \bibinfo{pages}{163}.

\bibitem[{\citenamefont{Hoffmann} \emph{et~al.}(2009)\citenamefont{Hoffmann,
  Nilsson, Matthews, Nakpathomkun, Persson, Samuelson, and
  Linke}}]{Hoffmann:779}
\bibinfo{author}{\bibnamefont{Hoffmann}, \bibfnamefont{E.~A.}},
  \bibinfo{author}{\bibfnamefont{H.~A.} \bibnamefont{Nilsson}},
  \bibinfo{author}{\bibfnamefont{J.~E.} \bibnamefont{Matthews}},
  \bibinfo{author}{\bibfnamefont{N.}~\bibnamefont{Nakpathomkun}},
  \bibinfo{author}{\bibfnamefont{A.~I.} \bibnamefont{Persson}},
  \bibinfo{author}{\bibfnamefont{L.}~\bibnamefont{Samuelson}}, and
  \bibinfo{author}{\bibfnamefont{H.}~\bibnamefont{Linke}},
  \bibinfo{year}{2009}, \bibinfo{journal}{Nano Lett.}
  \textbf{\bibinfo{volume}{9}}(\bibinfo{number}{2}), \bibinfo{pages}{779}.

\bibitem[{\citenamefont{Hoover}(1985)}]{Hoover:1695}
\bibinfo{author}{\bibnamefont{Hoover}, \bibfnamefont{W.~G.}},
  \bibinfo{year}{1985}, \bibinfo{journal}{Phys. Rev. A}
  \textbf{\bibinfo{volume}{31}}(\bibinfo{number}{3}), \bibinfo{pages}{1695}.

\bibitem[{\citenamefont{van Houten}
  \emph{et~al.}(1992{\natexlab{a}})\citenamefont{van Houten, Molenkamp,
  Beenakker, and Foxon}}]{vanHouten:215}
\bibinfo{author}{\bibnamefont{van Houten}, \bibfnamefont{H.}},
  \bibinfo{author}{\bibfnamefont{L.~W.} \bibnamefont{Molenkamp}},
  \bibinfo{author}{\bibfnamefont{C.~W.~J.} \bibnamefont{Beenakker}}, and
  \bibinfo{author}{\bibfnamefont{C.~T.} \bibnamefont{Foxon}},
  \bibinfo{year}{1992}{\natexlab{a}}, \bibinfo{journal}{Semiconductor Science
  and Technology} \textbf{\bibinfo{volume}{7}}(\bibinfo{number}{3B}),
  \bibinfo{pages}{B215}.

\bibitem[{\citenamefont{van Houten}
  \emph{et~al.}(1992{\natexlab{b}})\citenamefont{van Houten, Molenkamp,
  Beenakker, and Foxon}}]{Houten:B215}
\bibinfo{author}{\bibnamefont{van Houten}, \bibfnamefont{H.}},
  \bibinfo{author}{\bibfnamefont{L.~W.} \bibnamefont{Molenkamp}},
  \bibinfo{author}{\bibfnamefont{C.~W.~J.} \bibnamefont{Beenakker}}, and
  \bibinfo{author}{\bibfnamefont{C.~T.} \bibnamefont{Foxon}},
  \bibinfo{year}{1992}{\natexlab{b}}, \bibinfo{journal}{Semiconductor Science
  and Technology} \textbf{\bibinfo{volume}{7}}(\bibinfo{number}{3B}),
  \bibinfo{pages}{B215}.

\bibitem[{\citenamefont{Hsu} \emph{et~al.}(2009)\citenamefont{Hsu, Pettes,
  Bushmaker, Aykol, Shi, and Cronin}}]{Hsu:590}
\bibinfo{author}{\bibnamefont{Hsu}, \bibfnamefont{I.-K.}},
  \bibinfo{author}{\bibfnamefont{M.~T.} \bibnamefont{Pettes}},
  \bibinfo{author}{\bibfnamefont{A.}~\bibnamefont{Bushmaker}},
  \bibinfo{author}{\bibfnamefont{M.}~\bibnamefont{Aykol}},
  \bibinfo{author}{\bibfnamefont{L.}~\bibnamefont{Shi}}, and
  \bibinfo{author}{\bibfnamefont{S.~B.} \bibnamefont{Cronin}},
  \bibinfo{year}{2009}, \bibinfo{journal}{Nano Lett.}
  \textbf{\bibinfo{volume}{9}}(\bibinfo{number}{2}), \bibinfo{pages}{590}.

\bibitem[{\citenamefont{Hu} \emph{et~al.}(2008)\citenamefont{Hu, Keblinski,
  Wang, and Raravikar}}]{Hu:083503}
\bibinfo{author}{\bibnamefont{Hu}, \bibfnamefont{M.}},
  \bibinfo{author}{\bibfnamefont{P.}~\bibnamefont{Keblinski}},
  \bibinfo{author}{\bibfnamefont{J.-S.} \bibnamefont{Wang}}, and
  \bibinfo{author}{\bibfnamefont{N.}~\bibnamefont{Raravikar}},
  \bibinfo{year}{2008}, \bibinfo{journal}{Journal of Applied Physics}
  \textbf{\bibinfo{volume}{104}}(\bibinfo{number}{8}), \bibinfo{eid}{083503}.

\bibitem[{\citenamefont{Huang} \emph{et~al.}(2007)\citenamefont{Huang, Chen,
  D'Agosta, Bennett, {Di Ventra}, and Tao}}]{Huang:698}
\bibinfo{author}{\bibnamefont{Huang}, \bibfnamefont{Z.}},
  \bibinfo{author}{\bibfnamefont{F.}~\bibnamefont{Chen}},
  \bibinfo{author}{\bibfnamefont{R.}~\bibnamefont{D'Agosta}},
  \bibinfo{author}{\bibfnamefont{P.~A.} \bibnamefont{Bennett}},
  \bibinfo{author}{\bibfnamefont{M.}~\bibnamefont{{Di Ventra}}}, and
  \bibinfo{author}{\bibfnamefont{N.}~\bibnamefont{Tao}}, \bibinfo{year}{2007},
  \bibinfo{journal}{Nature Nanotechnology}
  \textbf{\bibinfo{volume}{2}}(\bibinfo{number}{11}), \bibinfo{pages}{698}.

\bibitem[{\citenamefont{Huang} \emph{et~al.}(2006)\citenamefont{Huang, Xu,
  Chen, {Di Ventra}, and Tao}}]{Huang:1240}
\bibinfo{author}{\bibnamefont{Huang}, \bibfnamefont{Z.}},
  \bibinfo{author}{\bibfnamefont{B.}~\bibnamefont{Xu}},
  \bibinfo{author}{\bibfnamefont{Y.}~\bibnamefont{Chen}},
  \bibinfo{author}{\bibfnamefont{M.}~\bibnamefont{{Di Ventra}}}, and
  \bibinfo{author}{\bibfnamefont{N.}~\bibnamefont{Tao}}, \bibinfo{year}{2006},
  \bibinfo{journal}{Nano Lett.}
  \textbf{\bibinfo{volume}{6}}(\bibinfo{number}{6}), \bibinfo{pages}{1240}.

\bibitem[{\citenamefont{Imry}(1997)}]{Imry:1997}
\bibinfo{author}{\bibnamefont{Imry}, \bibfnamefont{Y.}}, \bibinfo{year}{1997},
  \emph{\bibinfo{title}{Introduction to Mesoscopic Physics}}
  (\bibinfo{publisher}{Oxford University Press}).

\bibitem[{\citenamefont{Ioffe} \emph{et~al.}(2008)\citenamefont{Ioffe, Shamai,
  Ophir, Noy, Yutsis, Kfir, Cheshnovsky, and Selzer}}]{Ioffe:727}
\bibinfo{author}{\bibnamefont{Ioffe}, \bibfnamefont{Z.}},
  \bibinfo{author}{\bibfnamefont{T.}~\bibnamefont{Shamai}},
  \bibinfo{author}{\bibfnamefont{A.}~\bibnamefont{Ophir}},
  \bibinfo{author}{\bibfnamefont{G.}~\bibnamefont{Noy}},
  \bibinfo{author}{\bibfnamefont{I.}~\bibnamefont{Yutsis}},
  \bibinfo{author}{\bibfnamefont{K.}~\bibnamefont{Kfir}},
  \bibinfo{author}{\bibfnamefont{O.}~\bibnamefont{Cheshnovsky}}, and
  \bibinfo{author}{\bibfnamefont{Y.}~\bibnamefont{Selzer}},
  \bibinfo{year}{2008}, \bibinfo{journal}{Nature Nanotechnology}
  \textbf{\bibinfo{volume}{3}}(\bibinfo{number}{12}), \bibinfo{pages}{727}.

\bibitem[{\citenamefont{Jacquet}(2009)}]{Jacquet:709}
\bibinfo{author}{\bibnamefont{Jacquet}, \bibfnamefont{P.}},
  \bibinfo{year}{2009}, \bibinfo{journal}{J. Stat. Phys.}
  \textbf{\bibinfo{volume}{134}}(\bibinfo{number}{4}), \bibinfo{pages}{709}.

\bibitem[{\citenamefont{Kadanoff and Baym}(1962)}]{KadanoffBaym}
\bibinfo{author}{\bibnamefont{Kadanoff}, \bibfnamefont{L.~P.}}, and
  \bibinfo{author}{\bibfnamefont{G.}~\bibnamefont{Baym}}, \bibinfo{year}{1962},
  \emph{\bibinfo{title}{Quantum Statistical Mechanics}}
  (\bibinfo{publisher}{Benjamin, New York}).

\bibitem[{\citenamefont{Kambili} \emph{et~al.}(1999)\citenamefont{Kambili,
  Fagas, Fal'ko, and Lambert}}]{Kambili:15593}
\bibinfo{author}{\bibnamefont{Kambili}, \bibfnamefont{A.}},
  \bibinfo{author}{\bibfnamefont{G.}~\bibnamefont{Fagas}},
  \bibinfo{author}{\bibfnamefont{V.~I.} \bibnamefont{Fal'ko}}, and
  \bibinfo{author}{\bibfnamefont{C.~J.} \bibnamefont{Lambert}},
  \bibinfo{year}{1999}, \bibinfo{journal}{Phys. Rev. B}
  \textbf{\bibinfo{volume}{60}}(\bibinfo{number}{23}), \bibinfo{pages}{15593}.

\bibitem[{\citenamefont{Kane and Fisher}(1996)}]{Kane:3192}
\bibinfo{author}{\bibnamefont{Kane}, \bibfnamefont{C.~L.}}, and
  \bibinfo{author}{\bibfnamefont{M.~P.~A.} \bibnamefont{Fisher}},
  \bibinfo{year}{1996}, \bibinfo{journal}{Phys. Rev. Lett.}
  \textbf{\bibinfo{volume}{76}}(\bibinfo{number}{17}), \bibinfo{pages}{3192}.

\bibitem[{\citenamefont{Ke} \emph{et~al.}(2009)\citenamefont{Ke, Yang,
  Curtarolo, and Baranger}}]{Ke:1011}
\bibinfo{author}{\bibnamefont{Ke}, \bibfnamefont{S.-H.}},
  \bibinfo{author}{\bibfnamefont{W.}~\bibnamefont{Yang}},
  \bibinfo{author}{\bibfnamefont{S.}~\bibnamefont{Curtarolo}}, and
  \bibinfo{author}{\bibfnamefont{H.~U.} \bibnamefont{Baranger}},
  \bibinfo{year}{2009}, \bibinfo{journal}{Nano Lett.}
  \textbf{\bibinfo{volume}{9}}(\bibinfo{number}{3}), \bibinfo{pages}{1011}.

\bibitem[{\citenamefont{Keldysh}(1964)}]{Keldysh:1515}
\bibinfo{author}{\bibnamefont{Keldysh}, \bibfnamefont{L.~V.}},
  \bibinfo{year}{1964}, \bibinfo{journal}{Zh. Eksp. Teor. Fiz}
  \textbf{\bibinfo{volume}{47}}, \bibinfo{pages}{1515}.

\bibitem[{\citenamefont{Khomyakov and Brocks}(2004)}]{Khomyakov:195402}
\bibinfo{author}{\bibnamefont{Khomyakov}, \bibfnamefont{P.~A.}}, and
  \bibinfo{author}{\bibfnamefont{G.}~\bibnamefont{Brocks}},
  \bibinfo{year}{2004}, \bibinfo{journal}{Phys. Rev. B}
  \textbf{\bibinfo{volume}{70}}(\bibinfo{number}{19}), \bibinfo{pages}{195402}.

\bibitem[{\citenamefont{Kim} \emph{et~al.}(2008)\citenamefont{Kim, Chung, Won,
  Kwon, Lee, Park, and Choi}}]{Kim:203115}
\bibinfo{author}{\bibnamefont{Kim}, \bibfnamefont{K.}},
  \bibinfo{author}{\bibfnamefont{J.}~\bibnamefont{Chung}},
  \bibinfo{author}{\bibfnamefont{J.}~\bibnamefont{Won}},
  \bibinfo{author}{\bibfnamefont{O.}~\bibnamefont{Kwon}},
  \bibinfo{author}{\bibfnamefont{J.~S.} \bibnamefont{Lee}},
  \bibinfo{author}{\bibfnamefont{S.~H.} \bibnamefont{Park}}, and
  \bibinfo{author}{\bibfnamefont{Y.~K.} \bibnamefont{Choi}},
  \bibinfo{year}{2008}, \bibinfo{journal}{Appl. Phys. Lett.}
  \textbf{\bibinfo{volume}{93}}(\bibinfo{number}{20}), \bibinfo{eid}{203115}.

\bibitem[{\citenamefont{Kim} \emph{et~al.}(2001)\citenamefont{Kim, Shi,
  Majumdar, and McEuen}}]{Kim:215502}
\bibinfo{author}{\bibnamefont{Kim}, \bibfnamefont{P.}},
  \bibinfo{author}{\bibfnamefont{L.}~\bibnamefont{Shi}},
  \bibinfo{author}{\bibfnamefont{A.}~\bibnamefont{Majumdar}}, and
  \bibinfo{author}{\bibfnamefont{P.~L.} \bibnamefont{McEuen}},
  \bibinfo{year}{2001}, \bibinfo{journal}{Phys. Rev. Lett.}
  \textbf{\bibinfo{volume}{87}}(\bibinfo{number}{21}), \bibinfo{pages}{215502}.

\bibitem[{\citenamefont{Koch} \emph{et~al.}(2004)\citenamefont{Koch, von Oppen,
  Oreg, and Sela}}]{Koch:195107}
\bibinfo{author}{\bibnamefont{Koch}, \bibfnamefont{J.}},
  \bibinfo{author}{\bibfnamefont{F.}~\bibnamefont{von Oppen}},
  \bibinfo{author}{\bibfnamefont{Y.}~\bibnamefont{Oreg}}, and
  \bibinfo{author}{\bibfnamefont{E.}~\bibnamefont{Sela}}, \bibinfo{year}{2004},
  \bibinfo{journal}{Phys. Rev. B}
  \textbf{\bibinfo{volume}{70}}(\bibinfo{number}{19}), \bibinfo{pages}{195107}.

\bibitem[{\citenamefont{Koch} \emph{et~al.}(2006)\citenamefont{Koch,
  Semmelhack, von Oppen, and Nitzan}}]{Koch:155306}
\bibinfo{author}{\bibnamefont{Koch}, \bibfnamefont{J.}},
  \bibinfo{author}{\bibfnamefont{M.}~\bibnamefont{Semmelhack}},
  \bibinfo{author}{\bibfnamefont{F.}~\bibnamefont{von Oppen}}, and
  \bibinfo{author}{\bibfnamefont{A.}~\bibnamefont{Nitzan}},
  \bibinfo{year}{2006}, \bibinfo{journal}{Phys. Rev. B}
  \textbf{\bibinfo{volume}{73}}(\bibinfo{number}{15}), \bibinfo{eid}{155306}.

\bibitem[{\citenamefont{Kong} \emph{et~al.}(2005)\citenamefont{Kong, Lu, Zhu,
  Wei, and Wu}}]{Kong:1923}
\bibinfo{author}{\bibnamefont{Kong}, \bibfnamefont{W.~J.}},
  \bibinfo{author}{\bibfnamefont{L.}~\bibnamefont{Lu}},
  \bibinfo{author}{\bibfnamefont{H.~W.} \bibnamefont{Zhu}},
  \bibinfo{author}{\bibfnamefont{B.~Q.} \bibnamefont{Wei}}, and
  \bibinfo{author}{\bibfnamefont{D.~H.} \bibnamefont{Wu}},
  \bibinfo{year}{2005}, \bibinfo{journal}{Journal of Physics: Condensed Matter}
  \textbf{\bibinfo{volume}{17}}(\bibinfo{number}{12}), \bibinfo{pages}{1923}.

\bibitem[{\citenamefont{Kubala} \emph{et~al.}(2008)\citenamefont{Kubala,
  K\"{o}nig, and Pekola}}]{Kubala:066801}
\bibinfo{author}{\bibnamefont{Kubala}, \bibfnamefont{B.}},
  \bibinfo{author}{\bibfnamefont{J.}~\bibnamefont{K\"{o}nig}}, and
  \bibinfo{author}{\bibfnamefont{J.}~\bibnamefont{Pekola}},
  \bibinfo{year}{2008}, \bibinfo{journal}{Phys. Rev. Lett.}
  \textbf{\bibinfo{volume}{100}}(\bibinfo{number}{6}), \bibinfo{eid}{066801}.

\bibitem[{\citenamefont{Kundu} \emph{et~al.}(2009)\citenamefont{Kundu, Dhar,
  and Narayan}}]{Kundu2009}
\bibinfo{author}{\bibnamefont{Kundu}, \bibfnamefont{A.}},
  \bibinfo{author}{\bibfnamefont{A.}~\bibnamefont{Dhar}}, and
  \bibinfo{author}{\bibfnamefont{O.}~\bibnamefont{Narayan}},
  \bibinfo{year}{2009}, \bibinfo{journal}{Journal of Statistical Mechanics:
  Theory and Experiment} \textbf{\bibinfo{volume}{3}}(\bibinfo{number}{45}),
  \bibinfo{pages}{L03001}.

\bibitem[{\citenamefont{Lan} \emph{et~al.}(2009)\citenamefont{Lan, Wang, Gan,
  and Chin}}]{Lan:115401}
\bibinfo{author}{\bibnamefont{Lan}, \bibfnamefont{J.}},
  \bibinfo{author}{\bibfnamefont{J.-S.} \bibnamefont{Wang}},
  \bibinfo{author}{\bibfnamefont{C.~K.} \bibnamefont{Gan}}, and
  \bibinfo{author}{\bibfnamefont{S.~K.} \bibnamefont{Chin}},
  \bibinfo{year}{2009}, \bibinfo{journal}{Phys. Rev. B}
  \textbf{\bibinfo{volume}{79}}(\bibinfo{number}{11}), \bibinfo{eid}{115401}.

\bibitem[{\citenamefont{Landauer}(1957)}]{LANDAUER:223}
\bibinfo{author}{\bibnamefont{Landauer}, \bibfnamefont{R.}},
  \bibinfo{year}{1957}, \bibinfo{journal}{IBM journal of research and
  development} \textbf{\bibinfo{volume}{1}}(\bibinfo{number}{3}),
  \bibinfo{pages}{223}.

\bibitem[{\citenamefont{Landauer}(1970)}]{LANDAUER:863}
\bibinfo{author}{\bibnamefont{Landauer}, \bibfnamefont{R.}},
  \bibinfo{year}{1970}, \bibinfo{journal}{Philosophical Magazine}
  \textbf{\bibinfo{volume}{21}}(\bibinfo{number}{172}), \bibinfo{pages}{863}.

\bibitem[{\citenamefont{Lee} \emph{et~al.}(2009)\citenamefont{Lee, Yi, Zuev,
  and Kim}}]{Lee:022106}
\bibinfo{author}{\bibnamefont{Lee}, \bibfnamefont{C.-H.}},
  \bibinfo{author}{\bibfnamefont{G.-C.} \bibnamefont{Yi}},
  \bibinfo{author}{\bibfnamefont{Y.~M.} \bibnamefont{Zuev}}, and
  \bibinfo{author}{\bibfnamefont{P.}~\bibnamefont{Kim}}, \bibinfo{year}{2009},
  \bibinfo{journal}{Appl. Phys. Lett.}
  \textbf{\bibinfo{volume}{94}}(\bibinfo{number}{2}), \bibinfo{eid}{022106}.

\bibitem[{\citenamefont{Lepri} \emph{et~al.}(1997)\citenamefont{Lepri, Livi,
  and Politi}}]{Lepri:1896}
\bibinfo{author}{\bibnamefont{Lepri}, \bibfnamefont{S.}},
  \bibinfo{author}{\bibfnamefont{R.}~\bibnamefont{Livi}}, and
  \bibinfo{author}{\bibfnamefont{A.}~\bibnamefont{Politi}},
  \bibinfo{year}{1997}, \bibinfo{journal}{Phys. Rev. Lett.}
  \textbf{\bibinfo{volume}{78}}(\bibinfo{number}{10}), \bibinfo{pages}{1896}.

\bibitem[{\citenamefont{Lepri} \emph{et~al.}(2003)\citenamefont{Lepri, Livi,
  and Politi}}]{Lepri:1}
\bibinfo{author}{\bibnamefont{Lepri}, \bibfnamefont{S.}},
  \bibinfo{author}{\bibfnamefont{R.}~\bibnamefont{Livi}}, and
  \bibinfo{author}{\bibfnamefont{A.}~\bibnamefont{Politi}},
  \bibinfo{year}{2003}, \bibinfo{journal}{Physics Reports}
  \textbf{\bibinfo{volume}{377}}(\bibinfo{number}{1}), \bibinfo{pages}{1 }.

\bibitem[{\citenamefont{Li} \emph{et~al.}(2004{\natexlab{a}})\citenamefont{Li,
  Wang, and Casati}}]{Li:184301}
\bibinfo{author}{\bibnamefont{Li}, \bibfnamefont{B.}},
  \bibinfo{author}{\bibfnamefont{L.}~\bibnamefont{Wang}}, and
  \bibinfo{author}{\bibfnamefont{G.}~\bibnamefont{Casati}},
  \bibinfo{year}{2004}{\natexlab{a}}, \bibinfo{journal}{Phys. Rev. Lett.}
  \textbf{\bibinfo{volume}{93}}(\bibinfo{number}{18}), \bibinfo{pages}{184301}.

\bibitem[{\citenamefont{Li} \emph{et~al.}(2006)\citenamefont{Li, Wang, and
  Casati}}]{Li:143501}
\bibinfo{author}{\bibnamefont{Li}, \bibfnamefont{B.}},
  \bibinfo{author}{\bibfnamefont{L.}~\bibnamefont{Wang}}, and
  \bibinfo{author}{\bibfnamefont{G.}~\bibnamefont{Casati}},
  \bibinfo{year}{2006}, \bibinfo{journal}{Applied Physics Letters}
  \textbf{\bibinfo{volume}{88}}(\bibinfo{number}{14}), \bibinfo{eid}{143501},
  \urlprefix\url{http://link.aip.org/link/?APL/88/143501/1}.

\bibitem[{\citenamefont{Li} \emph{et~al.}(2002)\citenamefont{Li, Wang, and
  Hu}}]{Li:223901}
\bibinfo{author}{\bibnamefont{Li}, \bibfnamefont{B.}},
  \bibinfo{author}{\bibfnamefont{L.}~\bibnamefont{Wang}}, and
  \bibinfo{author}{\bibfnamefont{B.}~\bibnamefont{Hu}}, \bibinfo{year}{2002},
  \bibinfo{journal}{Phys. Rev. Lett.}
  \textbf{\bibinfo{volume}{88}}(\bibinfo{number}{22}), \bibinfo{pages}{223901}.

\bibitem[{\citenamefont{Li} \emph{et~al.}(2004{\natexlab{b}})\citenamefont{Li,
  Wang, and Casati}}]{Casati:184301}
\bibinfo{author}{\bibnamefont{Li}, \bibfnamefont{B.~W.}},
  \bibinfo{author}{\bibfnamefont{L.}~\bibnamefont{Wang}}, and
  \bibinfo{author}{\bibfnamefont{G.}~\bibnamefont{Casati}},
  \bibinfo{year}{2004}{\natexlab{b}}, \bibinfo{journal}{Phys. Rev. Lett.}
  \textbf{\bibinfo{volume}{93}}(\bibinfo{number}{18}), \bibinfo{pages}{184301}.

\bibitem[{\citenamefont{Li} \emph{et~al.}(2003)\citenamefont{Li, Wu, Kim, Shi,
  Yang, and Majumdar}}]{Li:2934}
\bibinfo{author}{\bibnamefont{Li}, \bibfnamefont{D.}},
  \bibinfo{author}{\bibfnamefont{Y.}~\bibnamefont{Wu}},
  \bibinfo{author}{\bibfnamefont{P.}~\bibnamefont{Kim}},
  \bibinfo{author}{\bibfnamefont{L.}~\bibnamefont{Shi}},
  \bibinfo{author}{\bibfnamefont{P.}~\bibnamefont{Yang}}, and
  \bibinfo{author}{\bibfnamefont{A.}~\bibnamefont{Majumdar}},
  \bibinfo{year}{2003}, \bibinfo{journal}{Appl. Phys. Lett.}
  \textbf{\bibinfo{volume}{83}}(\bibinfo{number}{14}), \bibinfo{pages}{2934}.

\bibitem[{\citenamefont{Li and Orignac}(2002)}]{Li:432}
\bibinfo{author}{\bibnamefont{Li}, \bibfnamefont{M.-R.}}, and
  \bibinfo{author}{\bibfnamefont{E.}~\bibnamefont{Orignac}},
  \bibinfo{year}{2002}, \bibinfo{journal}{EPL (Europhys. Lett.)}
  \textbf{\bibinfo{volume}{60}}(\bibinfo{number}{3}), \bibinfo{pages}{432}.

\bibitem[{\citenamefont{Liu and Yi}(2006)}]{Liu:3194}
\bibinfo{author}{\bibnamefont{Liu}, \bibfnamefont{H.-P.}}, and
  \bibinfo{author}{\bibfnamefont{L.}~\bibnamefont{Yi}}, \bibinfo{year}{2006},
  \bibinfo{journal}{Chin. Phys. Lett}
  \textbf{\bibinfo{volume}{23}}(\bibinfo{number}{12}), \bibinfo{pages}{3194}.

\bibitem[{\citenamefont{Liu}
  \emph{et~al.}(2009{\natexlab{a}})\citenamefont{Liu, Song, Sun, and
  Xie}}]{Liu:161309}
\bibinfo{author}{\bibnamefont{Liu}, \bibfnamefont{J.}},
  \bibinfo{author}{\bibfnamefont{J.}~\bibnamefont{Song}},
  \bibinfo{author}{\bibfnamefont{Q.-F.} \bibnamefont{Sun}}, and
  \bibinfo{author}{\bibfnamefont{X.~C.} \bibnamefont{Xie}},
  \bibinfo{year}{2009}{\natexlab{a}}, \bibinfo{journal}{Phys. Rev. B}
  \textbf{\bibinfo{volume}{79}}(\bibinfo{number}{16}), \bibinfo{eid}{161309}.

\bibitem[{\citenamefont{Liu and Chen}(2009)}]{Liu:193101}
\bibinfo{author}{\bibnamefont{Liu}, \bibfnamefont{Y.-S.}}, and
  \bibinfo{author}{\bibfnamefont{Y.-C.} \bibnamefont{Chen}},
  \bibinfo{year}{2009}, \bibinfo{journal}{Phys. Rev. B}
  \textbf{\bibinfo{volume}{79}}(\bibinfo{number}{19}), \bibinfo{eid}{193101}.

\bibitem[{\citenamefont{Liu}
  \emph{et~al.}(2009{\natexlab{b}})\citenamefont{Liu, Chen, and
  Chen}}]{Liu:3497}
\bibinfo{author}{\bibnamefont{Liu}, \bibfnamefont{Y.-S.}},
  \bibinfo{author}{\bibfnamefont{Y.-R.} \bibnamefont{Chen}}, and
  \bibinfo{author}{\bibfnamefont{Y.-C.} \bibnamefont{Chen}},
  \bibinfo{year}{2009}{\natexlab{b}}, \bibinfo{journal}{ACS NANO}
  \textbf{\bibinfo{volume}{3}}(\bibinfo{number}{11}), \bibinfo{pages}{3497}.

\bibitem[{\citenamefont{Lo} \emph{et~al.}(2008)\citenamefont{Lo, Wang, and
  Li}}]{Lo:054402}
\bibinfo{author}{\bibnamefont{Lo}, \bibfnamefont{W.~C.}},
  \bibinfo{author}{\bibfnamefont{L.}~\bibnamefont{Wang}}, and
  \bibinfo{author}{\bibfnamefont{B.}~\bibnamefont{Li}}, \bibinfo{year}{2008},
  \bibinfo{journal}{J. Phys. Soc. Jpn.} \textbf{\bibinfo{volume}{77}},
  \bibinfo{pages}{054402}.

\bibitem[{\citenamefont{L\"{u}} \emph{et~al.}(2010)\citenamefont{L\"{u}, Zhu,
  Zu, and Zhang}}]{Lu:123111}
\bibinfo{author}{\bibnamefont{L\"{u}}, \bibfnamefont{H.-F.}},
  \bibinfo{author}{\bibfnamefont{L.-C.} \bibnamefont{Zhu}},
  \bibinfo{author}{\bibfnamefont{X.-T.} \bibnamefont{Zu}}, and
  \bibinfo{author}{\bibfnamefont{H.-W.} \bibnamefont{Zhang}},
  \bibinfo{year}{2010}, \bibinfo{journal}{Applied Physics Letters}
  \textbf{\bibinfo{volume}{96}}(\bibinfo{number}{12}), \bibinfo{pages}{123111}.

\bibitem[{\citenamefont{L\"{u} and Wang}(2007)}]{Lu:165418}
\bibinfo{author}{\bibnamefont{L\"{u}}, \bibfnamefont{J.~T.}}, and
  \bibinfo{author}{\bibfnamefont{J.-S.} \bibnamefont{Wang}},
  \bibinfo{year}{2007}, \bibinfo{journal}{Phys. Rev. B}
  \textbf{\bibinfo{volume}{76}}(\bibinfo{number}{16}), \bibinfo{eid}{165418}.

\bibitem[{\citenamefont{L\"{u} and Wang}(2008)}]{Lu:235436}
\bibinfo{author}{\bibnamefont{L\"{u}}, \bibfnamefont{J.~T.}}, and
  \bibinfo{author}{\bibfnamefont{J.-S.} \bibnamefont{Wang}},
  \bibinfo{year}{2008}, \bibinfo{journal}{Phys. Rev. B}
  \textbf{\bibinfo{volume}{78}}(\bibinfo{number}{23}), \bibinfo{eid}{235436}.

\bibitem[{\citenamefont{Lu} \emph{et~al.}(2001)\citenamefont{Lu, Yi, and
  Zhang}}]{Lu:2996}
\bibinfo{author}{\bibnamefont{Lu}, \bibfnamefont{L.}},
  \bibinfo{author}{\bibfnamefont{W.}~\bibnamefont{Yi}}, and
  \bibinfo{author}{\bibfnamefont{D.~L.} \bibnamefont{Zhang}},
  \bibinfo{year}{2001}, \bibinfo{journal}{Review of Scientific Instruments}
  \textbf{\bibinfo{volume}{72}}(\bibinfo{number}{7}), \bibinfo{pages}{2996}.

\bibitem[{\citenamefont{Ludoph and Ruitenbeek}(1999)}]{Ludoph:12290}
\bibinfo{author}{\bibnamefont{Ludoph}, \bibfnamefont{B.}}, and
  \bibinfo{author}{\bibfnamefont{J.~M.~v.} \bibnamefont{Ruitenbeek}},
  \bibinfo{year}{1999}, \bibinfo{journal}{Phys. Rev. B}
  \textbf{\bibinfo{volume}{59}}(\bibinfo{number}{19}), \bibinfo{pages}{12290}.

\bibitem[{\citenamefont{Lunde and Flensberg}(2005)}]{Lunde:3879}
\bibinfo{author}{\bibnamefont{Lunde}, \bibfnamefont{A.~M.}}, and
  \bibinfo{author}{\bibfnamefont{K.}~\bibnamefont{Flensberg}},
  \bibinfo{year}{2005}, \bibinfo{journal}{Journal of Physics: Condensed Matter}
  \textbf{\bibinfo{volume}{17}}(\bibinfo{number}{25}), \bibinfo{pages}{3879}.

\bibitem[{\citenamefont{Lunde} \emph{et~al.}(2006)\citenamefont{Lunde,
  Flensberg, and Glazman}}]{Lunde:256802}
\bibinfo{author}{\bibnamefont{Lunde}, \bibfnamefont{A.~M.}},
  \bibinfo{author}{\bibfnamefont{K.}~\bibnamefont{Flensberg}}, and
  \bibinfo{author}{\bibfnamefont{L.~I.} \bibnamefont{Glazman}},
  \bibinfo{year}{2006}, \bibinfo{journal}{Phys. Rev. Lett.}
  \textbf{\bibinfo{volume}{97}}(\bibinfo{number}{25}), \bibinfo{eid}{256802}.

\bibitem[{\citenamefont{Luttinger}(1964)}]{Luttinger:1505}
\bibinfo{author}{\bibnamefont{Luttinger}, \bibfnamefont{J.~M.}},
  \bibinfo{year}{1964}, \bibinfo{journal}{Phys. Rev.}
  \textbf{\bibinfo{volume}{135}}(\bibinfo{number}{6A}), \bibinfo{pages}{A1505}.

\bibitem[{\citenamefont{Maci\'{a}}(2005)}]{Macia:254}
\bibinfo{author}{\bibnamefont{Maci\'{a}}, \bibfnamefont{E.}},
  \bibinfo{year}{2005}, \bibinfo{journal}{Nanotechnology}
  \textbf{\bibinfo{volume}{16}}(\bibinfo{number}{5}), \bibinfo{pages}{S254}.

\bibitem[{\citenamefont{Maci\'{a}}(2007)}]{Macia:035130}
\bibinfo{author}{\bibnamefont{Maci\'{a}}, \bibfnamefont{E.}},
  \bibinfo{year}{2007}, \bibinfo{journal}{Phys. Rev. B}
  \textbf{\bibinfo{volume}{75}}(\bibinfo{number}{3}), \bibinfo{eid}{035130}.

\bibitem[{\citenamefont{Mahan and Sofo}(1996)}]{Mahan:7436}
\bibinfo{author}{\bibnamefont{Mahan}, \bibfnamefont{G.~D.}}, and
  \bibinfo{author}{\bibfnamefont{J.~O.} \bibnamefont{Sofo}},
  \bibinfo{year}{1996}, \bibinfo{journal}{Proceedings of the National Academy
  of Sciences of the United States of America}
  \textbf{\bibinfo{volume}{93}}(\bibinfo{number}{15}), \bibinfo{pages}{7436}.

\bibitem[{\citenamefont{Majumdar}(2004)}]{Majumdar:02062004}
\bibinfo{author}{\bibnamefont{Majumdar}, \bibfnamefont{A.}},
  \bibinfo{year}{2004}, \bibinfo{journal}{Science}
  \textbf{\bibinfo{volume}{303}}(\bibinfo{number}{5659}), \bibinfo{pages}{777}.

\bibitem[{\citenamefont{Malen}
  \emph{et~al.}(2009{\natexlab{a}})\citenamefont{Malen, Doak, Baheti, Tilley,
  Majumdar, and Segalman}}]{Malen:3406}
\bibinfo{author}{\bibnamefont{Malen}, \bibfnamefont{J.~A.}},
  \bibinfo{author}{\bibfnamefont{P.}~\bibnamefont{Doak}},
  \bibinfo{author}{\bibfnamefont{K.}~\bibnamefont{Baheti}},
  \bibinfo{author}{\bibfnamefont{T.~D.} \bibnamefont{Tilley}},
  \bibinfo{author}{\bibfnamefont{A.}~\bibnamefont{Majumdar}}, and
  \bibinfo{author}{\bibfnamefont{R.~A.} \bibnamefont{Segalman}},
  \bibinfo{year}{2009}{\natexlab{a}}, \bibinfo{journal}{Nano Letters}
  \textbf{\bibinfo{volume}{9}}(\bibinfo{number}{10}), \bibinfo{pages}{3406}.

\bibitem[{\citenamefont{Malen}
  \emph{et~al.}(2009{\natexlab{b}})\citenamefont{Malen, Doak, Baheti, Tilley,
  Segalman, and Majumdar}}]{Malen:1164}
\bibinfo{author}{\bibnamefont{Malen}, \bibfnamefont{J.~A.}},
  \bibinfo{author}{\bibfnamefont{P.}~\bibnamefont{Doak}},
  \bibinfo{author}{\bibfnamefont{K.}~\bibnamefont{Baheti}},
  \bibinfo{author}{\bibfnamefont{T.~D.} \bibnamefont{Tilley}},
  \bibinfo{author}{\bibfnamefont{R.~A.} \bibnamefont{Segalman}}, and
  \bibinfo{author}{\bibfnamefont{A.}~\bibnamefont{Majumdar}},
  \bibinfo{year}{2009}{\natexlab{b}}, \bibinfo{journal}{Nano Lett.}
  \textbf{\bibinfo{volume}{9}}(\bibinfo{number}{3}), \bibinfo{pages}{1164}.

\bibitem[{\citenamefont{Maynard and Akkermans}(1985)}]{Maynard:5440}
\bibinfo{author}{\bibnamefont{Maynard}, \bibfnamefont{R.}}, and
  \bibinfo{author}{\bibfnamefont{E.}~\bibnamefont{Akkermans}},
  \bibinfo{year}{1985}, \bibinfo{journal}{Phys. Rev. B}
  \textbf{\bibinfo{volume}{32}}(\bibinfo{number}{8}), \bibinfo{pages}{5440}.

\bibitem[{\citenamefont{McEniry} \emph{et~al.}(2002)\citenamefont{McEniry,
  Todorov, and Dundas}}]{McEniry:195304}
\bibinfo{author}{\bibnamefont{McEniry}, \bibfnamefont{E.~J.}},
  \bibinfo{author}{\bibfnamefont{T.~N.} \bibnamefont{Todorov}}, and
  \bibinfo{author}{\bibfnamefont{D.}~\bibnamefont{Dundas}},
  \bibinfo{year}{2002}, \bibinfo{journal}{Journal of Physics: Condensed Matter}
  \textbf{\bibinfo{volume}{21}}(\bibinfo{number}{19}), \bibinfo{pages}{195304}.

\bibitem[{\citenamefont{Meir and Wingreen}(1992)}]{Meir:2512}
\bibinfo{author}{\bibnamefont{Meir}, \bibfnamefont{Y.}}, and
  \bibinfo{author}{\bibfnamefont{N.~S.} \bibnamefont{Wingreen}},
  \bibinfo{year}{1992}, \bibinfo{journal}{Phys. Rev. Lett.}
  \textbf{\bibinfo{volume}{68}}(\bibinfo{number}{16}), \bibinfo{pages}{2512}.

\bibitem[{\citenamefont{Mejia-Monasterio and Wichterich}(2007)}]{Mejia:113}
\bibinfo{author}{\bibnamefont{Mejia-Monasterio}, \bibfnamefont{C.}}, and
  \bibinfo{author}{\bibfnamefont{H.}~\bibnamefont{Wichterich}},
  \bibinfo{year}{2007}, \bibinfo{journal}{European Physical Journal-Special
  Topics} \textbf{\bibinfo{volume}{151}}, \bibinfo{pages}{113}.

\bibitem[{\citenamefont{Meschke} \emph{et~al.}(2006)\citenamefont{Meschke,
  Guichard, and Pekola}}]{Meschke:187}
\bibinfo{author}{\bibnamefont{Meschke}, \bibfnamefont{M.}},
  \bibinfo{author}{\bibfnamefont{W.}~\bibnamefont{Guichard}}, and
  \bibinfo{author}{\bibfnamefont{J.~P.} \bibnamefont{Pekola}},
  \bibinfo{year}{2006}, \bibinfo{journal}{Nature}
  \textbf{\bibinfo{volume}{444}}(\bibinfo{number}{7116}), \bibinfo{pages}{187}.

\bibitem[{\citenamefont{Michel} \emph{et~al.}(2006)\citenamefont{Michel,
  Gemmer, and Mahler}}]{Michel:4855}
\bibinfo{author}{\bibnamefont{Michel}, \bibfnamefont{M.}},
  \bibinfo{author}{\bibfnamefont{J.}~\bibnamefont{Gemmer}}, and
  \bibinfo{author}{\bibfnamefont{G.}~\bibnamefont{Mahler}},
  \bibinfo{year}{2006}, \bibinfo{journal}{International Journal of Modern
  Physics B} \textbf{\bibinfo{volume}{20}}(\bibinfo{number}{29}),
  \bibinfo{eid}{4855}.

\bibitem[{\citenamefont{Michel} \emph{et~al.}(2003)\citenamefont{Michel,
  Hartmann, Gemmer, and Mahler}}]{Michel:325}
\bibinfo{author}{\bibnamefont{Michel}, \bibfnamefont{M.}},
  \bibinfo{author}{\bibfnamefont{M.}~\bibnamefont{Hartmann}},
  \bibinfo{author}{\bibfnamefont{J.}~\bibnamefont{Gemmer}}, and
  \bibinfo{author}{\bibfnamefont{G.}~\bibnamefont{Mahler}},
  \bibinfo{year}{2003}, \bibinfo{journal}{European Physical Journal B}
  \textbf{\bibinfo{volume}{34}}(\bibinfo{number}{3}), \bibinfo{pages}{325}.

\bibitem[{\citenamefont{Mingo}(2006)}]{Mingo:125402}
\bibinfo{author}{\bibnamefont{Mingo}, \bibfnamefont{N.}}, \bibinfo{year}{2006},
  \bibinfo{journal}{Phys. Rev. B}
  \textbf{\bibinfo{volume}{74}}(\bibinfo{number}{12}), \bibinfo{eid}{125402}.

\bibitem[{\citenamefont{Mingo and Broido}(2005{\natexlab{a}})}]{Mingo:096105}
\bibinfo{author}{\bibnamefont{Mingo}, \bibfnamefont{N.}}, and
  \bibinfo{author}{\bibfnamefont{D.~A.} \bibnamefont{Broido}},
  \bibinfo{year}{2005}{\natexlab{a}}, \bibinfo{journal}{Phys. Rev. Lett.}
  \textbf{\bibinfo{volume}{95}}(\bibinfo{number}{9}), \bibinfo{pages}{096105}.

\bibitem[{\citenamefont{Mingo and Broido}(2005{\natexlab{b}})}]{Mingo:1221}
\bibinfo{author}{\bibnamefont{Mingo}, \bibfnamefont{N.}}, and
  \bibinfo{author}{\bibfnamefont{D.~A.} \bibnamefont{Broido}},
  \bibinfo{year}{2005}{\natexlab{b}}, \bibinfo{journal}{Nano Lett.}
  \textbf{\bibinfo{volume}{5}}(\bibinfo{number}{7}), \bibinfo{pages}{1221}.

\bibitem[{\citenamefont{Mingo and Yang}(2003)}]{Mingo:245406}
\bibinfo{author}{\bibnamefont{Mingo}, \bibfnamefont{N.}}, and
  \bibinfo{author}{\bibfnamefont{L.}~\bibnamefont{Yang}}, \bibinfo{year}{2003},
  \bibinfo{journal}{Phys. Rev. B}
  \textbf{\bibinfo{volume}{68}}(\bibinfo{number}{24}), \bibinfo{pages}{245406}.

\bibitem[{\citenamefont{Mingo} \emph{et~al.}(2003)\citenamefont{Mingo, Yang,
  Li, and Majumdar}}]{Mingo:1713}
\bibinfo{author}{\bibnamefont{Mingo}, \bibfnamefont{N.}},
  \bibinfo{author}{\bibfnamefont{L.}~\bibnamefont{Yang}},
  \bibinfo{author}{\bibfnamefont{D.}~\bibnamefont{Li}}, and
  \bibinfo{author}{\bibfnamefont{A.}~\bibnamefont{Majumdar}},
  \bibinfo{year}{2003}, \bibinfo{journal}{Nano Lett.}
  \textbf{\bibinfo{volume}{3}}(\bibinfo{number}{12}), \bibinfo{pages}{1713}.

\bibitem[{\citenamefont{Molenkamp} \emph{et~al.}(1994)\citenamefont{Molenkamp,
  Staring, Alphenaar, van Houten, and Beenakker}}]{Molenkamp:903}
\bibinfo{author}{\bibnamefont{Molenkamp}, \bibfnamefont{L.}},
  \bibinfo{author}{\bibfnamefont{A.~A.~M.} \bibnamefont{Staring}},
  \bibinfo{author}{\bibfnamefont{B.~W.} \bibnamefont{Alphenaar}},
  \bibinfo{author}{\bibfnamefont{H.}~\bibnamefont{van Houten}}, and
  \bibinfo{author}{\bibfnamefont{C.~W.~J.} \bibnamefont{Beenakker}},
  \bibinfo{year}{1994}, \bibinfo{journal}{Semiconductor Science and Technology}
  \textbf{\bibinfo{volume}{9}}(\bibinfo{number}{5S}), \bibinfo{pages}{903}.

\bibitem[{\citenamefont{Molenkamp} \emph{et~al.}(1990)\citenamefont{Molenkamp,
  van Houten, Beenakker, Eppenga, and Foxon}}]{Molenkamp:1052}
\bibinfo{author}{\bibnamefont{Molenkamp}, \bibfnamefont{L.~W.}},
  \bibinfo{author}{\bibfnamefont{H.}~\bibnamefont{van Houten}},
  \bibinfo{author}{\bibfnamefont{C.~W.~J.} \bibnamefont{Beenakker}},
  \bibinfo{author}{\bibfnamefont{R.}~\bibnamefont{Eppenga}}, and
  \bibinfo{author}{\bibfnamefont{C.~T.} \bibnamefont{Foxon}},
  \bibinfo{year}{1990}, \bibinfo{journal}{Phys. Rev. Lett.}
  \textbf{\bibinfo{volume}{65}}(\bibinfo{number}{8}), \bibinfo{pages}{1052}.

\bibitem[{\citenamefont{Mozyrsky} \emph{et~al.}(2006)\citenamefont{Mozyrsky,
  Hastings, and Martin}}]{Mozyrsky:035104}
\bibinfo{author}{\bibnamefont{Mozyrsky}, \bibfnamefont{D.}},
  \bibinfo{author}{\bibfnamefont{M.~B.} \bibnamefont{Hastings}}, and
  \bibinfo{author}{\bibfnamefont{I.}~\bibnamefont{Martin}},
  \bibinfo{year}{2006}, \bibinfo{journal}{Phys. Rev. B}
  \textbf{\bibinfo{volume}{73}}(\bibinfo{number}{3}), \bibinfo{eid}{035104}.

\bibitem[{\citenamefont{M\"{u}ller}(2008)}]{Muller:044708}
\bibinfo{author}{\bibnamefont{M\"{u}ller}, \bibfnamefont{K.-H.}},
  \bibinfo{year}{2008}, \bibinfo{journal}{The Journal of Chemical Physics}
  \textbf{\bibinfo{volume}{129}}(\bibinfo{number}{4}), \bibinfo{eid}{044708}.

\bibitem[{\citenamefont{Murphy} \emph{et~al.}(2008)\citenamefont{Murphy,
  Mukerjee, and Moore}}]{Murphy:161406}
\bibinfo{author}{\bibnamefont{Murphy}, \bibfnamefont{P.}},
  \bibinfo{author}{\bibfnamefont{S.}~\bibnamefont{Mukerjee}}, and
  \bibinfo{author}{\bibfnamefont{J.}~\bibnamefont{Moore}},
  \bibinfo{year}{2008}, \bibinfo{journal}{Phys. Rev. B}
  \textbf{\bibinfo{volume}{78}}(\bibinfo{number}{16}), \bibinfo{eid}{161406}.

\bibitem[{\citenamefont{Murphy and Moore}(2007)}]{Murphy:155313}
\bibinfo{author}{\bibnamefont{Murphy}, \bibfnamefont{P.~G.}}, and
  \bibinfo{author}{\bibfnamefont{J.~E.} \bibnamefont{Moore}},
  \bibinfo{year}{2007}, \bibinfo{journal}{Phys. Rev. B}
  \textbf{\bibinfo{volume}{76}}(\bibinfo{number}{15}), \bibinfo{eid}{155313}.

\bibitem[{\citenamefont{Nicholls and Chiatti}(2008)}]{Nicholls:164210}
\bibinfo{author}{\bibnamefont{Nicholls}, \bibfnamefont{J.~T.}}, and
  \bibinfo{author}{\bibfnamefont{O.}~\bibnamefont{Chiatti}},
  \bibinfo{year}{2008}, \bibinfo{journal}{Journal of Physics: Condensed Matter}
  \textbf{\bibinfo{volume}{20}}(\bibinfo{number}{16}), \bibinfo{pages}{164210}.

\bibitem[{\citenamefont{Nitzan and Ratner}(2003)}]{Nitzan:1384}
\bibinfo{author}{\bibnamefont{Nitzan}, \bibfnamefont{A.}}, and
  \bibinfo{author}{\bibfnamefont{M.~A.} \bibnamefont{Ratner}},
  \bibinfo{year}{2003}, \bibinfo{journal}{Science}
  \textbf{\bibinfo{volume}{300}}(\bibinfo{number}{5624}),
  \bibinfo{pages}{1384}.

\bibitem[{\citenamefont{Nos\'{e}}(1984)}]{Nose:511}
\bibinfo{author}{\bibnamefont{Nos\'{e}}, \bibfnamefont{S.}},
  \bibinfo{year}{1984}, \bibinfo{journal}{The Journal of Chemical Physics}
  \textbf{\bibinfo{volume}{81}}(\bibinfo{number}{1}), \bibinfo{pages}{511}.

\bibitem[{\citenamefont{Ozpineci and Ciraci}(2001)}]{Ozpineci:125415}
\bibinfo{author}{\bibnamefont{Ozpineci}, \bibfnamefont{A.}}, and
  \bibinfo{author}{\bibfnamefont{S.}~\bibnamefont{Ciraci}},
  \bibinfo{year}{2001}, \bibinfo{journal}{Phys. Rev. B}
  \textbf{\bibinfo{volume}{63}}(\bibinfo{number}{12}), \bibinfo{pages}{125415}.

\bibitem[{\citenamefont{Padgett and Brenner}(2004)}]{Padgett:1051}
\bibinfo{author}{\bibnamefont{Padgett}, \bibfnamefont{C.~W.}}, and
  \bibinfo{author}{\bibfnamefont{D.~W.} \bibnamefont{Brenner}},
  \bibinfo{year}{2004}, \bibinfo{journal}{Nano Lett.}
  \textbf{\bibinfo{volume}{4}}(\bibinfo{number}{6}), \bibinfo{pages}{1051}.

\bibitem[{\citenamefont{Padgett} \emph{et~al.}(2006)\citenamefont{Padgett,
  Shenderova, and Brenner}}]{Padgett:1827}
\bibinfo{author}{\bibnamefont{Padgett}, \bibfnamefont{C.~W.}},
  \bibinfo{author}{\bibfnamefont{O.}~\bibnamefont{Shenderova}}, and
  \bibinfo{author}{\bibfnamefont{D.~W.} \bibnamefont{Brenner}},
  \bibinfo{year}{2006}, \bibinfo{journal}{Nano Lett.}
  \textbf{\bibinfo{volume}{6}}(\bibinfo{number}{8}), \bibinfo{pages}{1827}.

\bibitem[{\citenamefont{Paulsson and Datta}(2003)}]{Paulsson:241403}
\bibinfo{author}{\bibnamefont{Paulsson}, \bibfnamefont{M.}}, and
  \bibinfo{author}{\bibfnamefont{S.}~\bibnamefont{Datta}},
  \bibinfo{year}{2003}, \bibinfo{journal}{Phys. Rev. B}
  \textbf{\bibinfo{volume}{67}}(\bibinfo{number}{24}), \bibinfo{pages}{241403}.

\bibitem[{\citenamefont{Pauly} \emph{et~al.}(2008)\citenamefont{Pauly, Viljas,
  and Cuevas}}]{Pauly:035315}
\bibinfo{author}{\bibnamefont{Pauly}, \bibfnamefont{F.}},
  \bibinfo{author}{\bibfnamefont{J.~K.} \bibnamefont{Viljas}}, and
  \bibinfo{author}{\bibfnamefont{J.~C.} \bibnamefont{Cuevas}},
  \bibinfo{year}{2008}, \bibinfo{journal}{Phys. Rev. B}
  \textbf{\bibinfo{volume}{78}}(\bibinfo{number}{3}), \bibinfo{eid}{035315}.

\bibitem[{\citenamefont{Peierls}(1929)}]{Peierls:1055}
\bibinfo{author}{\bibnamefont{Peierls}, \bibfnamefont{R.}},
  \bibinfo{year}{1929}, \bibinfo{journal}{Ann. Phys. (Berlin)}
  \textbf{\bibinfo{volume}{395}}, \bibinfo{pages}{1055}.

\bibitem[{\citenamefont{Peierls}(1955)}]{Peierls:1955}
\bibinfo{author}{\bibnamefont{Peierls}, \bibfnamefont{R.~E.}},
  \bibinfo{year}{1955}, \emph{\bibinfo{title}{Quantum Theory of Solids}}
  (\bibinfo{publisher}{Oxford University Press}).

\bibitem[{\citenamefont{Pekola} \emph{et~al.}(2004)\citenamefont{Pekola,
  Heikkil\"a, Savin, Flyktman, Giazotto, and Hekking}}]{Pekola:056804}
\bibinfo{author}{\bibnamefont{Pekola}, \bibfnamefont{J.~P.}},
  \bibinfo{author}{\bibfnamefont{T.~T.} \bibnamefont{Heikkil\"a}},
  \bibinfo{author}{\bibfnamefont{A.~M.} \bibnamefont{Savin}},
  \bibinfo{author}{\bibfnamefont{J.~T.} \bibnamefont{Flyktman}},
  \bibinfo{author}{\bibfnamefont{F.}~\bibnamefont{Giazotto}}, and
  \bibinfo{author}{\bibfnamefont{F.~W.~J.} \bibnamefont{Hekking}},
  \bibinfo{year}{2004}, \bibinfo{journal}{Phys. Rev. Lett.}
  \textbf{\bibinfo{volume}{92}}(\bibinfo{number}{5}), \bibinfo{pages}{056804}.

\bibitem[{\citenamefont{Pendry}(1983)}]{Pendry:2161}
\bibinfo{author}{\bibnamefont{Pendry}, \bibfnamefont{J.~B.}},
  \bibinfo{year}{1983}, \bibinfo{journal}{Journal of Physics A: Mathematical
  and General} \textbf{\bibinfo{volume}{16}}(\bibinfo{number}{10}),
  \bibinfo{pages}{2161}.

\bibitem[{\citenamefont{Peng} \emph{et~al.}(2007)\citenamefont{Peng, Chen, Zou,
  and Zhang}}]{Peng:193502}
\bibinfo{author}{\bibnamefont{Peng}, \bibfnamefont{X.-F.}},
  \bibinfo{author}{\bibfnamefont{K.-Q.} \bibnamefont{Chen}},
  \bibinfo{author}{\bibfnamefont{B.~S.} \bibnamefont{Zou}}, and
  \bibinfo{author}{\bibfnamefont{Y.}~\bibnamefont{Zhang}},
  \bibinfo{year}{2007}, \bibinfo{journal}{Appl. Phys. Lett.}
  \textbf{\bibinfo{volume}{90}}(\bibinfo{number}{19}), \bibinfo{eid}{193502}.

\bibitem[{\citenamefont{Pershin} \emph{et~al.}(2008)\citenamefont{Pershin,
  Dubi, and {Di Ventra}}}]{Pershin:054302}
\bibinfo{author}{\bibnamefont{Pershin}, \bibfnamefont{Y.~V.}},
  \bibinfo{author}{\bibfnamefont{Y.}~\bibnamefont{Dubi}}, and
  \bibinfo{author}{\bibfnamefont{M.}~\bibnamefont{{Di Ventra}}},
  \bibinfo{year}{2008}, \bibinfo{journal}{Phys. Rev. B}
  \textbf{\bibinfo{volume}{78}}(\bibinfo{number}{5}), \bibinfo{eid}{054302}.

\bibitem[{\citenamefont{Pistolesi}(2009)}]{Pistolesi:199}
\bibinfo{author}{\bibnamefont{Pistolesi}, \bibfnamefont{F.}},
  \bibinfo{year}{2009}, \bibinfo{journal}{Journal of Low Temperature Physics}
  \textbf{\bibinfo{volume}{154}}(\bibinfo{number}{5-6}), \bibinfo{pages}{199}.

\bibitem[{\citenamefont{Ponomareva}
  \emph{et~al.}(2007)\citenamefont{Ponomareva, Srivastava, and
  Menon}}]{Ponomareva:1155}
\bibinfo{author}{\bibnamefont{Ponomareva}, \bibfnamefont{I.}},
  \bibinfo{author}{\bibfnamefont{D.}~\bibnamefont{Srivastava}}, and
  \bibinfo{author}{\bibfnamefont{M.}~\bibnamefont{Menon}},
  \bibinfo{year}{2007}, \bibinfo{journal}{Nano Lett.}
  \textbf{\bibinfo{volume}{7}}(\bibinfo{number}{5}), \bibinfo{pages}{1155}.

\bibitem[{\citenamefont{Pop} \emph{et~al.}(2006)\citenamefont{Pop, Mann, Wang,
  Goodson, and Dai}}]{Pop:96}
\bibinfo{author}{\bibnamefont{Pop}, \bibfnamefont{E.}},
  \bibinfo{author}{\bibfnamefont{D.}~\bibnamefont{Mann}},
  \bibinfo{author}{\bibfnamefont{Q.}~\bibnamefont{Wang}},
  \bibinfo{author}{\bibfnamefont{K.}~\bibnamefont{Goodson}}, and
  \bibinfo{author}{\bibfnamefont{H.}~\bibnamefont{Dai}}, \bibinfo{year}{2006},
  \bibinfo{journal}{Nano Lett.}
  \textbf{\bibinfo{volume}{6}}(\bibinfo{number}{1}), \bibinfo{pages}{96}.

\bibitem[{\citenamefont{Reddy} \emph{et~al.}(2007)\citenamefont{Reddy, Jang,
  Segalman, and Majumdar}}]{Reddy:1568}
\bibinfo{author}{\bibnamefont{Reddy}, \bibfnamefont{P.}},
  \bibinfo{author}{\bibfnamefont{S.-Y.} \bibnamefont{Jang}},
  \bibinfo{author}{\bibfnamefont{R.~A.} \bibnamefont{Segalman}}, and
  \bibinfo{author}{\bibfnamefont{A.}~\bibnamefont{Majumdar}},
  \bibinfo{year}{2007}, \bibinfo{journal}{Science}
  \textbf{\bibinfo{volume}{315}}(\bibinfo{number}{5818}),
  \bibinfo{pages}{1568}.

\bibitem[{\citenamefont{Rego}(2001)}]{Rego:239}
\bibinfo{author}{\bibnamefont{Rego}, \bibfnamefont{L.~G.~C.}},
  \bibinfo{year}{2001}, \bibinfo{journal}{{Physica Status Solidi A-applied
  research}} \textbf{\bibinfo{volume}{187}}(\bibinfo{number}{1}),
  \bibinfo{pages}{239}.

\bibitem[{\citenamefont{Rego and Kirczenow}(1998)}]{Rego:232}
\bibinfo{author}{\bibnamefont{Rego}, \bibfnamefont{L.~G.~C.}}, and
  \bibinfo{author}{\bibfnamefont{G.}~\bibnamefont{Kirczenow}},
  \bibinfo{year}{1998}, \bibinfo{journal}{Phys. Rev. Lett.}
  \textbf{\bibinfo{volume}{81}}(\bibinfo{number}{1}), \bibinfo{pages}{232}.

\bibitem[{\citenamefont{Rejec} \emph{et~al.}(2002)\citenamefont{Rejec,
  Ram\ifmmode~\check{s}\else \v{s}\fi{}ak, and Jefferson}}]{Rejec:235301}
\bibinfo{author}{\bibnamefont{Rejec}, \bibfnamefont{T.}},
  \bibinfo{author}{\bibfnamefont{A.}~\bibnamefont{Ram\ifmmode~\check{s}\else
  \v{s}\fi{}ak}}, and \bibinfo{author}{\bibfnamefont{J.~H.}
  \bibnamefont{Jefferson}}, \bibinfo{year}{2002}, \bibinfo{journal}{Phys. Rev.
  B} \textbf{\bibinfo{volume}{65}}(\bibinfo{number}{23}),
  \bibinfo{pages}{235301}.

\bibitem[{\citenamefont{Ren} \emph{et~al.}(2006)\citenamefont{Ren, Cheng, and
  Chen}}]{Ren:103505}
\bibinfo{author}{\bibnamefont{Ren}, \bibfnamefont{S.-F.}},
  \bibinfo{author}{\bibfnamefont{W.}~\bibnamefont{Cheng}}, and
  \bibinfo{author}{\bibfnamefont{G.}~\bibnamefont{Chen}}, \bibinfo{year}{2006},
  \bibinfo{journal}{Journal of Applied Physics}
  \textbf{\bibinfo{volume}{100}}(\bibinfo{number}{10}),
  \bibinfo{pages}{103505}.

\bibitem[{\citenamefont{Rodgers}(2008)}]{Rodgers:1748}
\bibinfo{author}{\bibnamefont{Rodgers}, \bibfnamefont{P.}},
  \bibinfo{year}{2008}, \bibinfo{journal}{Nature Nanotechnology}
  \textbf{\bibinfo{volume}{3}}(\bibinfo{number}{2}), \bibinfo{pages}{76}.

\bibitem[{\citenamefont{Roy}(2008)}]{Roy:062102}
\bibinfo{author}{\bibnamefont{Roy}, \bibfnamefont{D.}}, \bibinfo{year}{2008},
  \bibinfo{journal}{Phys. Rev. E}
  \textbf{\bibinfo{volume}{77}}(\bibinfo{number}{6}), \bibinfo{eid}{062102}.

\bibitem[{\citenamefont{Roy and Dhar}(2008)}]{Dhar:2008}
\bibinfo{author}{\bibnamefont{Roy}, \bibfnamefont{D.}}, and
  \bibinfo{author}{\bibfnamefont{A.}~\bibnamefont{Dhar}}, \bibinfo{year}{2008},
  \bibinfo{journal}{Phys. Rev. E}
  \textbf{\bibinfo{volume}{78}}(\bibinfo{number}{4}), \bibinfo{pages}{051112}.

\bibitem[{\citenamefont{Runge and Gross}(1984)}]{Runge:997}
\bibinfo{author}{\bibnamefont{Runge}, \bibfnamefont{E.}}, and
  \bibinfo{author}{\bibfnamefont{E.~K.~U.} \bibnamefont{Gross}},
  \bibinfo{year}{1984}, \bibinfo{journal}{Phys. Rev. Lett.}
  \textbf{\bibinfo{volume}{52}}(\bibinfo{number}{12}), \bibinfo{pages}{997}.

\bibitem[{\citenamefont{Rurali}(2010)}]{Rurali:427}
\bibinfo{author}{\bibnamefont{Rurali}, \bibfnamefont{R.}},
  \bibinfo{year}{2010}, \bibinfo{journal}{Rev. Mod. Phys.}
  \textbf{\bibinfo{volume}{82}}(\bibinfo{number}{1}), \bibinfo{pages}{427}.

\bibitem[{\citenamefont{S.~Volz}(2009)}]{Volz:2009}
\bibinfo{author}{\bibnamefont{S.~Volz}, \bibfnamefont{e.}},
  \bibinfo{year}{2009}, \emph{\bibinfo{title}{Thermal Nanosystems and
  Nanomaterials (topics in applied physics)}} (\bibinfo{publisher}{Springer}).

\bibitem[{\citenamefont{Saira} \emph{et~al.}(2007)\citenamefont{Saira, Meschke,
  Giazotto, Savin, M\"{o}tt\"{o}nen, and Pekola}}]{saira:027203}
\bibinfo{author}{\bibnamefont{Saira}, \bibfnamefont{O.-P.}},
  \bibinfo{author}{\bibfnamefont{M.}~\bibnamefont{Meschke}},
  \bibinfo{author}{\bibfnamefont{F.}~\bibnamefont{Giazotto}},
  \bibinfo{author}{\bibfnamefont{A.~M.} \bibnamefont{Savin}},
  \bibinfo{author}{\bibfnamefont{M.}~\bibnamefont{M\"{o}tt\"{o}nen}}, and
  \bibinfo{author}{\bibfnamefont{J.~P.} \bibnamefont{Pekola}},
  \bibinfo{year}{2007}, \bibinfo{journal}{Phys. Rev. Lett.}
  \textbf{\bibinfo{volume}{99}}(\bibinfo{number}{2}), \bibinfo{eid}{027203}.

\bibitem[{\citenamefont{Saito}(2003)}]{Saito:34}
\bibinfo{author}{\bibnamefont{Saito}, \bibfnamefont{K.}}, \bibinfo{year}{2003},
  \bibinfo{journal}{Europhys. Lett.}
  \textbf{\bibinfo{volume}{61}}(\bibinfo{number}{1}), \bibinfo{pages}{34}.

\bibitem[{\citenamefont{Santamore and Cross}(2001)}]{Santamore:184306}
\bibinfo{author}{\bibnamefont{Santamore}, \bibfnamefont{D.~H.}}, and
  \bibinfo{author}{\bibfnamefont{M.~C.} \bibnamefont{Cross}},
  \bibinfo{year}{2001}, \bibinfo{journal}{Phys. Rev. B}
  \textbf{\bibinfo{volume}{63}}(\bibinfo{number}{18}), \bibinfo{pages}{184306}.

\bibitem[{\citenamefont{Savi\'{c}}
  \emph{et~al.}(2008{\natexlab{a}})\citenamefont{Savi\'{c}, Mingo, and
  Stewart}}]{Savic:165502}
\bibinfo{author}{\bibnamefont{Savi\'{c}}, \bibfnamefont{I.}},
  \bibinfo{author}{\bibfnamefont{N.}~\bibnamefont{Mingo}}, and
  \bibinfo{author}{\bibfnamefont{D.~A.} \bibnamefont{Stewart}},
  \bibinfo{year}{2008}{\natexlab{a}}, \bibinfo{journal}{Phys. Rev. Lett.}
  \textbf{\bibinfo{volume}{101}}(\bibinfo{number}{16}), \bibinfo{eid}{165502}.

\bibitem[{\citenamefont{Savi\'{c}}
  \emph{et~al.}(2008{\natexlab{b}})\citenamefont{Savi\'{c}, Stewart, and
  Mingo}}]{Savic:235434}
\bibinfo{author}{\bibnamefont{Savi\'{c}}, \bibfnamefont{I.}},
  \bibinfo{author}{\bibfnamefont{D.~A.} \bibnamefont{Stewart}}, and
  \bibinfo{author}{\bibfnamefont{N.}~\bibnamefont{Mingo}},
  \bibinfo{year}{2008}{\natexlab{b}}, \bibinfo{journal}{Phys. Rev. B}
  \textbf{\bibinfo{volume}{78}}(\bibinfo{number}{23}), \bibinfo{eid}{235434}.

\bibitem[{\citenamefont{Scheibner} \emph{et~al.}(2005)\citenamefont{Scheibner,
  Buhmann, Reuter, Kiselev, and Molenkamp}}]{Scheibner:176602}
\bibinfo{author}{\bibnamefont{Scheibner}, \bibfnamefont{R.}},
  \bibinfo{author}{\bibfnamefont{H.}~\bibnamefont{Buhmann}},
  \bibinfo{author}{\bibfnamefont{D.}~\bibnamefont{Reuter}},
  \bibinfo{author}{\bibfnamefont{M.~N.} \bibnamefont{Kiselev}}, and
  \bibinfo{author}{\bibfnamefont{L.~W.} \bibnamefont{Molenkamp}},
  \bibinfo{year}{2005}, \bibinfo{journal}{Phys. Rev. Lett.}
  \textbf{\bibinfo{volume}{95}}(\bibinfo{number}{17}), \bibinfo{pages}{176602}.

\bibitem[{\citenamefont{Scheibner} \emph{et~al.}(2007)\citenamefont{Scheibner,
  Novik, Borzenko, K\"{o}nig, Reuter, Wieck, Buhmann, and
  Molenkamp}}]{scheibner:041301}
\bibinfo{author}{\bibnamefont{Scheibner}, \bibfnamefont{R.}},
  \bibinfo{author}{\bibfnamefont{E.~G.} \bibnamefont{Novik}},
  \bibinfo{author}{\bibfnamefont{T.}~\bibnamefont{Borzenko}},
  \bibinfo{author}{\bibfnamefont{M.}~\bibnamefont{K\"{o}nig}},
  \bibinfo{author}{\bibfnamefont{D.}~\bibnamefont{Reuter}},
  \bibinfo{author}{\bibfnamefont{A.~D.} \bibnamefont{Wieck}},
  \bibinfo{author}{\bibfnamefont{H.}~\bibnamefont{Buhmann}}, and
  \bibinfo{author}{\bibfnamefont{L.~W.} \bibnamefont{Molenkamp}},
  \bibinfo{year}{2007}, \bibinfo{journal}{Phys. Rev. B}
  \textbf{\bibinfo{volume}{75}}(\bibinfo{number}{4}), \bibinfo{pages}{041301}.

\bibitem[{\citenamefont{Schmidt} \emph{et~al.}(2004)\citenamefont{Schmidt,
  Schoelkopf, and Cleland}}]{Schmidt:045901}
\bibinfo{author}{\bibnamefont{Schmidt}, \bibfnamefont{D.~R.}},
  \bibinfo{author}{\bibfnamefont{R.~J.} \bibnamefont{Schoelkopf}}, and
  \bibinfo{author}{\bibfnamefont{A.~N.} \bibnamefont{Cleland}},
  \bibinfo{year}{2004}, \bibinfo{journal}{Phys. Rev. Lett.}
  \textbf{\bibinfo{volume}{93}}(\bibinfo{number}{4}), \bibinfo{pages}{045901}.

\bibitem[{\citenamefont{Schulze} \emph{et~al.}(2008)\citenamefont{Schulze,
  Franke, Gagliardi, Romano, Lin, Rosa, Niehaus, Frauenheim, {Di Carlo},
  Pecchia, and Pascual}}]{Schulze:136801}
\bibinfo{author}{\bibnamefont{Schulze}, \bibfnamefont{G.}},
  \bibinfo{author}{\bibfnamefont{K.~J.} \bibnamefont{Franke}},
  \bibinfo{author}{\bibfnamefont{A.}~\bibnamefont{Gagliardi}},
  \bibinfo{author}{\bibfnamefont{G.}~\bibnamefont{Romano}},
  \bibinfo{author}{\bibfnamefont{C.~S.} \bibnamefont{Lin}},
  \bibinfo{author}{\bibfnamefont{A.~L.} \bibnamefont{Rosa}},
  \bibinfo{author}{\bibfnamefont{T.~A.} \bibnamefont{Niehaus}},
  \bibinfo{author}{\bibfnamefont{T.}~\bibnamefont{Frauenheim}},
  \bibinfo{author}{\bibfnamefont{A.}~\bibnamefont{{Di Carlo}}},
  \bibinfo{author}{\bibfnamefont{A.}~\bibnamefont{Pecchia}}, and
  \bibinfo{author}{\bibfnamefont{J.~I.} \bibnamefont{Pascual}},
  \bibinfo{year}{2008}, \bibinfo{journal}{Phys. Rev. Lett.}
  \textbf{\bibinfo{volume}{100}}(\bibinfo{number}{13}), \bibinfo{eid}{136801}.

\bibitem[{\citenamefont{Schwab} \emph{et~al.}(2000)\citenamefont{Schwab,
  Henriksen, Worlock, and Roukes}}]{Schwab:974}
\bibinfo{author}{\bibnamefont{Schwab}, \bibfnamefont{K.}},
  \bibinfo{author}{\bibfnamefont{E.}~\bibnamefont{Henriksen}},
  \bibinfo{author}{\bibfnamefont{J.}~\bibnamefont{Worlock}}, and
  \bibinfo{author}{\bibfnamefont{M.}~\bibnamefont{Roukes}},
  \bibinfo{year}{2000}, \bibinfo{journal}{Nature}
  \textbf{\bibinfo{volume}{404}}(\bibinfo{number}{6781}), \bibinfo{pages}{974}.

\bibitem[{\citenamefont{Segal}(2005)}]{Segal:165426}
\bibinfo{author}{\bibnamefont{Segal}, \bibfnamefont{D.}}, \bibinfo{year}{2005},
  \bibinfo{journal}{Phys. Rev. B}
  \textbf{\bibinfo{volume}{72}}(\bibinfo{number}{16}), \bibinfo{pages}{165426}.

\bibitem[{\citenamefont{Segal and Nitzan}(2005)}]{segal:034301}
\bibinfo{author}{\bibnamefont{Segal}, \bibfnamefont{D.}}, and
  \bibinfo{author}{\bibfnamefont{A.}~\bibnamefont{Nitzan}},
  \bibinfo{year}{2005}, \bibinfo{journal}{Phys. Rev. Lett.}
  \textbf{\bibinfo{volume}{94}}(\bibinfo{number}{3}), \bibinfo{eid}{034301}.

\bibitem[{\citenamefont{Segal} \emph{et~al.}(2003)\citenamefont{Segal, Nitzan,
  and H\"{a}nggi}}]{Segal:6840}
\bibinfo{author}{\bibnamefont{Segal}, \bibfnamefont{D.}},
  \bibinfo{author}{\bibfnamefont{A.}~\bibnamefont{Nitzan}}, and
  \bibinfo{author}{\bibfnamefont{P.}~\bibnamefont{H\"{a}nggi}},
  \bibinfo{year}{2003}, \bibinfo{journal}{The Journal of Chemical Physics}
  \textbf{\bibinfo{volume}{119}}(\bibinfo{number}{13}), \bibinfo{pages}{6840}.

\bibitem[{\citenamefont{Seol} \emph{et~al.}(2007)\citenamefont{Seol, Moore,
  Saha, Zhou, Shi, Ye, Scheffler, Mingo, and Yamada}}]{Seol:023706}
\bibinfo{author}{\bibnamefont{Seol}, \bibfnamefont{J.~H.}},
  \bibinfo{author}{\bibfnamefont{A.~L.} \bibnamefont{Moore}},
  \bibinfo{author}{\bibfnamefont{S.~K.} \bibnamefont{Saha}},
  \bibinfo{author}{\bibfnamefont{F.}~\bibnamefont{Zhou}},
  \bibinfo{author}{\bibfnamefont{L.}~\bibnamefont{Shi}},
  \bibinfo{author}{\bibfnamefont{Q.~L.} \bibnamefont{Ye}},
  \bibinfo{author}{\bibfnamefont{R.}~\bibnamefont{Scheffler}},
  \bibinfo{author}{\bibfnamefont{N.}~\bibnamefont{Mingo}}, and
  \bibinfo{author}{\bibfnamefont{T.}~\bibnamefont{Yamada}},
  \bibinfo{year}{2007}, \bibinfo{journal}{Journal of Applied Physics}
  \textbf{\bibinfo{volume}{101}}(\bibinfo{number}{2}), \bibinfo{eid}{023706}.

\bibitem[{\citenamefont{Shakouri}(2006)}]{Shakouri:1705146}
\bibinfo{author}{\bibnamefont{Shakouri}, \bibfnamefont{A.}},
  \bibinfo{year}{2006}, \bibinfo{journal}{Proceedings of the IEEE}
  \textbf{\bibinfo{volume}{94}}(\bibinfo{number}{8}), \bibinfo{pages}{1613}.

\bibitem[{\citenamefont{Shi} \emph{et~al.}(2003)\citenamefont{Shi, Li, Yu,
  Jang, Kim, Yao, Kim, and Majumdar}}]{Shi:881}
\bibinfo{author}{\bibnamefont{Shi}, \bibfnamefont{L.}},
  \bibinfo{author}{\bibfnamefont{D.}~\bibnamefont{Li}},
  \bibinfo{author}{\bibfnamefont{C.}~\bibnamefont{Yu}},
  \bibinfo{author}{\bibfnamefont{W.}~\bibnamefont{Jang}},
  \bibinfo{author}{\bibfnamefont{D.}~\bibnamefont{Kim}},
  \bibinfo{author}{\bibfnamefont{Z.}~\bibnamefont{Yao}},
  \bibinfo{author}{\bibfnamefont{P.}~\bibnamefont{Kim}}, and
  \bibinfo{author}{\bibfnamefont{A.}~\bibnamefont{Majumdar}},
  \bibinfo{year}{2003}, \bibinfo{journal}{Journal of Heat Transfer}
  \textbf{\bibinfo{volume}{125}}(\bibinfo{number}{5}), \bibinfo{pages}{881}.

\bibitem[{\citenamefont{Small} \emph{et~al.}(2003)\citenamefont{Small, Perez,
  and Kim}}]{Small:256801}
\bibinfo{author}{\bibnamefont{Small}, \bibfnamefont{J.~P.}},
  \bibinfo{author}{\bibfnamefont{K.~M.} \bibnamefont{Perez}}, and
  \bibinfo{author}{\bibfnamefont{P.}~\bibnamefont{Kim}}, \bibinfo{year}{2003},
  \bibinfo{journal}{Phys. Rev. Lett.}
  \textbf{\bibinfo{volume}{91}}(\bibinfo{number}{25}), \bibinfo{pages}{256801}.

\bibitem[{\citenamefont{Staring} \emph{et~al.}(1993)\citenamefont{Staring,
  Molenkamp, Alphenaar, van Houten, Buyk, Mabesoone, Beenakker, and
  Foxon}}]{Staring:57}
\bibinfo{author}{\bibnamefont{Staring}, \bibfnamefont{A.~A.~M.}},
  \bibinfo{author}{\bibfnamefont{L.~W.} \bibnamefont{Molenkamp}},
  \bibinfo{author}{\bibfnamefont{B.~W.} \bibnamefont{Alphenaar}},
  \bibinfo{author}{\bibfnamefont{H.}~\bibnamefont{van Houten}},
  \bibinfo{author}{\bibfnamefont{O.~J.~A.} \bibnamefont{Buyk}},
  \bibinfo{author}{\bibfnamefont{M.~A.~A.} \bibnamefont{Mabesoone}},
  \bibinfo{author}{\bibfnamefont{C.~W.~J.} \bibnamefont{Beenakker}}, and
  \bibinfo{author}{\bibfnamefont{C.~T.} \bibnamefont{Foxon}},
  \bibinfo{year}{1993}, \bibinfo{journal}{Europhys. Lett.}
  \textbf{\bibinfo{volume}{22}}(\bibinfo{number}{1}), \bibinfo{pages}{57}.

\bibitem[{\citenamefont{Stewart} \emph{et~al.}(2009)\citenamefont{Stewart,
  Savi\'{c}, and Mingo}}]{Stewart:81}
\bibinfo{author}{\bibnamefont{Stewart}, \bibfnamefont{D.~A.}},
  \bibinfo{author}{\bibfnamefont{I.}~\bibnamefont{Savi\'{c}}}, and
  \bibinfo{author}{\bibfnamefont{N.}~\bibnamefont{Mingo}},
  \bibinfo{year}{2009}, \bibinfo{journal}{Nano Lett.}
  \textbf{\bibinfo{volume}{9}}(\bibinfo{number}{1}), \bibinfo{pages}{81}.

\bibitem[{\citenamefont{Sumanasekera}
  \emph{et~al.}(2002)\citenamefont{Sumanasekera, Pradhan, Romero, Adu, and
  Eklund}}]{Sumanasekera:166801}
\bibinfo{author}{\bibnamefont{Sumanasekera}, \bibfnamefont{G.~U.}},
  \bibinfo{author}{\bibfnamefont{B.~K.} \bibnamefont{Pradhan}},
  \bibinfo{author}{\bibfnamefont{H.~E.} \bibnamefont{Romero}},
  \bibinfo{author}{\bibfnamefont{K.~W.} \bibnamefont{Adu}}, and
  \bibinfo{author}{\bibfnamefont{P.~C.} \bibnamefont{Eklund}},
  \bibinfo{year}{2002}, \bibinfo{journal}{Phys. Rev. Lett.}
  \textbf{\bibinfo{volume}{89}}(\bibinfo{number}{16}), \bibinfo{pages}{166801}.

\bibitem[{\citenamefont{Tan} \emph{et~al.}(2010)\citenamefont{Tan, Sadat, and
  Reddy}}]{tan:013110}
\bibinfo{author}{\bibnamefont{Tan}, \bibfnamefont{A.}},
  \bibinfo{author}{\bibfnamefont{S.}~\bibnamefont{Sadat}}, and
  \bibinfo{author}{\bibfnamefont{P.}~\bibnamefont{Reddy}},
  \bibinfo{year}{2010}, \bibinfo{journal}{Applied Physics Letters}
  \textbf{\bibinfo{volume}{96}}(\bibinfo{number}{1}), \bibinfo{eid}{013110}.

\bibitem[{\citenamefont{Tanaka} \emph{et~al.}(2005)\citenamefont{Tanaka,
  Yoshida, and Tamura}}]{Tanaka:205308}
\bibinfo{author}{\bibnamefont{Tanaka}, \bibfnamefont{Y.}},
  \bibinfo{author}{\bibfnamefont{F.}~\bibnamefont{Yoshida}}, and
  \bibinfo{author}{\bibfnamefont{S.}~\bibnamefont{Tamura}},
  \bibinfo{year}{2005}, \bibinfo{journal}{Phys. Rev. B}
  \textbf{\bibinfo{volume}{71}}(\bibinfo{number}{20}), \bibinfo{pages}{205308}.

\bibitem[{\citenamefont{Tang} \emph{et~al.}(2007)\citenamefont{Tang, Wang,
  Huang, Zou, and Chen}}]{Tang:1497}
\bibinfo{author}{\bibnamefont{Tang}, \bibfnamefont{L.-M.}},
  \bibinfo{author}{\bibfnamefont{L.}~\bibnamefont{Wang}},
  \bibinfo{author}{\bibfnamefont{W.-Q.} \bibnamefont{Huang}},
  \bibinfo{author}{\bibfnamefont{B.~S.} \bibnamefont{Zou}}, and
  \bibinfo{author}{\bibfnamefont{K.-Q.} \bibnamefont{Chen}},
  \bibinfo{year}{2007}, \bibinfo{journal}{Journal of Physics D: Applied
  Physics} \textbf{\bibinfo{volume}{40}}(\bibinfo{number}{5}),
  \bibinfo{pages}{1497}.

\bibitem[{\citenamefont{Tang} \emph{et~al.}(2006)\citenamefont{Tang, Wang,
  Chen, Huang, and Zou}}]{Tang:163505}
\bibinfo{author}{\bibnamefont{Tang}, \bibfnamefont{L.-M.}},
  \bibinfo{author}{\bibfnamefont{L.-L.} \bibnamefont{Wang}},
  \bibinfo{author}{\bibfnamefont{K.-Q.} \bibnamefont{Chen}},
  \bibinfo{author}{\bibfnamefont{W.-Q.} \bibnamefont{Huang}}, and
  \bibinfo{author}{\bibfnamefont{B.~S.} \bibnamefont{Zou}},
  \bibinfo{year}{2006}, \bibinfo{journal}{Appl. Phys. Lett.}
  \textbf{\bibinfo{volume}{88}}(\bibinfo{number}{16}), \bibinfo{eid}{163505}.

\bibitem[{\citenamefont{Teramae} \emph{et~al.}(2008)\citenamefont{Teramae,
  Horiguchi, Hashimoto, Tsutsui, Kurokawa, and Sakai}}]{Teramae:083121}
\bibinfo{author}{\bibnamefont{Teramae}, \bibfnamefont{Y.}},
  \bibinfo{author}{\bibfnamefont{K.}~\bibnamefont{Horiguchi}},
  \bibinfo{author}{\bibfnamefont{S.}~\bibnamefont{Hashimoto}},
  \bibinfo{author}{\bibfnamefont{M.}~\bibnamefont{Tsutsui}},
  \bibinfo{author}{\bibfnamefont{S.}~\bibnamefont{Kurokawa}}, and
  \bibinfo{author}{\bibfnamefont{A.}~\bibnamefont{Sakai}},
  \bibinfo{year}{2008}, \bibinfo{journal}{Appl. Phys. Lett.}
  \textbf{\bibinfo{volume}{93}}(\bibinfo{number}{8}), \bibinfo{eid}{083121}.

\bibitem[{\citenamefont{Terraneo} \emph{et~al.}(2002)\citenamefont{Terraneo,
  Peyrard, and Casati}}]{Terraneo:094302}
\bibinfo{author}{\bibnamefont{Terraneo}, \bibfnamefont{M.}},
  \bibinfo{author}{\bibfnamefont{M.}~\bibnamefont{Peyrard}}, and
  \bibinfo{author}{\bibfnamefont{G.}~\bibnamefont{Casati}},
  \bibinfo{year}{2002}, \bibinfo{journal}{Phys. Rev. Lett.}
  \textbf{\bibinfo{volume}{88}}(\bibinfo{number}{9}), \bibinfo{pages}{094302}.

\bibitem[{\citenamefont{Ting} \emph{et~al.}(1992)\citenamefont{Ting, Yu, and
  McGill}}]{Ting:3583}
\bibinfo{author}{\bibnamefont{Ting}, \bibfnamefont{D.~Z.~Y.}},
  \bibinfo{author}{\bibfnamefont{E.~T.} \bibnamefont{Yu}}, and
  \bibinfo{author}{\bibfnamefont{T.~C.} \bibnamefont{McGill}},
  \bibinfo{year}{1992}, \bibinfo{journal}{Phys. Rev. B}
  \textbf{\bibinfo{volume}{45}}(\bibinfo{number}{7}), \bibinfo{pages}{3583}.

\bibitem[{\citenamefont{Todorov}(1998)}]{Todorov:965}
\bibinfo{author}{\bibnamefont{Todorov}, \bibfnamefont{T.}},
  \bibinfo{year}{1998}, \bibinfo{journal}{Philosophical Magazine B}
  \textbf{\bibinfo{volume}{77}}(\bibinfo{number}{4}), \bibinfo{pages}{965}.

\bibitem[{\citenamefont{Tong} \emph{et~al.}(1999)\citenamefont{Tong, Li, and
  Hu}}]{Tong:8639}
\bibinfo{author}{\bibnamefont{Tong}, \bibfnamefont{P.}},
  \bibinfo{author}{\bibfnamefont{B.}~\bibnamefont{Li}}, and
  \bibinfo{author}{\bibfnamefont{B.}~\bibnamefont{Hu}}, \bibinfo{year}{1999},
  \bibinfo{journal}{Phys. Rev. B}
  \textbf{\bibinfo{volume}{59}}(\bibinfo{number}{13}), \bibinfo{pages}{8639}.

\bibitem[{\citenamefont{Tsutsui} \emph{et~al.}(2007)\citenamefont{Tsutsui,
  Kurokawa, and Sakai}}]{Tsutsui:133121}
\bibinfo{author}{\bibnamefont{Tsutsui}, \bibfnamefont{M.}},
  \bibinfo{author}{\bibfnamefont{S.}~\bibnamefont{Kurokawa}}, and
  \bibinfo{author}{\bibfnamefont{A.}~\bibnamefont{Sakai}},
  \bibinfo{year}{2007}, \bibinfo{journal}{Appl. Phys. Lett.}
  \textbf{\bibinfo{volume}{90}}(\bibinfo{number}{8}), \bibinfo{eid}{133121}.

\bibitem[{\citenamefont{Tsutsui}
  \emph{et~al.}(2008{\natexlab{a}})\citenamefont{Tsutsui, Shoji, Morimoto,
  Taniguchi, and Kawai}}]{Tsutsui:223110}
\bibinfo{author}{\bibnamefont{Tsutsui}, \bibfnamefont{M.}},
  \bibinfo{author}{\bibfnamefont{K.}~\bibnamefont{Shoji}},
  \bibinfo{author}{\bibfnamefont{K.}~\bibnamefont{Morimoto}},
  \bibinfo{author}{\bibfnamefont{M.}~\bibnamefont{Taniguchi}}, and
  \bibinfo{author}{\bibfnamefont{T.}~\bibnamefont{Kawai}},
  \bibinfo{year}{2008}{\natexlab{a}}, \bibinfo{journal}{Appl. Phys. Lett.}
  \textbf{\bibinfo{volume}{92}}(\bibinfo{number}{22}), \bibinfo{eid}{223110}.

\bibitem[{\citenamefont{Tsutsui}
  \emph{et~al.}(2008{\natexlab{b}})\citenamefont{Tsutsui, Taniguchi, and
  Kawai}}]{Tsutsui:3293}
\bibinfo{author}{\bibnamefont{Tsutsui}, \bibfnamefont{M.}},
  \bibinfo{author}{\bibfnamefont{M.}~\bibnamefont{Taniguchi}}, and
  \bibinfo{author}{\bibfnamefont{T.}~\bibnamefont{Kawai}},
  \bibinfo{year}{2008}{\natexlab{b}}, \bibinfo{journal}{Nano Lett.}
  \textbf{\bibinfo{volume}{8}}(\bibinfo{number}{10}), \bibinfo{pages}{3293}.

\bibitem[{\citenamefont{Turek and Matveev}(2002)}]{Turek:115332}
\bibinfo{author}{\bibnamefont{Turek}, \bibfnamefont{M.}}, and
  \bibinfo{author}{\bibfnamefont{K.~A.} \bibnamefont{Matveev}},
  \bibinfo{year}{2002}, \bibinfo{journal}{Phys. Rev. B}
  \textbf{\bibinfo{volume}{65}}(\bibinfo{number}{11}), \bibinfo{pages}{115332}.

\bibitem[{\citenamefont{Turek} \emph{et~al.}(2005)\citenamefont{Turek, Siewert,
  and Richter}}]{Turek:220503}
\bibinfo{author}{\bibnamefont{Turek}, \bibfnamefont{M.}},
  \bibinfo{author}{\bibfnamefont{J.}~\bibnamefont{Siewert}}, and
  \bibinfo{author}{\bibfnamefont{K.}~\bibnamefont{Richter}},
  \bibinfo{year}{2005}, \bibinfo{journal}{Phys. Rev. B}
  \textbf{\bibinfo{volume}{71}}(\bibinfo{number}{22}), \bibinfo{pages}{220503}.

\bibitem[{\citenamefont{Turney} \emph{et~al.}(2009)\citenamefont{Turney,
  Landry, McGaughey, and Amon}}]{Turney:064301}
\bibinfo{author}{\bibnamefont{Turney}, \bibfnamefont{J.~E.}},
  \bibinfo{author}{\bibfnamefont{E.~S.} \bibnamefont{Landry}},
  \bibinfo{author}{\bibfnamefont{A.~J.~H.} \bibnamefont{McGaughey}}, and
  \bibinfo{author}{\bibfnamefont{C.~H.} \bibnamefont{Amon}},
  \bibinfo{year}{2009}, \bibinfo{journal}{Phys. Rev. B}
  \textbf{\bibinfo{volume}{79}}(\bibinfo{number}{6}), \bibinfo{pages}{064301}.

\bibitem[{\citenamefont{Uchida} \emph{et~al.}(2008)\citenamefont{Uchida,
  Takahashi, Harii, Ieda, Koshibae, Ando, Maekawa, and Saitoh}}]{Uchida:778}
\bibinfo{author}{\bibnamefont{Uchida}, \bibfnamefont{K.}},
  \bibinfo{author}{\bibfnamefont{S.}~\bibnamefont{Takahashi}},
  \bibinfo{author}{\bibfnamefont{K.}~\bibnamefont{Harii}},
  \bibinfo{author}{\bibfnamefont{J.}~\bibnamefont{Ieda}},
  \bibinfo{author}{\bibfnamefont{W.}~\bibnamefont{Koshibae}},
  \bibinfo{author}{\bibfnamefont{K.}~\bibnamefont{Ando}},
  \bibinfo{author}{\bibfnamefont{S.}~\bibnamefont{Maekawa}}, and
  \bibinfo{author}{\bibfnamefont{E.}~\bibnamefont{Saitoh}},
  \bibinfo{year}{2008}, \bibinfo{journal}{Nature}
  \textbf{\bibinfo{volume}{455}}(\bibinfo{number}{7214}), \bibinfo{pages}{778}.

\bibitem[{\citenamefont{Uchida} \emph{et~al.}(2009)\citenamefont{Uchida,
  Takahashi, Ieda, Harii, Ikeda, Koshibae, Maekawa, and
  Saitoh}}]{Uchida:07C908}
\bibinfo{author}{\bibnamefont{Uchida}, \bibfnamefont{K.}},
  \bibinfo{author}{\bibfnamefont{S.}~\bibnamefont{Takahashi}},
  \bibinfo{author}{\bibfnamefont{J.}~\bibnamefont{Ieda}},
  \bibinfo{author}{\bibfnamefont{K.}~\bibnamefont{Harii}},
  \bibinfo{author}{\bibfnamefont{K.}~\bibnamefont{Ikeda}},
  \bibinfo{author}{\bibfnamefont{W.}~\bibnamefont{Koshibae}},
  \bibinfo{author}{\bibfnamefont{S.}~\bibnamefont{Maekawa}}, and
  \bibinfo{author}{\bibfnamefont{E.}~\bibnamefont{Saitoh}},
  \bibinfo{year}{2009} (\bibinfo{publisher}{AIP}), volume
  \bibinfo{volume}{105}, p. \bibinfo{pages}{07C908}.

\bibitem[{\citenamefont{USDOE}(2009)}]{DOE}
\bibinfo{author}{\bibnamefont{USDOE}}, \bibinfo{year}{2009},
  \bibinfo{title}{Advanced thermoelectric materials for efficient waste heat
  recovery in process industries},
  \urlprefix\url{http://www1.eere.energy.gov/}.

\bibitem[{\citenamefont{{Van Kampen}}(2001)}]{VanKampen:2001}
\bibinfo{author}{\bibnamefont{{Van Kampen}}, \bibfnamefont{N.~G.}},
  \bibinfo{year}{2001}, \emph{\bibinfo{title}{Stochastic Processes in Physics
  and Chemistry}} (\bibinfo{publisher}{North Holland}).

\bibitem[{\citenamefont{Vignale and {Di Ventra}}(2009)}]{Vignale2009}
\bibinfo{author}{\bibnamefont{Vignale}, \bibfnamefont{G.}}, and
  \bibinfo{author}{\bibfnamefont{M.}~\bibnamefont{{Di Ventra}}},
  \bibinfo{year}{2009}, \bibinfo{journal}{Phys. Rev. B}
  \textbf{\bibinfo{volume}{79}}(\bibinfo{number}{1}), \bibinfo{eid}{014201}.

\bibitem[{\citenamefont{Wang} \emph{et~al.}(2005)\citenamefont{Wang, Xing, Wan,
  Wei, and Wang}}]{Wang:233406}
\bibinfo{author}{\bibnamefont{Wang}, \bibfnamefont{B.}},
  \bibinfo{author}{\bibfnamefont{Y.}~\bibnamefont{Xing}},
  \bibinfo{author}{\bibfnamefont{L.}~\bibnamefont{Wan}},
  \bibinfo{author}{\bibfnamefont{Y.}~\bibnamefont{Wei}}, and
  \bibinfo{author}{\bibfnamefont{J.}~\bibnamefont{Wang}}, \bibinfo{year}{2005},
  \bibinfo{journal}{Phys. Rev. B}
  \textbf{\bibinfo{volume}{71}}(\bibinfo{number}{23}), \bibinfo{pages}{233406}.

\bibitem[{\citenamefont{Wang and Wang}(2006)}]{wang:054303}
\bibinfo{author}{\bibnamefont{Wang}, \bibfnamefont{J.}}, and
  \bibinfo{author}{\bibfnamefont{J.-S.} \bibnamefont{Wang}},
  \bibinfo{year}{2006}, \bibinfo{journal}{Phys. Rev. B}
  \textbf{\bibinfo{volume}{74}}(\bibinfo{number}{5}), \bibinfo{eid}{054303}.

\bibitem[{\citenamefont{Wang and Wang}(2007)}]{Wang:241908}
\bibinfo{author}{\bibnamefont{Wang}, \bibfnamefont{J.}}, and
  \bibinfo{author}{\bibfnamefont{J.-S.} \bibnamefont{Wang}},
  \bibinfo{year}{2007}, \bibinfo{journal}{Appl. Phys. Lett.}
  \textbf{\bibinfo{volume}{90}}(\bibinfo{number}{24}), \bibinfo{pages}{241908}.

\bibitem[{\citenamefont{Wang} \emph{et~al.}(2008)\citenamefont{Wang, Wang, and
  Lü}}]{Wang:381}
\bibinfo{author}{\bibnamefont{Wang}, \bibfnamefont{J.-S.}},
  \bibinfo{author}{\bibfnamefont{J.}~\bibnamefont{Wang}}, and
  \bibinfo{author}{\bibfnamefont{J.~T.} \bibnamefont{Lü}},
  \bibinfo{year}{2008}, \bibinfo{journal}{The European Physical Journal B -
  Condensed Matter and Complex Systems}
  \textbf{\bibinfo{volume}{62}}(\bibinfo{number}{4}), \bibinfo{pages}{381}.

\bibitem[{\citenamefont{Wang} \emph{et~al.}(2006)\citenamefont{Wang, Wang, and
  Zeng}}]{Wang:033408}
\bibinfo{author}{\bibnamefont{Wang}, \bibfnamefont{J.-S.}},
  \bibinfo{author}{\bibfnamefont{J.}~\bibnamefont{Wang}}, and
  \bibinfo{author}{\bibfnamefont{N.}~\bibnamefont{Zeng}}, \bibinfo{year}{2006},
  \bibinfo{journal}{Phys. Rev. B}
  \textbf{\bibinfo{volume}{74}}(\bibinfo{number}{3}), \bibinfo{eid}{033408}.

\bibitem[{\citenamefont{Wang} \emph{et~al.}(2007)\citenamefont{Wang, Zeng,
  Wang, and Gan}}]{Wang:061128}
\bibinfo{author}{\bibnamefont{Wang}, \bibfnamefont{J.-S.}},
  \bibinfo{author}{\bibfnamefont{N.}~\bibnamefont{Zeng}},
  \bibinfo{author}{\bibfnamefont{J.}~\bibnamefont{Wang}}, and
  \bibinfo{author}{\bibfnamefont{C.~K.} \bibnamefont{Gan}},
  \bibinfo{year}{2007}, \bibinfo{journal}{Phys. Rev. E}
  \textbf{\bibinfo{volume}{75}}(\bibinfo{number}{6}), \bibinfo{eid}{061128}.

\bibitem[{\citenamefont{Wang and Li}(2007)}]{Wang:177208}
\bibinfo{author}{\bibnamefont{Wang}, \bibfnamefont{L.}}, and
  \bibinfo{author}{\bibfnamefont{B.}~\bibnamefont{Li}}, \bibinfo{year}{2007},
  \bibinfo{journal}{Phys. Rev. Lett.}
  \textbf{\bibinfo{volume}{99}}(\bibinfo{number}{17}), \bibinfo{pages}{177208}.

\bibitem[{\citenamefont{Wang and Li}(2008)}]{Wang:267203}
\bibinfo{author}{\bibnamefont{Wang}, \bibfnamefont{L.}}, and
  \bibinfo{author}{\bibfnamefont{B.}~\bibnamefont{Li}}, \bibinfo{year}{2008},
  \bibinfo{journal}{Phys. Rev. Lett.}
  \textbf{\bibinfo{volume}{101}}(\bibinfo{number}{26}),
  \bibinfo{pages}{267203}.

\bibitem[{\citenamefont{Wang} \emph{et~al.}(2010)\citenamefont{Wang, Sheng,
  Shen, Wang, and Xing}}]{Wang:057202}
\bibinfo{author}{\bibnamefont{Wang}, \bibfnamefont{R.-Q.}},
  \bibinfo{author}{\bibfnamefont{L.}~\bibnamefont{Sheng}},
  \bibinfo{author}{\bibfnamefont{R.}~\bibnamefont{Shen}},
  \bibinfo{author}{\bibfnamefont{B.}~\bibnamefont{Wang}}, and
  \bibinfo{author}{\bibfnamefont{D.~Y.} \bibnamefont{Xing}},
  \bibinfo{year}{2010}, \bibinfo{journal}{Phys. Rev. Lett.}
  \textbf{\bibinfo{volume}{105}}(\bibinfo{number}{5}), \bibinfo{pages}{057202}.

\bibitem[{\citenamefont{Ward} \emph{et~al.}(2008)\citenamefont{Ward, Halas, and
  Natelson}}]{Ward:213108}
\bibinfo{author}{\bibnamefont{Ward}, \bibfnamefont{D.~R.}},
  \bibinfo{author}{\bibfnamefont{N.~J.} \bibnamefont{Halas}}, and
  \bibinfo{author}{\bibfnamefont{D.}~\bibnamefont{Natelson}},
  \bibinfo{year}{2008}, \bibinfo{journal}{Applied Physics Letters}
  \textbf{\bibinfo{volume}{93}}(\bibinfo{number}{21}), \bibinfo{eid}{213108}.

\bibitem[{\citenamefont{van Wees} \emph{et~al.}(1988)\citenamefont{van Wees,
  van Houten, Beenakker, Williamson, Kouwenhoven, van~der Marel, and
  Foxon}}]{Wees:848}
\bibinfo{author}{\bibnamefont{van Wees}, \bibfnamefont{B.~J.}},
  \bibinfo{author}{\bibfnamefont{H.}~\bibnamefont{van Houten}},
  \bibinfo{author}{\bibfnamefont{C.~W.~J.} \bibnamefont{Beenakker}},
  \bibinfo{author}{\bibfnamefont{J.~G.} \bibnamefont{Williamson}},
  \bibinfo{author}{\bibfnamefont{L.~P.} \bibnamefont{Kouwenhoven}},
  \bibinfo{author}{\bibfnamefont{D.}~\bibnamefont{van~der Marel}}, and
  \bibinfo{author}{\bibfnamefont{C.~T.} \bibnamefont{Foxon}},
  \bibinfo{year}{1988}, \bibinfo{journal}{Phys. Rev. Lett.}
  \textbf{\bibinfo{volume}{60}}(\bibinfo{number}{9}), \bibinfo{pages}{848}.

\bibitem[{\citenamefont{Wu and Li}(2007)}]{Wu:085424}
\bibinfo{author}{\bibnamefont{Wu}, \bibfnamefont{G.}}, and
  \bibinfo{author}{\bibfnamefont{B.}~\bibnamefont{Li}}, \bibinfo{year}{2007},
  \bibinfo{journal}{Phys. Rev. B}
  \textbf{\bibinfo{volume}{76}}(\bibinfo{number}{8}), \bibinfo{pages}{085424}.

\bibitem[{\citenamefont{Wu and Segal}(2008)}]{Wu:060101}
\bibinfo{author}{\bibnamefont{Wu}, \bibfnamefont{L.-A.}}, and
  \bibinfo{author}{\bibfnamefont{D.}~\bibnamefont{Segal}},
  \bibinfo{year}{2008}, \bibinfo{journal}{Physical Review E (Statistical,
  Nonlinear, and Soft Matter Physics)}
  \textbf{\bibinfo{volume}{77}}(\bibinfo{number}{6}), \bibinfo{pages}{060101}.

\bibitem[{\citenamefont{Wu and Segal}(2009)}]{Wu:025302}
\bibinfo{author}{\bibnamefont{Wu}, \bibfnamefont{L.-A.}}, and
  \bibinfo{author}{\bibfnamefont{D.}~\bibnamefont{Segal}},
  \bibinfo{year}{2009}, \bibinfo{journal}{J. Phys. A: Math. Theor.}
  \textbf{\bibinfo{volume}{42}}(\bibinfo{number}{2}), \bibinfo{pages}{025302}.

\bibitem[{\citenamefont{Xie} \emph{et~al.}(2008)\citenamefont{Xie, Chen, Wang,
  and Zhang}}]{Xie:084501}
\bibinfo{author}{\bibnamefont{Xie}, \bibfnamefont{F.}},
  \bibinfo{author}{\bibfnamefont{K.-Q.} \bibnamefont{Chen}},
  \bibinfo{author}{\bibfnamefont{Y.~G.} \bibnamefont{Wang}}, and
  \bibinfo{author}{\bibfnamefont{Y.}~\bibnamefont{Zhang}},
  \bibinfo{year}{2008}, \bibinfo{journal}{Journal of Applied Physics}
  \textbf{\bibinfo{volume}{103}}(\bibinfo{number}{8}), \bibinfo{eid}{084501}.

\bibitem[{\citenamefont{Xu} \emph{et~al.}(2008)\citenamefont{Xu, Wang, Duan,
  Gu, and Li}}]{Xu:224303}
\bibinfo{author}{\bibnamefont{Xu}, \bibfnamefont{Y.}},
  \bibinfo{author}{\bibfnamefont{J.-S.} \bibnamefont{Wang}},
  \bibinfo{author}{\bibfnamefont{W.}~\bibnamefont{Duan}},
  \bibinfo{author}{\bibfnamefont{B.-L.} \bibnamefont{Gu}}, and
  \bibinfo{author}{\bibfnamefont{B.}~\bibnamefont{Li}}, \bibinfo{year}{2008},
  \bibinfo{journal}{Phys. Rev. B}
  \textbf{\bibinfo{volume}{78}}(\bibinfo{number}{22}), \bibinfo{eid}{224303}.

\bibitem[{\citenamefont{Yamamoto and Watanabe}(2006)}]{Yamamoto:255503}
\bibinfo{author}{\bibnamefont{Yamamoto}, \bibfnamefont{T.}}, and
  \bibinfo{author}{\bibfnamefont{K.}~\bibnamefont{Watanabe}},
  \bibinfo{year}{2006}, \bibinfo{journal}{Phys. Rev. Lett.}
  \textbf{\bibinfo{volume}{96}}(\bibinfo{number}{25}), \bibinfo{eid}{255503}.

\bibitem[{\citenamefont{Yamamoto} \emph{et~al.}(2004)\citenamefont{Yamamoto,
  Watanabe, and Watanabe}}]{Yamamoto:075502}
\bibinfo{author}{\bibnamefont{Yamamoto}, \bibfnamefont{T.}},
  \bibinfo{author}{\bibfnamefont{S.}~\bibnamefont{Watanabe}}, and
  \bibinfo{author}{\bibfnamefont{K.}~\bibnamefont{Watanabe}},
  \bibinfo{year}{2004}, \bibinfo{journal}{Phys. Rev. Lett.}
  \textbf{\bibinfo{volume}{92}}(\bibinfo{number}{7}), \bibinfo{pages}{075502}.

\bibitem[{\citenamefont{Yang} \emph{et~al.}(2008)\citenamefont{Yang, Zhang, and
  Li}}]{Yang:276}
\bibinfo{author}{\bibnamefont{Yang}, \bibfnamefont{N.}},
  \bibinfo{author}{\bibfnamefont{G.}~\bibnamefont{Zhang}}, and
  \bibinfo{author}{\bibfnamefont{B.}~\bibnamefont{Li}}, \bibinfo{year}{2008},
  \bibinfo{journal}{Nano Lett.}
  \textbf{\bibinfo{volume}{8}}(\bibinfo{number}{1}), \bibinfo{pages}{276}.

\bibitem[{\citenamefont{Yang} \emph{et~al.}(2009)\citenamefont{Yang, Zhang, and
  Li}}]{Yang:033107}
\bibinfo{author}{\bibnamefont{Yang}, \bibfnamefont{N.}},
  \bibinfo{author}{\bibfnamefont{G.}~\bibnamefont{Zhang}}, and
  \bibinfo{author}{\bibfnamefont{B.}~\bibnamefont{Li}}, \bibinfo{year}{2009},
  \bibinfo{journal}{Applied Physics Letters}
  \textbf{\bibinfo{volume}{95}}(\bibinfo{number}{3}), \bibinfo{eid}{033107}.

\bibitem[{\citenamefont{Yang} \emph{et~al.}(2010)\citenamefont{Yang, Zhang, and
  Li}}]{Yang:201085}
\bibinfo{author}{\bibnamefont{Yang}, \bibfnamefont{N.}},
  \bibinfo{author}{\bibfnamefont{G.}~\bibnamefont{Zhang}}, and
  \bibinfo{author}{\bibfnamefont{B.}~\bibnamefont{Li}}, \bibinfo{year}{2010},
  \bibinfo{journal}{Nano Today}
  \textbf{\bibinfo{volume}{5}}(\bibinfo{number}{2}), \bibinfo{pages}{85 }.

\bibitem[{\citenamefont{Yang} \emph{et~al.}(2005)\citenamefont{Yang, Chshiev,
  Zwolak, Chen, and {Di Ventra}}}]{Yang:041402}
\bibinfo{author}{\bibnamefont{Yang}, \bibfnamefont{Z.}},
  \bibinfo{author}{\bibfnamefont{M.}~\bibnamefont{Chshiev}},
  \bibinfo{author}{\bibfnamefont{M.}~\bibnamefont{Zwolak}},
  \bibinfo{author}{\bibfnamefont{Y.-C.} \bibnamefont{Chen}}, and
  \bibinfo{author}{\bibfnamefont{M.}~\bibnamefont{{Di Ventra}}},
  \bibinfo{year}{2005}, \bibinfo{journal}{Phys. Rev. B}
  \textbf{\bibinfo{volume}{71}}(\bibinfo{number}{12}), \bibinfo{pages}{041402}.

\bibitem[{\citenamefont{Ying and Jin}(2010)}]{Ying:093104}
\bibinfo{author}{\bibnamefont{Ying}, \bibfnamefont{Y.}}, and
  \bibinfo{author}{\bibfnamefont{G.}~\bibnamefont{Jin}}, \bibinfo{year}{2010},
  \bibinfo{journal}{Applied Physics Letters}
  \textbf{\bibinfo{volume}{96}}(\bibinfo{number}{9}), \bibinfo{pages}{093104}.

\bibitem[{\citenamefont{Yu} \emph{et~al.}(2005)\citenamefont{Yu, Shi, Yao, Li,
  and Majumdar}}]{Yu:1842}
\bibinfo{author}{\bibnamefont{Yu}, \bibfnamefont{C.}},
  \bibinfo{author}{\bibfnamefont{L.}~\bibnamefont{Shi}},
  \bibinfo{author}{\bibfnamefont{Z.}~\bibnamefont{Yao}},
  \bibinfo{author}{\bibfnamefont{D.}~\bibnamefont{Li}}, and
  \bibinfo{author}{\bibfnamefont{A.}~\bibnamefont{Majumdar}},
  \bibinfo{year}{2005}, \bibinfo{journal}{Nano Lett.}
  \textbf{\bibinfo{volume}{5}}(\bibinfo{number}{9}), \bibinfo{pages}{1842}.

\bibitem[{\citenamefont{Yudson and Kravtsov}(2003)}]{Yudson:155310}
\bibinfo{author}{\bibnamefont{Yudson}, \bibfnamefont{V.~I.}}, and
  \bibinfo{author}{\bibfnamefont{V.~E.} \bibnamefont{Kravtsov}},
  \bibinfo{year}{2003}, \bibinfo{journal}{Phys. Rev. B}
  \textbf{\bibinfo{volume}{67}}(\bibinfo{number}{15}), \bibinfo{pages}{155310}.

\bibitem[{\citenamefont{Zhang and Li}(2005)}]{Zhang:114714}
\bibinfo{author}{\bibnamefont{Zhang}, \bibfnamefont{G.}}, and
  \bibinfo{author}{\bibfnamefont{B.}~\bibnamefont{Li}}, \bibinfo{year}{2005},
  \bibinfo{journal}{The Journal of Chemical Physics}
  \textbf{\bibinfo{volume}{123}}(\bibinfo{number}{11}),
  \bibinfo{pages}{114714}.

\bibitem[{\citenamefont{Zhang}(2007)}]{Zhang:086214}
\bibinfo{author}{\bibnamefont{Zhang}, \bibfnamefont{Z.-Y.}},
  \bibinfo{year}{2007}, \bibinfo{journal}{Journal of Physics: Condensed Matter}
  \textbf{\bibinfo{volume}{19}}(\bibinfo{number}{8}), \bibinfo{pages}{086214}.

\bibitem[{\citenamefont{Zheng} \emph{et~al.}(2004)\citenamefont{Zheng, Zheng,
  Wei, Zeng, and Wang}}]{Zheng:8537}
\bibinfo{author}{\bibnamefont{Zheng}, \bibfnamefont{X.}},
  \bibinfo{author}{\bibfnamefont{W.}~\bibnamefont{Zheng}},
  \bibinfo{author}{\bibfnamefont{Y.}~\bibnamefont{Wei}},
  \bibinfo{author}{\bibfnamefont{Z.}~\bibnamefont{Zeng}}, and
  \bibinfo{author}{\bibfnamefont{J.}~\bibnamefont{Wang}}, \bibinfo{year}{2004},
  \bibinfo{journal}{The Journal of Chemical Physics}
  \textbf{\bibinfo{volume}{121}}(\bibinfo{number}{17}), \bibinfo{pages}{8537}.

\bibitem[{\citenamefont{Zhernov and Chulkin}(2000)}]{Zhernov:308}
\bibinfo{author}{\bibnamefont{Zhernov}, \bibfnamefont{A.}}, and
  \bibinfo{author}{\bibfnamefont{E.}~\bibnamefont{Chulkin}},
  \bibinfo{year}{2000}, \bibinfo{journal}{Journal of Experimental and
  Theoretical Physics} \textbf{\bibinfo{volume}{90}}, \bibinfo{pages}{308}.

\bibitem[{\citenamefont{Zhong and Lukes}(2006)}]{Zhong:125403}
\bibinfo{author}{\bibnamefont{Zhong}, \bibfnamefont{H.}}, and
  \bibinfo{author}{\bibfnamefont{J.~R.} \bibnamefont{Lukes}},
  \bibinfo{year}{2006}, \bibinfo{journal}{Phys. Rev. B}
  \textbf{\bibinfo{volume}{74}}(\bibinfo{number}{12}), \bibinfo{pages}{125403}.

\bibitem[{\citenamefont{Zimmermann}
  \emph{et~al.}(2008)\citenamefont{Zimmermann, Pavone, and
  Cuniberti}}]{Zimmermann:045410}
\bibinfo{author}{\bibnamefont{Zimmermann}, \bibfnamefont{J.}},
  \bibinfo{author}{\bibfnamefont{P.}~\bibnamefont{Pavone}}, and
  \bibinfo{author}{\bibfnamefont{G.}~\bibnamefont{Cuniberti}},
  \bibinfo{year}{2008}, \bibinfo{journal}{Phys. Rev. B}
  \textbf{\bibinfo{volume}{78}}(\bibinfo{number}{4}), \bibinfo{eid}{045410}.

\bibitem[{\citenamefont{Zippilli} \emph{et~al.}(2009)\citenamefont{Zippilli,
  Morigi, and Bachtold}}]{Zippilli:096804}
\bibinfo{author}{\bibnamefont{Zippilli}, \bibfnamefont{S.}},
  \bibinfo{author}{\bibfnamefont{G.}~\bibnamefont{Morigi}}, and
  \bibinfo{author}{\bibfnamefont{A.}~\bibnamefont{Bachtold}},
  \bibinfo{year}{2009}, \bibinfo{journal}{Phys. Rev. Lett.}
  \textbf{\bibinfo{volume}{102}}(\bibinfo{number}{9}), \bibinfo{eid}{096804}.

\bibitem[{\citenamefont{\^Zuti\'c} \emph{et~al.}(2004)\citenamefont{\^Zuti\'c,
  Fabian, and {Das Sarma}}}]{Zutic}
\bibinfo{author}{\bibnamefont{\^Zuti\'c}, \bibfnamefont{I.}},
  \bibinfo{author}{\bibfnamefont{J.}~\bibnamefont{Fabian}}, and
  \bibinfo{author}{\bibfnamefont{S.}~\bibnamefont{{Das Sarma}}},
  \bibinfo{year}{2004}, \bibinfo{journal}{Rev. Mod. Phys.}
  \textbf{\bibinfo{volume}{76}}(\bibinfo{number}{1}), \bibinfo{pages}{323}.

\bibitem[{\citenamefont{Zwolak and {Di Ventra}}(2008)}]{Zwolak:141}
\bibinfo{author}{\bibnamefont{Zwolak}, \bibfnamefont{M.}}, and
  \bibinfo{author}{\bibfnamefont{M.}~\bibnamefont{{Di Ventra}}},
  \bibinfo{year}{2008}, \bibinfo{journal}{Rev. Mod. Phys.}
  \textbf{\bibinfo{volume}{80}}(\bibinfo{number}{1}), \bibinfo{pages}{141}.

\end{thebibliography}

\end{document}